\documentclass[twocolumn, reprint, amsmath, amssymb, prb ,superscriptaddress, floatfix, sort, compress, numbers, linenumbers]{revtex4-1}

\usepackage{graphicx}
\usepackage{dcolumn}
\usepackage{bm}
\usepackage[colorlinks=true,
 linkcolor=black,
 anchorcolor=black,
 citecolor=black,
 urlcolor=blue
 ]{hyperref}

\usepackage{verbatim}
\usepackage{braket}
\usepackage{siunitx}
\usepackage{makecell}
\usepackage{xcolor}
\usepackage{lineno}

\newcommand{\be}{\begin{equation}}
\newcommand{\ee}{\end{equation}}
\newcommand{\bs}{\begin{split}}
\newcommand{\es}{\end{split}}

\newcommand{\LL}[1]{\textcolor[RGB]{0,0,0}{#1}}

\renewcommand{\figurename}{Figure}

\begin{document}

\setlength{\abovecaptionskip}{0pt}
\setlength{\belowcaptionskip}{0pt}

\nolinenumbers
\title{Observation of quantum information collapse-and-revival in a strongly-interacting Rydberg atom array}

\author
{De-Sheng Xiang$^{1,\ast}$, Yao-Wen Zhang$^{1,\ast}$, Hao-Xiang Liu$^{1,\ast}$, Peng Zhou$^{1,\ast}$, Dong Yuan$^{2,\ast}$, Kuan Zhang$^{1,\ast}$, Shun-Yao Zhang$^{3}$, Biao Xu$^{1}$, Lu Liu$^{1}$, Yitong Li$^{1}$, and Lin Li$^{1,\dagger}$\\
\normalsize{$^{1}$\textit{MOE Key Laboratory of Fundamental Physical Quantities Measurement,}}
\normalsize{\textit{Hubei Key Laboratory of Gravitation and Quantum Physics, PGMF,}}\\
\normalsize{\textit{Institute for Quantum Science and Engineering, School of Physics,}}\\
\normalsize{\textit{Huazhong University of Science and Technology, Wuhan 430074, China}}\\
\normalsize{$^{2}$\textit{Center for Quantum Information, IIIS, Tsinghua University, Beijing 100084, China}}\\
\normalsize{$^{3}$\textit{Lingang Laboratory, Shanghai 200031, China}}\\
\normalsize{$^\ast$ These authors contributed equally to this work.}\\
\normalsize{$^\dagger$ E-mail: li\_lin@hust.edu.cn}
}

\noindent

\maketitle

\nolinenumbers

\textbf{Interactions of isolated quantum many-body systems typically scramble local information into the entire system and make it unrecoverable~\cite{deutsch1991quantum,srednicki1994chaos,rigol2008thermalization,lewis2019dynamics, swingle2018unscrambling}.
Ergodicity-breaking systems possess the potential to exhibit fundamentally different information scrambling dynamics beyond this paradigm. For many-body localized systems with strong ergodicity breaking~\cite{nandkishore2015many,abanin2019colloquium,lukin2019probing}, local transport vanishes and information scrambles logarithmically slowly~\cite{huang2017out}.
Whereas in Rydberg atom arrays, local qubit flips induce dynamical retardation on surrounding qubits through the Rydberg blockade effect, giving rise to quantum many-body scars that weakly break ergodicity~\cite{bernien2017probing,turner2018weak,serbyn2021quantum}, 
and resulting in the predicted unconventional quantum information spreading behaviours~\cite{yuan2022quantum}.
Here, we present the first measurements of out-of-time-ordered correlators and Holevo information in a Rydberg atom array, enabling us to precisely track quantum information scrambling and transport dynamics. By leveraging these tools, we observe a novel spatio-temporal \textit{collapse-and-revival} behaviour of quantum information, which differs from both typical chaotic and many-body localized systems. 
Our experiment sheds light on the unique information dynamics in many-body systems with kinetic constraints, and demonstrates an effective digital-analogue approach to coherently reverse time evolution and steer information propagation in near-term quantum devices.}\\

\begin{figure*} [t]
  \centering
  \includegraphics[width=0.9\textwidth]{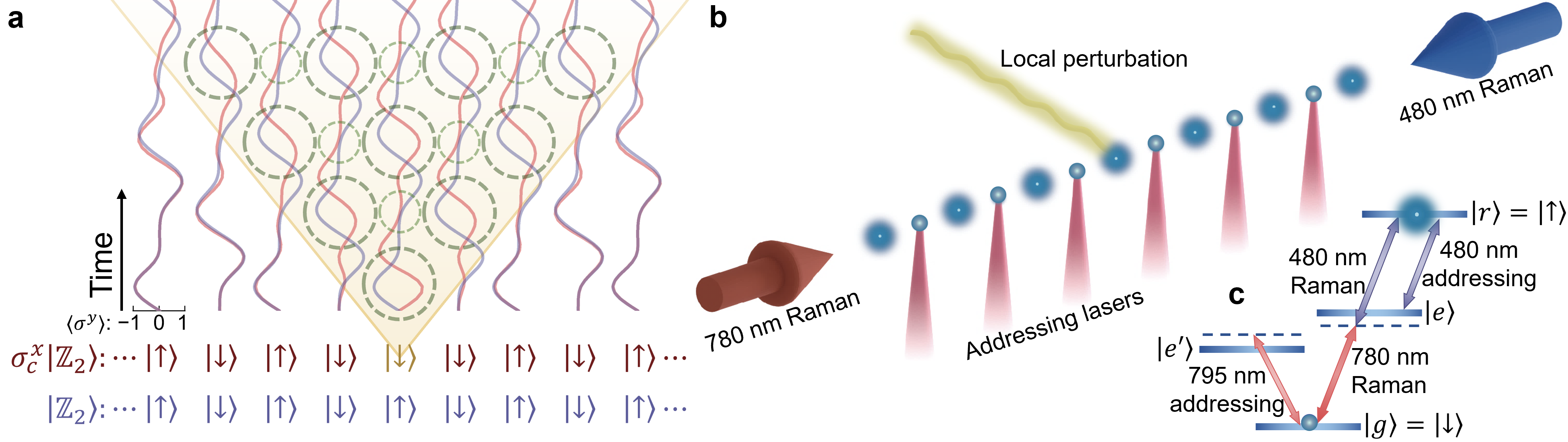}
  \caption{
	\textbf{Illustrations of kinetically constrained qubit dynamics in the PXP model and its experimental realization.}
\textbf{a},
The blue and red curves display the Hamiltonian evolution dynamics of $\sigma^y$ expectation values for each qubit, starting from the initial states $\ket{\mathbb{Z}_2}$ and $\sigma^x_c \ket{\mathbb{Z}_2}$ (with the central spin being flipped), respectively.
The initial rotation retardation near the central spin will propagate outwards in later dynamics. 
The yellow solid lines indicate where these two evolution dynamics begin to diverge, outlining a linear light cone (the yellow-shaded area). Within the light cone spins resume constrained rotations, yet the distinguishability of spin states periodically vanishes (small dashed circles) and reappears (large dashed circles). See Supplementary Information section 5.1 for more details on kinetically constrained qubit dynamics.
\textbf{b}, 
A one-dimensional, defect-free array of $^{87}$Rb atoms with approximately \SI{7}{\micro\meter} spacing between neighbouring atoms.
The spin rotation between the states $\ket{\uparrow}$ and $\ket{\downarrow}$ is achieved by driving the ground-Rydberg state transition, via an off-resonant two-photon Raman process with 780-\SI{}{\nano\meter} and 480-\SI{}{\nano\meter} global excitation laser beams. Single-site qubit operations are realized by 795-\SI{}{\nano\meter} and 480-\SI{}{\nano\meter} addressing laser beams.
\textbf{c}, 
The inset shows relevant atomic levels: the ground state $\ket{\downarrow} = \ket{g} = \ket{5S_{1/2}, F=2, m_F=2}$, the Rydberg state $\ket{\uparrow}=\ket{r} = \ket{68D_{5/2}, m_J=5/2}$, and the intermediate states $\ket{e} = \ket{5P_{3/2}, F=3, m_F=3}$ and $\ket{e'} = \ket{5P_{1/2}}$.
See Extended Data Fig. 1 and Supplementary Information section 1 for more experimental details.
     } 
  \label{Fig:1}
\end{figure*}

Quantum information scrambling---the spread of initial local information across the interacting many-body systems~\cite{lewis2019dynamics,swingle2018unscrambling}---illuminates fundamental aspects in quantum thermalization and chaos~\cite{sekino2008fast,hosur2016chaos,maldacena2016bound},
the black hole information paradox~\cite{hayden2007black,shenker2014black}, 
and quantum machine learning~\cite{Shen2020Information,garcia2022Quantifying}. 
Understanding scrambling dynamics is also practically crucial for advancing quantum technologies for quantum simulation, benchmarking, and metrology~\cite{garttner2017measuring,mi2021information,choi2023preparing,li2023improving}.
Recent works have revealed multiple information dynamics in quantum many-body systems, from rapid scrambling caused by chaotic non-integrable Hamiltonians~\cite{lewis2019dynamics,swingle2018unscrambling} to logarithmically slow information propagation in many-body localized systems~\cite{nandkishore2015many,abanin2019colloquium,lukin2019probing}.
Yet, the rich landscape of many-body physics, characterized by its diverse interaction forms and emergent phenomena, promises further unexplored regimes of information scrambling dynamics.

An intriguing frontier in this context is the quantum information dynamics within kinetically constrained systems, such as Rydberg atom arrays~\cite{browaeys2020many}. The strong van der Waals interaction between Rydberg atoms leads to the Rydberg blockade mechanism~\cite{lukin2001dipole,urban2009observation,gaetan2009observation} and constrains the local qubit flips.
This essential physics can be approximately described by the PXP Hamiltonian~\cite{lesanovsky2012interacting,turner2018weak}:
\begin{equation}
H_\text{PXP} = \sum_i P_i \sigma^x_{i+1} P_{i+2}.\label{Eq:PXP}
\end{equation}
Here, $P_i = (1 - \sigma^z_i)/2$ is the projector onto the spin-down $\ket{\downarrow}$ state at site $i$, and $\sigma^{x,y,z}_i$ are Pauli matrices for the $i$-th spin.
Notably, studies of these systems have uncovered quantum many-body scars\cite{bernien2017probing,turner2018weak,serbyn2021quantum,bluvstein2021controlling}---special eigenstates that exhibit weak ergodicity breaking and support coherent non-thermal dynamics.

Figure~\ref{Fig:1}a depicts an intuitive understanding of the novel information spreading dynamics in this kinetically constrained system. In the PXP model, flipping the central spin in the antiferromagnetic N\'eel state $\ket{\mathbb{Z}_2}= \ket{\uparrow \downarrow\uparrow\downarrow\uparrow...}$ encodes one bit of information. This local perturbation induces a retardation in the rotations of neighbouring spins, which then propagate ballistically, forming a linear light-cone-like wavefront. Within this light cone, spins resume their constrained rotations, yet the distinguishability of spin states periodically vanishes and reappears, as signified by differences in their $\sigma^y$ expectation values (dashed circles). This behaviour deviates significantly from typical chaotic quantum systems, where spin states inside the light cone quickly become indistinguishable due to thermalization. As suggested by previous theoretical work~\cite{yuan2022quantum}, these kinetically constrained dynamics may lead to persistent information backflow and unusual breakdown of quantum chaos, hinting at new avenues for probing distinct information scrambling dynamics.

Here, we explore quantum information dynamics in a kinetically constrained system using a programmable Rydberg atom array. By developing novel experimental techniques for high-fidelity $\ket{\mathbb{Z}_2}$ state preparation, many-body time-reversed evolution, and quantum state tomography under qubit rotation constraints, we successfully measure constrained spin dynamics, out-of-time-ordered correlators (OTOCs), and Holevo information, thus probing the mechanisms of information scrambling and transport. Notably, we observe the spatio-temporal \textit{collapse-and-revival} of quantum information in a many-body system, revealing unique coherence properties where locally encoded information spreads across distant sites and remains periodically accessible despite strong many-body interactions. This behaviour differs significantly from typical chaotic and many-body localized systems, offering new insights into quantum scrambling, and highlighting the potential of Rydberg atom arrays for novel quantum information processing, with dynamical quantum protection inherently provided by kinetic constraints.

\begin{figure*}
  \centering
  \includegraphics[width=0.9\textwidth]{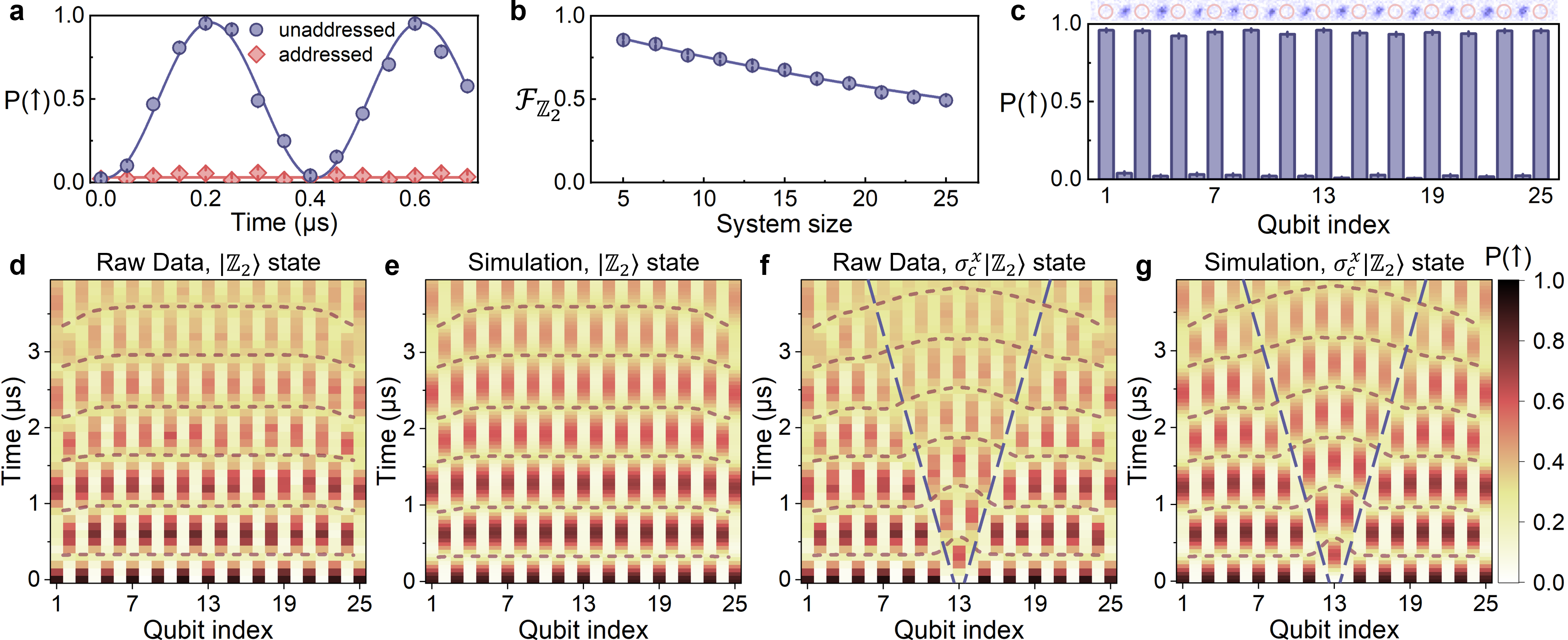}  
  \caption{
\textbf{State preparation and kinetically constrained many-body state evolution.
} 
\textbf{a}, Under the global coherent driving, the 795-\SI{}{\nano\meter}-laser addressed atoms (red) remain static, while the unaddressed atoms (blue) exhibit ground-Rydberg-state Rabi oscillations. 
\textbf{b}, $\ket{\mathbb{Z}_2}$ state preparation fidelity $\mathcal{F}_{\mathbb{Z}_2}$ as a function of the system size.
\textbf{c}, Measured site-resolved Rydberg probability $P_i(\uparrow)$ for a 25-qubit $\ket{\mathbb{Z}_2}$ state. Upper panel: Atomic fluorescence illustration of the $\ket{\mathbb{Z}_2}$ state, where red circles denote Rydberg atoms.
\textbf{d}--\textbf{g}, Kinetically constrained many-body dynamics. 
\textbf{d} and \textbf{f}, Measured site-resolved Rydberg state probability $P_i(\uparrow)$ as a function of evolution time from the initial state $\ket{\mathbb{Z}_2}$ and $\sigma^x_c\ket{\mathbb{Z}_2}$, respectively. \textbf{e} and \textbf{g} are the corresponding results from numerical simulations.
Red dashed curves in \textbf{d}--\textbf{g} illustrate the propagating wavefronts, while blue dashed lines in \textbf{f} and \textbf{g} show the linear light cones.
}
  \label{Fig:2}
\end{figure*}

\begin{figure*}
  \centering
  \includegraphics[width=\textwidth]{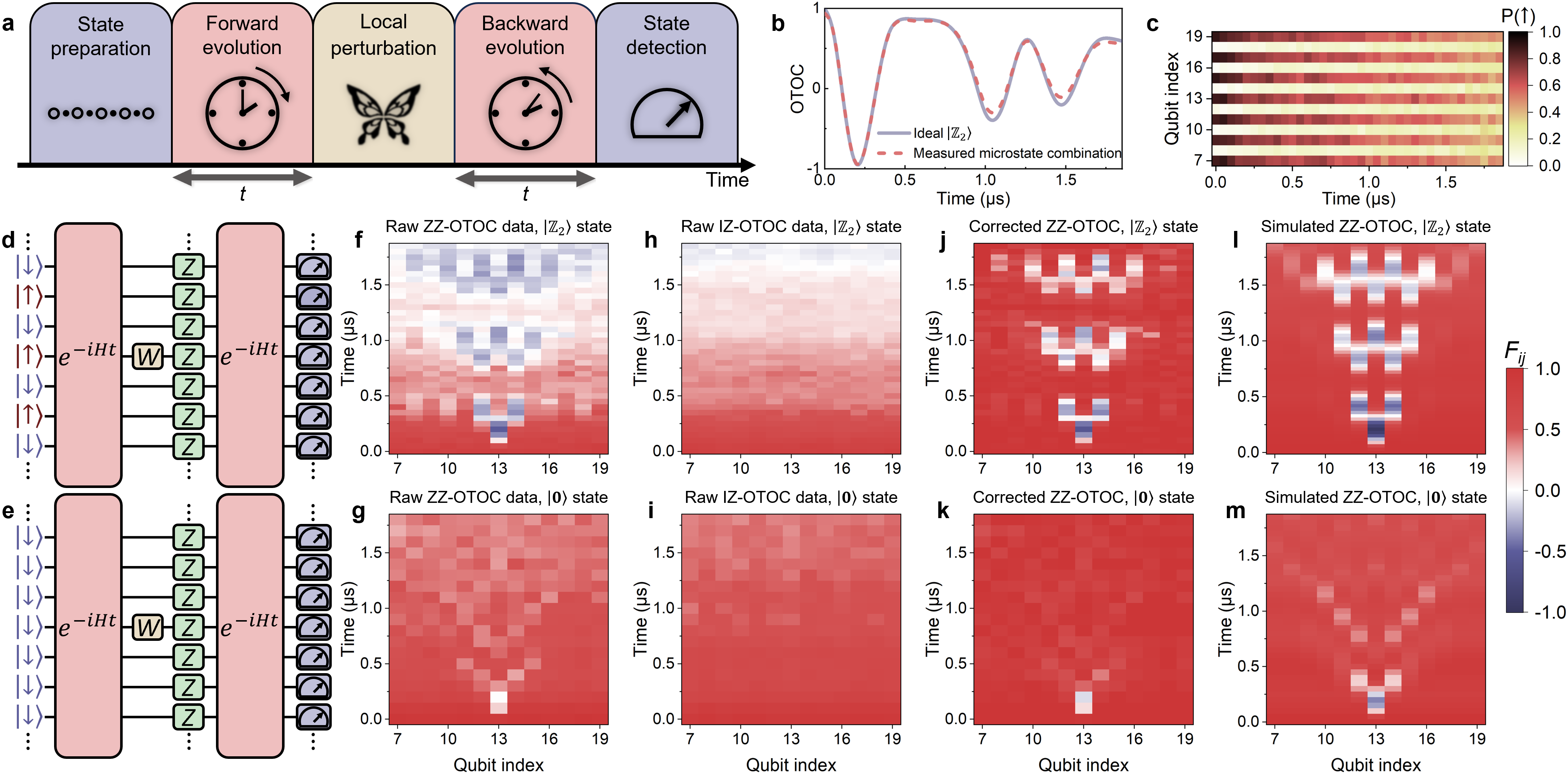} 
  \caption{
\textbf{Probing quantum information scrambling dynamics via OTOCs.
} 
\textbf{a}, An illustration of the OTOC measurement protocol.
\textbf{b}, The solid and dashed lines correspond to numerically simulated OTOC dynamics of the central qubit with the perfect $\ket{\mathbb{Z}_2}$ state and the experimentally measured microstate combination (with $\mathcal{F}_{\mathbb{Z}_2}=78(1)\%$) as inputs, respectively.
\textbf{c}, Forward-and-backward Hamiltonian evolution of the experimentally prepared $\ket{\mathbb{Z}_2}$ state. 
\textbf{d} and \textbf{e} illustrate the digital-analogue scheme for OTOC measurements with $\ket{\mathbb{Z}_2}$ and $\ket{\mathbf{0}}$ states, using $W=\sigma_c^z$ for ZZ-OTOC and $W=I$ for IZ-OTOC. 
\textbf{f}--\textbf{m}, Spatio-temporal OTOC dynamics. The colour bar corresponds to the values of the OTOC, $F_{ij}(t)$. 
\textbf{f} and \textbf{g}, Measured ZZ-OTOC for $\ket{\mathbb{Z}_2}$ and $\ket{\mathbf{0}}$ states, respectively. 
\textbf{h} and \textbf{i}, Measured IZ-OTOC for $\ket{\mathbb{Z}_2}$ and $\ket{\mathbf{0}}$ states, respectively.
\textbf{j} and \textbf{k}, Corrected ZZ-OTOC for $\ket{\mathbb{Z}_2}$ and $\ket{\mathbf{0}}$ states.
\textbf{l} and \textbf{m}, Numerically simulated ZZ-OTOC for $\ket{\mathbb{Z}_2}$ and $\ket{\mathbf{0}}$ states.
The OTOC features in \textbf{j} and \textbf{k} (corrected experimental data) are slightly less pronounced than in \textbf{l} and \textbf{m} (simulations). This is because the simulations, based on the ideal Rydberg Hamiltonian, do not account for the local perturbation imperfections in the experiments. These imperfections degrade the OTOC features but cannot be corrected using IZ-OTOC. For a detailed comparison between the corrected experimental data and simulations that include these imperfections, see Extended Data Fig. 2 and Fig. S19 (Supplementary Information section 4).
}
  \label{Fig:3}
\end{figure*}

\begin{figure} [t]
  \centering
  \includegraphics[width=\columnwidth]{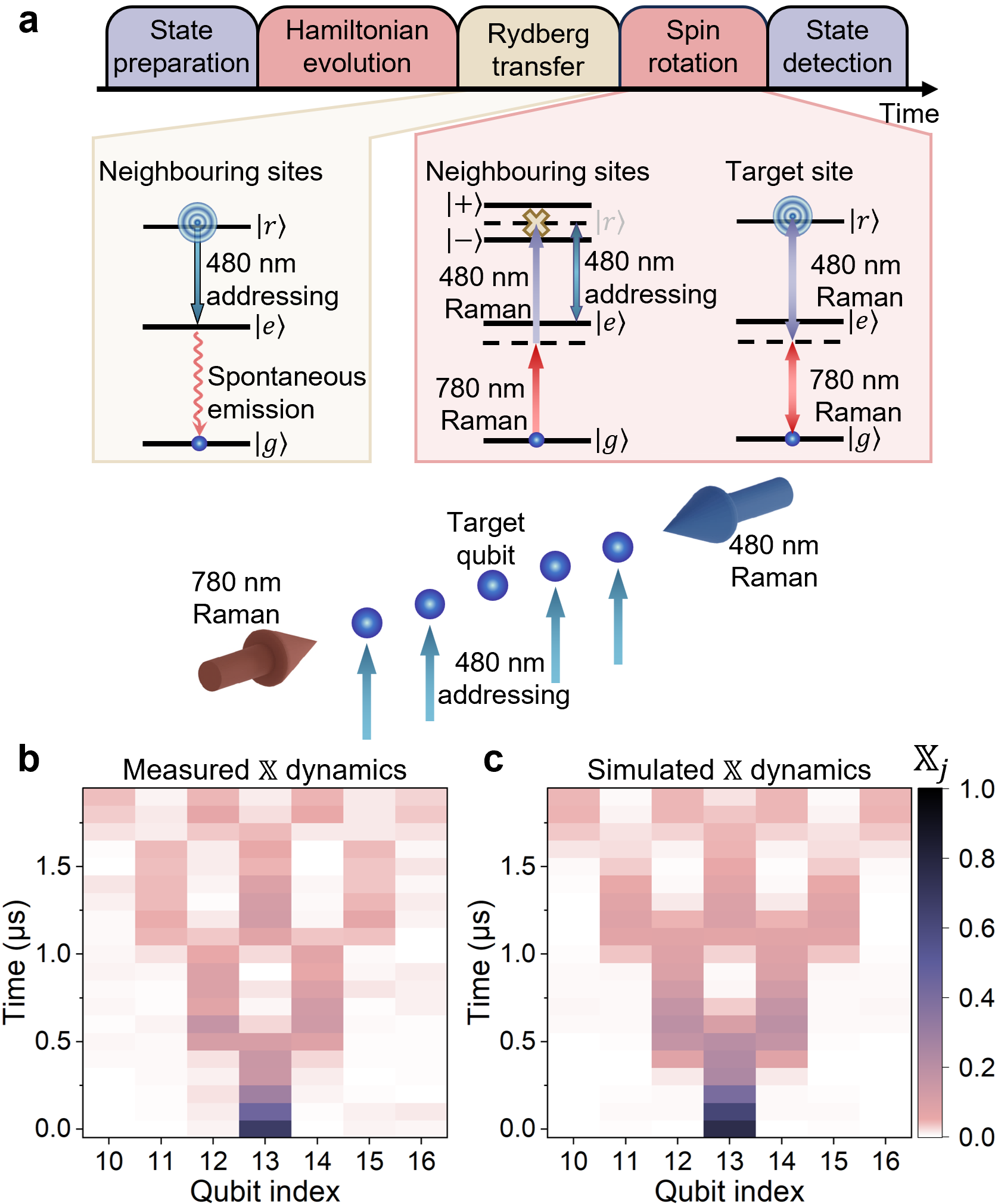}  
  \caption{
	\textbf{Probing quantum information transport dynamics via Holevo information.
 } 
\textbf{a}, 
    Illustration of the Holevo information measurement protocol. In order to measure the off-diagonal elements of the reduced density matrices, the 480-\SI{}{\nano\meter} addressing lasers first transfer Rydberg populations at the neighbouring sites to the ground state $\ket{g}$, and then split the $\ket{r}$ state into two dressed states $\ket{+}=(\ket{r}+\ket{e})/\sqrt{2}$ and $\ket{-}=(\ket{r}-\ket{e})/\sqrt{2}$, thus shielding these sites from spin rotation under global driving. See Supplementary Information section 3.2 for further details on quantum state tomography under spin rotation constraints.
\textbf{b} and \textbf{c}, 
    Experimentally measured and numerically simulated spatio-temporal dynamics of the Holevo information, respectively.
    The colour bar corresponds to the values of the Holevo information $\mathbb{X}_{j}(t)$.
  }
  \label{Fig:4}
\end{figure}

\textbf{Kinetically constrained spin dynamics}

As shown in Fig.~\ref{Fig:1}b,c, our quantum simulator employs a linear array of up to 25 individual $^{87}$Rb atoms trapped in optical tweezers.
The qubit states $\ket{\downarrow}$ and $\ket{\uparrow}$ are encoded in the atomic ground state $\ket{g}$ and a Rydberg state $\ket{r}$, respectively.
The Hamiltonian governing the system reads:
\begin{equation}
\label{Rydberg Hamiltonian}
H = \frac{\Omega}{2} \sum_i \sigma^x_i - \Delta \sum_i n_i + \sum_{i<j} V_{ij} n_i n_j,
\end{equation}
where $\Omega$ is the Rabi frequency of the coherent driving between $\ket{\uparrow}$ and $\ket{\downarrow}$, $\Delta$ is the detuning of driving lasers from the ground-Rydberg transition, $n_i = \ket{\uparrow_i}\bra{\uparrow_i}$ is the projector towards the state $\ket{\uparrow_i}$, and $V_{ij} \propto 1/R^6_{ij}$ represents the van der Waals 
interaction between Rydberg atoms at sites $i$ and $j$ with the distance $R_{ij}$.
In the regime where \(V_{i,i+2} \ll \Omega \ll V_{i,i+1}\), the Rydberg blockade prevents configurations with adjacent qubits both in the \(\ket{\uparrow}\) state (e.g., \(\ket{\cdots \uparrow_i \uparrow_{i+1} \cdots}\)), making the system well-approximated by the PXP model in equation (\ref{Eq:PXP}).

High-fidelity state preparation is essential for the success of all subsequent experiments.
We prepare the $\ket{\mathbb{Z}_2}$ state by using the global Rydberg excitation lasers (\SI{780}{nm} and \SI{480}{nm}) combined with site-selective 795-\SI{}{\nano\meter} addressing lasers. The latter detune the addressed atoms from the ground-Rydberg transition, creating an alternating pattern of excitable and non-excitable atoms (Fig.~\ref{Fig:2}a).
As shown in Fig.~\ref{Fig:2}b,c, our method achieves high $\ket{\mathbb{Z}_2}$ state preparation fidelities: 70(1)\% for the 13 central qubits, and 49(3)\% for the entire 25-qubit chain. After correcting for detection errors, the fidelities improve to 78(1)\% for 13 qubits and 60(3)\% for 25 qubits.
Achieving such high preparation fidelity in a large Hilbert space ($2^{25}$ dimensions) is critical for accurately probing kinetically constrained many-body state evolution and quantum information dynamics.

Next, we investigate the kinetically constrained spin dynamics described in the introduction, focusing on the evolution of the spin configurations $\ket{\mathbb{Z}_2}$ and $\sigma^x_c \ket{\mathbb{Z}_2}$ under the PXP Hamiltonian. For the $\ket{\mathbb{Z}_2}$ state (Fig.~\ref{Fig:2}d), all spins rotate coherently under the PXP constraints, forming a uniform wavefront. This synchronized evolution persists until boundary effects gradually alter the spin rotations from the outer edge inwards.

In contrast, for the $\sigma^x_c \ket{\mathbb{Z}_2}$ state (Fig.~\ref{Fig:2}f), we observe a clear linear light cone structure. Spins outside the light cone remain unaffected by the flipped central spin, exhibiting dynamics similar to the $\ket{\mathbb{Z}_2}$ state. 
Within the light cone, the flipped central spin induces retardation in the evolution of neighbouring spins, creating a distinct evolving wavefront.
This arc-shaped bent wavefront propagates outwards, driven by the retarded spin rotations, which clearly illustrates the constrained spin dynamics in the PXP model.
The excellent agreement between experimental data and numerical results (Fig.~\ref{Fig:2}e,g) confirms the effectiveness of our Rydberg quantum simulator in capturing the behaviour of kinetically constrained systems, paving the way for further explorations of quantum information dynamics.

\textbf{OTOC dynamics}

The out-of-time-ordered correlator is a powerful tool for probing quantum information scrambling in many-body systems~\cite{lewis2019dynamics,swingle2018unscrambling}. However, due to considerable experimental challenges, OTOC measurements have been demonstrated only in several state-of-the-art physical platforms~\cite{
mi2021information,blok2021quantum,braumuller2022probing,zhao2022probing,wang2022information,
landsman2019verified,garttner2017measuring,joshi2020quantum,green2022experimental,
li2017measuring,wei2018exploring,niknam2020sensitivity,
chen2020detecting,pegahan2021energy,li2023improving}.
In the Rydberg atom array, we for the first time experimentally employ OTOCs to investigate the information scrambling dynamics under kinetic constraints. The OTOC is defined as follows, which quantifies how the operator growth 
 of a local perturbation $W$ (called the butterfly operator) spreads throughout the lattice sites:
\begin{equation}
F_{ij}(t) = \bra{\psi}W_i^\dagger(t)V_j^\dagger W_i(t)V_j\ket{\psi}.
\end{equation}
Our study focuses on the ZZ-OTOC, with local butterfly operator $W_i = \sigma^z_c$, and measurement operator $V_j = \sigma^z_j$, in order not to violate the kinetically constrained spin rotations. $W_i(t) = e^{iHt}W_i e^{-iHt}$ represents the Heisenberg evolution of $W_i$, and $\ket{\psi}$ is an initial state. The local perturbation, corresponding to a \(\sigma^z_c\) gate (a \(\pi\)-phase shift) on the central qubit, is experimentally implemented via the light shift from a 795-\SI{}{\nano\meter} addressing laser.
Fig.~\ref{Fig:3}a shows the five-step protocol for measuring the OTOC $F_{ij}(t)$: (1) state preparation; (2) forward evolution; (3) local perturbation; (4) backward (time-reversed) evolution; and (5) computational basis measurements.

Here, we employ two distinct initial states to study the OTOC dynamics, including the antiferromagnetic $\mathbb{Z}_2$-ordered N\'eel state $\ket{\mathbb{Z}_2} = \ket{\downarrow \uparrow\downarrow\uparrow\downarrow\cdots}$ (which has large overlap with the scarred eigenstate subspace~\cite{turner2018weak}) and a trivial product state $\ket{\mathbf{0}} = \ket{\downarrow \downarrow\downarrow\downarrow\downarrow\cdots}$ (which lies in the common thermal eigenstate subspace).
Our numerical simulations show that in a small-size spin chain (less than 10 qubits), boundary effects observed in Fig.~\ref{Fig:2} can lead to considerable deviations from the ideal PXP dynamics (Supplementary Information section 2.3).
To mitigate the boundary and finite-size effects, we particularly prepare a 25-qubit $\ket{\mathbb{Z}_2}$ ($\ket{\mathbf{\mathbf{0}}}$) state and perform the OTOC experiments on the central 13 qubits. This configuration shields the region of interest from boundary effects during evolution, better approximating the bulk PXP dynamics.
Moreover, our high $\ket{\mathbb{Z}_2}$ state preparation fidelity is crucial for accurately probing OTOC dynamics. Low preparation fidelity introduces numerous error states into the spin chain, which smear the expected OTOC patterns (Fig.~\ref{Fig:3}b and Supplementary Information section 4.1).
As a benchmark, the numerically simulated OTOC dynamics using the experimentally measured microstate combination as input (dashed lines in Fig.~\ref{Fig:3}b) show a negligible difference from the results obtained with the perfect $\ket{\mathbb{Z}_2}$ initial state (solid lines in Fig.~\ref{Fig:3}b), attesting to the accuracy of our experimental approach.

With the initial states well prepared, we address the next crucial and notoriously difficult step in OTOC measurements---implementing the time-reversed evolution under $-H$ in the many-body system.
While reversing single-particle Hamiltonians is relatively easy, reversing many-body interaction terms, such as switching between attractive and repulsive interactions, presents significantly more complex theoretical and experimental challenges.
For instance, the OTOC experiment on Google's Sycamore quantum computer\cite{mi2021information} highlights the intricate nature of achieving time reversal by the digital simulation approach, which requires sophisticated quantum circuits with cutting-edge gate fidelities and large circuit depths.

In contrast, our approach exploits the particle-hole symmetry of the PXP model, enabling us to implement the inverse Hamiltonian $-H$ through a global $\sigma^z$ gate on all qubits: $(\prod_i \sigma_i^z)H(\prod_i \sigma_i^z) = -H$. 
This is experimentally realized by using a far-detuned microwave field that induces a $\pi$-phase shift on the Rydberg states.
Figure~\ref{Fig:3}c shows the experimentally measured forward-and-backward evolution dynamics of the $\ket{\mathbb{Z}_2}$ state.
This novel digital-analogue protocol generally serves as a powerful technique in Rydberg atom arrays to coherently reverse many-body evolution and probe information scrambling dynamics.

Having achieved the high-fidelity state preparation and time reversal, we executed the complete OTOC measurement protocol (Fig.~\ref{Fig:3}d,e), demonstrating precise control over complex quantum many-body dynamics in our Rydberg atom array.
Figure~\ref{Fig:3}f and g present the experimentally measured spatio-temporal evolution of the ZZ-OTOCs for the initial states $\ket{\mathbb{Z}_2}$ and $\ket{\mathbf{0}}$, respectively.
The overall decay of the measured OTOC values primarily stems from atomic state losses and decoherence during the evolution (see Supplementary Information section 4), even in the absence of quantum information scrambling.
In order to distinguish the scrambling dynamics of interest from these extraneous decay mechanisms~\cite{landsman2019verified}, we additionally measure the evolution of both $\ket{\mathbb{Z}_2}$ and $\ket{\mathbf{0}}$ states under the action of $H$ and $-H$ but without any butterfly operator, i.e. the IZ-OTOCs, where I stands for the identity operator.
Based on the dynamics of the IZ-OTOCs (Fig.~\ref{Fig:3}h,i), we employ error mitigation techniques similar to those developed in previous work~\cite{mi2021information} to correct the raw ZZ-OTOC data (see Supplementary Information section 4).
The resulting corrected OTOC dynamics (Fig.~\ref{Fig:3}j,k) align well with the numerical simulations based on ideal Rydberg Hamiltonian evolution (Fig.~\ref{Fig:3}l,m).

The OTOC evolution dynamics reveal strikingly different behaviours for the $\ket{\mathbf{0}}$ and $\ket{\mathbb{Z}_2}$ initial states. For the $\ket{\mathbf{0}}$ state, the OTOCs quickly decay inside the linear light cone without discernible revivals.
This rapid information scrambling agrees with theoretical expectations for generic high-energy initial states in chaotic quantum systems.
In stark contrast, the $\ket{\mathbb{Z}_2}$ state exhibits a slower butterfly velocity, and remarkably a spatio-temporal \textit{collapse-and-revival} pattern of quantum information inside the linear light cone. 

The \textit{collapse-and-revival} phenomenon, previously mainly observed in single-particle systems such as the Jaynes-Cummings model~\cite{eberly1980periodic,rempe1987observation}, reflects the periodic dispersion and recurrence of quantum coherence. In contrast to conventional expectations for many-body systems where interactions rapidly scramble quantum information across an exponentially large Hilbert space, our observations reveal a striking preservation and refocusing of quantum coherence
inside the linear light cone. These unusual scrambling dynamics suggest that kinetically constrained systems allow for ballistic propagation and periodic recovery of quantum information, which differs significantly from both typical chaotic and many-body localized systems.
We note that the observed \textit{collapse-and-revival}
dynamics are related to quantum many-body scarred eigenstates of the PXP Hamiltonian, but can not be directly deduced from the oscillation of local observables reported in previous literature~\cite{bernien2017probing,turner2018weak}.
Moreover, the intrinsic differences between OTOC and local observable (such as the time-ordered correlators $\langle W_i(t) V_j\rangle$) dynamics have been pointed out in previous works for thermalizing and many-body localized systems~\cite{swingle2018unscrambling,huang2017out}.

\textbf{Holevo information dynamics}

\LL{
Although the OTOC measurements have successfully revealed the \textit{collapse-and-revival} phenomenon in quantum information dynamics, relying solely on OTOCs may not provide a complete picture. The nature of ZZ-OTOC measurements can lead to situations where the perturbation applied by $\sigma^z$ operator becomes ineffective (see Supplementary Information section 3.2 for details). For example, during certain time intervals (\SI{0.6}{\us}--\SI{0.7}{\us}, and \SI{1.2}{\us}--\SI{1.3}{\us}) in Fig.~\ref{Fig:3}j,l, the OTOC values of all qubits approach 1. This occurs because, at these times, the perturbed qubit is close to the $\ket{\uparrow}$ or $\ket{\downarrow}$ pure state, and therefore the butterfly operator $\sigma^z$ has no effect.
}

\LL{
To complement the insights provided by OTOCs and gain a more continuous view of how quantum information propagates in kinetically constrained systems, we also investigate the Holevo information dynamics, in which we leverage quantum state tomography to provide a complementary approach, especially in scenarios where the butterfly operator perturbations in ZZ-OTOC measurements become less effective.}
Holevo information was originally proposed to upper-bound the accessible information between two separate agents~\cite{holevo1973bounds}. 
Here, we treat the reduced Hamiltonian evolution on subsystems as quantum channels and use Holevo information to quantify the local distinguishability of quantum states evolving from slightly different initial conditions.

Our experiment explores two sets of Hamiltonian evolution applied to different initial states: \(\ket{\mathbb{Z}_2}\) and \(\sigma^x_{c}\ket{\mathbb{Z}_2}\). The \(\sigma^x_{c}\) operator acts on the central qubit in the one-dimensional chain, encoding one bit of local information in the initial states, as illustrated in Fig.~\ref{Fig:1}a.
During subsequent Hamiltonian evolution, we measure the Holevo information \(\mathbb{X}_j(t)\) at different sites \(j\), which quantifies how much information encoded on the central qubit can be retrieved through local measurements at distant sites:
\begin{equation}
\mathbb{X}_j(t) = S\left(\frac{\rho_j(t) + \rho'_j(t)}{2}\right) - \frac{S(\rho_j(t)) + S(\rho'_j(t))}{2},
\end{equation}
where $\rho_j(t)$ and $\rho'_j(t)$ are reduced density matrices of the $j$-th spin after Hamiltonian evolution for the initial states $\ket{\mathbb{Z}_2}$ and $\sigma^x_{c}\ket{\mathbb{Z}_2}$, respectively, and $S(\rho) = - \mathrm{Tr}(\rho \log \rho)$ is the von Neumann entropy.
The experimental measurement procedure is shown in Fig.~\ref{Fig:4}a, which includes: (1) preparation of initial states $\ket{\mathbb{Z}_2}$ and $\sigma^x_{c}\ket{\mathbb{Z}_2}$; (2) Hamiltonian evolution over time $t$; (3) quantum state tomography on the qubit $j$ to obtain $\rho_j(t)$ and $\rho'_j(t)$.

Performing full quantum state tomography under the PXP kinetic constraints is highly challenging. While the diagonal elements of $\rho_j(t)$ and $\rho'_j(t)$ can be directly extracted from Rydberg population measurements (Fig.~\ref{Fig:2}d,f), obtaining the off-diagonal elements is significantly more difficult. This requires spin rotations on the target qubit, a process complicated by the interactions with neighbouring qubits. In order to rotate the target qubit, its nearest and next-nearest qubits should be neither in the Rydberg state nor resonant with the ground-Rydberg transition.
To this end, we developed a new approach in which we apply 480-\SI{}{\nano\meter} Rydberg addressing lasers to the four neighbouring sites, as shown in Fig.~\ref{Fig:4}a.
The 480-\SI{}{\nano\meter} lasers will first transfer the Rydberg populations in these neighbouring sites to the ground state via spontaneous emission from the intermediate state $\ket{e}$. Then, these addressing lasers create an electromagnetically induced transparency (EIT) condition, preventing the neighbouring sites from being excited by the ground-Rydberg excitation lasers.
Finally, we apply the global excitation lasers to implement the spin rotations needed in state tomography onto the target qubit.

Figure~\ref{Fig:4}b(c) shows the experimentally measured (numerically simulated) spatio-temporal evolution of the Holevo information, which are in good agreement. We observe a clear linear light cone structure and a distinct \textit{collapse-and-revival} pattern of $\mathbb{X}_j(t)$, differing from the OTOC results in Fig.~\ref{Fig:3}j,l. Notably, during the time intervals when the ZZ-OTOC values of all qubits approach 1---indicating ineffective perturbations---we still observe non-uniform Holevo information values. In contrast to OTOCs, Holevo information allows for continuous tracking of quantum information dynamics.

We also mention that the observed \textit{collapse-and-revival} of Holevo information are related, but not equivalent to the oscillation of the scar state wavefunction under Hamiltonian evolution~\cite{bernien2017probing,turner2018weak}, because the $\sigma^x_{c}$ operation flips the central spin, and thus involves both the scarred subspace and the thermal eigenstate bath into the $\mathbb{X}_j(t)$ dynamics. We attribute the observed \textit{collapse-and-revival} pattern to the general kinetically constrained spin rotations in the PXP model. As illustrated in Fig.~\ref{Fig:1}a (and Supplementary Information section 5.2), the Rydberg blockade effect leads to retarded rotations of spins near the central flipped spin, which then propagate outwards and form the linear light cone structure. Within the light cone, spins continue their constrained rotations, and the distinguishability of the reduced density matrices $\rho_j(t)$ and $\rho'_j(t)$ periodically vanishes and reappears, manifesting as the \textit{collapse-and-revival} behaviour in the Holevo information.

Furthermore, our results reveal a unique information transport behaviour: The local information initially encoded at one site can be recovered at both the initial and other distant sites after the evolution under a many-body interacting Hamiltonian. These dynamics demonstrate persistent information backflow from the environment (surrounding qubits) to the system (qubits hosting encoded information), a hallmark of non-Markovian open quantum dynamics \cite{breuer2016colloquium}. This non-Markovianity can be quantified using the Holevo information and the trace distance approach (see Supplementary Information section 3.3). The observed non-Markovian quantum dynamics underscore the system's quantum memory effects for locally encoded information and its ability to transport the information across distant sites, which hold particular promise for advancing quantum memory technologies in kinetically constrained many-body systems.

\textbf{Discussions}

Our work reveals the emergence of quantum information \textit{collapse-and-revival} in a strongly-interacting Rydberg atom array. 
While typically observed in single-particle scenarios~\cite{rempe1987observation}, this phenomenon manifests in our many-body interacting system, offering a different picture of how quantum information propagates under kinetic constraints. We particularly mention that such novel information scrambling dynamics are distinct from those in generic chaotic or many-body localized systems, and are experimentally observed for the first time.
The underlying mechanism for these phenomena can be attributed to the kinetically constrained spin rotations induced by the Rydberg blockade effect. 
More specifically, due to the spin-flip in Holevo information dynamics, the periodic reassembly of quantum information emerges from dynamics involving both the scarred subspace and the thermal eigenstate bath. We emphasize that this distinguishes our observed \textit{collapse-and-revival} dynamics from the previously reported oscillation of scar state wavefunction~\cite{bernien2017probing,turner2018weak} (elaborated in Supplementary Information section 5.3).

The experimental techniques developed in this work are indispensable for the precise measurement of constrained spin dynamics, OTOCs, and Holevo information. Notably, we demonstrated a digital-analogue approach to effectively reverse the many-body evolution in Rydberg atom systems. The multi-transition, site-selective addressability of our platform allows us to achieve high-fidelity many-body state preparation, as well as perform full quantum state tomography, even in the presence of rotation constraints. These techniques enrich the toolbox for Rydberg atom quantum simulations and are poised for application in future experiments.

Our work opens up a range of exciting research directions and potential applications. For instance, one could investigate the propagation and interference dynamics caused by multiple butterfly operator perturbations, which will reveal novel collective behaviours arising from the interplay between kinetic constraints and quantum correlations in many-body systems~\cite{surace2020lattice}.
While the magnitude and persistence of the \textit{collapse-and-revival} effect observed here are currently limited by coherence, previous studies have suggested that techniques such as Floquet engineering can significantly extend coherence time and modify the interactions in Rydberg PXP systems\cite{bluvstein2021controlling,koyluouglu2024floquet}. Building on this, incorporating more single-site operations to introduce precise spatial modulation could enable controlled steering of quantum information propagation, which is dynamically protected by kinetic constraints. These advances hold promise for practical applications in near-term quantum technologies, including robust quantum memory, novel quantum state transfer protocols and quantum metrology in many-body interacting systems\cite{dooley2021robust,dong2023disorder}.

\nolinenumbers

\vspace*{0.5\baselineskip}

\textit{Note}: We became aware of related work from the group of Prof. Li You at Tsinghua University.

\nolinenumbers

\bibliographystyle{naturemag}
\bibliography{refs}

\clearpage
\newpage

\nolinenumbers

\section*{Methods}

\textbf{Experimental methods.} 
In our experiment, atoms are loaded from a magneto-optical trap and captured by optical tweezers, formed by focusing 808-nm laser beams diffracted through a spatial light modulator (SLM). A measurement and feedback process eliminates defects in the atom chain, enabling the rapid creation of fully filled arrays. Once the defect-free array is prepared, all atoms are optically pumped into the ground state.
The interatomic spacing is $\sim$ \SI{7}{\micro m}, resulting in nearest-neighbour and next-nearest-neighbour interactions of $V_{i,i+1} = 2\pi \times \SI{7.3}{MHz}$ and $V_{i,i+2} = 2\pi \times \SI{0.11}{MHz}$, respectively. To closely approximate the PXP model despite the fixed ratio of 64 between the vdW interactions $V_{i,i+1} / V_{i,i+2}$, the ground-Rydberg driving Rabi frequency $\Omega$ is set to $V_{i,i+1}/6 = 2\pi \times \SI{1.21(1)}{\MHz}$ during the state evolution, striking a balance between the two interaction strengths. In addition, a small detuning $\Delta =2\pi\times\SI{0.22(1)}{\MHz}$ is introduced to counteract the residual next-nearest-neighbour interactions $V_{i,i+2}$.
The 795-\SI{}{\nano\meter} and 480-\SI{}{\nano\meter} addressing laser beams are generated via SLM diffraction and focused onto the atoms through high numerical aperture objectives, enabling selective addressing by adjusting the SLM phase pattern. The 795-\SI{}{\nano\meter} addressing beams off-resonantly couple the $\ket{5S_{1/2}}$ ground state to the $\ket{5P_{1/2}}$ state, inducing a light shift on the addressed atoms. During $\ket{\mathbb{Z}_2}$ state preparation, the addressed atoms experience a light shift of $2\pi \times \SI{12.2(3)}{MHz}$, sufficient to shield the atom from Rydberg excitation. For local $\sigma^z$ gate, the light shift is about $2\pi \times \SI{5}{MHz}$, enabling a $\pi$-phase shift in $\sim$ \SI{110}{ns}.
The 480-\SI{}{\nano\meter} addressing lasers are resonant with the $\ket{e}$-$\ket{r}$ transition, with a Rabi frequency of $\Omega_{480} \sim 2\pi \times \SI{20}{MHz}$. These lasers enable rapid Rydberg-to-ground state transfer and create an EIT condition that prevents neighbouring qubits from participating in the ground-Rydberg transitions driven by the global excitation lasers.

\clearpage
\newpage
\setcounter{figure}{0}
\newcounter{EDfig}
\renewcommand{\figurename}{Extended Data Figure}

\begin{figure*}
  \centering
  \includegraphics[width=0.9\textwidth]{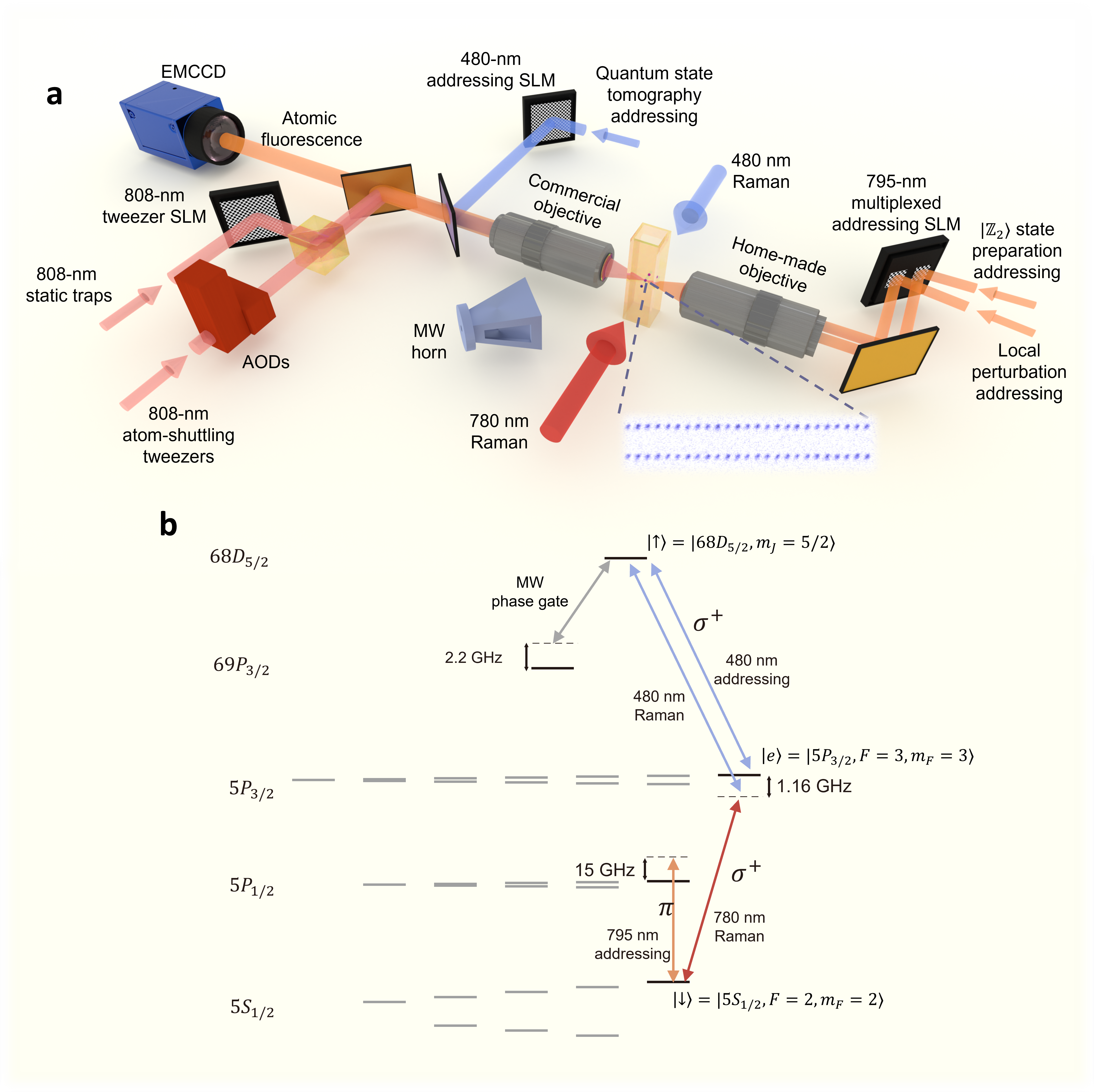}  
  \caption{
    \textbf{Experimental setup and atomic level diagram.} 
    \textbf{a}, Schematic of the essential elements in the experimental setup. The 808-nm static tweezer traps are generated by SLM, while the atom-shuttling tweezers are controlled by 2 acousto-optic deflectors (AODs). The 480-\SI{}{\nano\meter} addressing laser beams are generated by another SLM. A commercial objective (G Plan Apo 50$\times$, Mitutoyo, NA = 0.5) focuses the 808-nm tweezer beams and the 480-\SI{}{\nano\meter} addressing beams, while collecting the 795-\SI{}{\nano\meter} atomic fluorescence for imaging on an electron-multiplying CCD (EMCCD) camera. Counter-propagating 780-\SI{}{\nano\meter} and 480-\SI{}{\nano\meter} Raman beams enable coherent ground-Rydberg manipulation. The 795-\SI{}{\nano\meter} addressing laser for local perturbation \(\sigma_c^z\) and the laser array responsible for producing the alternating light shifts for $\ket{\mathbb{Z}_2}$ state preparation are targeted at separate areas of the same SLM. These distinct regions show different holograms, enabling varied addressing laser patterns. This spatial multiplexing technique allows for sub-microsecond switching between the \(\ket{\mathbb{Z}_2}\) state preparation and local perturbation laser configurations. The 795-\SI{}{\nano\meter} addressing laser beams are focused onto the atoms through a home-made objective (NA = 0.4). A microwave horn near the chamber generates MW pulses for Rydberg state manipulation (phase gate) and detection. The inset shows the atomic array.
    \textbf{b}, Level diagram showing the atomic levels of $^{87}$Rb involved in the experiment. Our Rydberg excitation scheme employs a two-photon Raman transition from the ground state \(\ket{\downarrow} = \ket{5S_{1/2}, F=2, m_F=2}\) to the Rydberg state \(\ket{\uparrow} = \ket{68D_{5/2}, m_J=5/2}\). The transition is driven by a \(\sigma^+\)-polarized 780-\SI{}{\nano\meter} laser, which couples the ground state to the intermediate state \(\ket{e}=\ket{5P_{3/2}, F=3, m_F=3}\), and a \(\sigma^+\)-polarized 480-\SI{}{\nano\meter} laser, which connects the intermediate state to the Rydberg state. Both lasers are detuned from the intermediate state by \(\Delta = 2\pi\times\SI{1.16}{\giga\hertz}\). A microwave field, red-detuned by $2\pi\times\SI{2.2}{\giga\hertz}$ from the \(\ket{\uparrow}\) to \(\ket{69P_{3/2}, m_J=3/2}\) transition, generates AC-Stark shifts on the Rydberg states and provides the phase gate for reversing the Hamiltonian evolution. For local single-qubit control, 795-\SI{}{\nano\meter} addressing beams, blue-detuned by $2\pi\times\SI{15}{\giga\hertz}$ from the \(\ket{\downarrow}\)-\(\ket{5P_{1/2}}\) transition, and 480-\SI{}{\nano\meter} addressing beams, resonant with the \(\ket{\uparrow}\)-\(\ket{e}\) transition, are employed.
    }
  \label{Fig:extended_1}
\end{figure*}

\begin{figure*}
\centering
\includegraphics[width=1\textwidth]{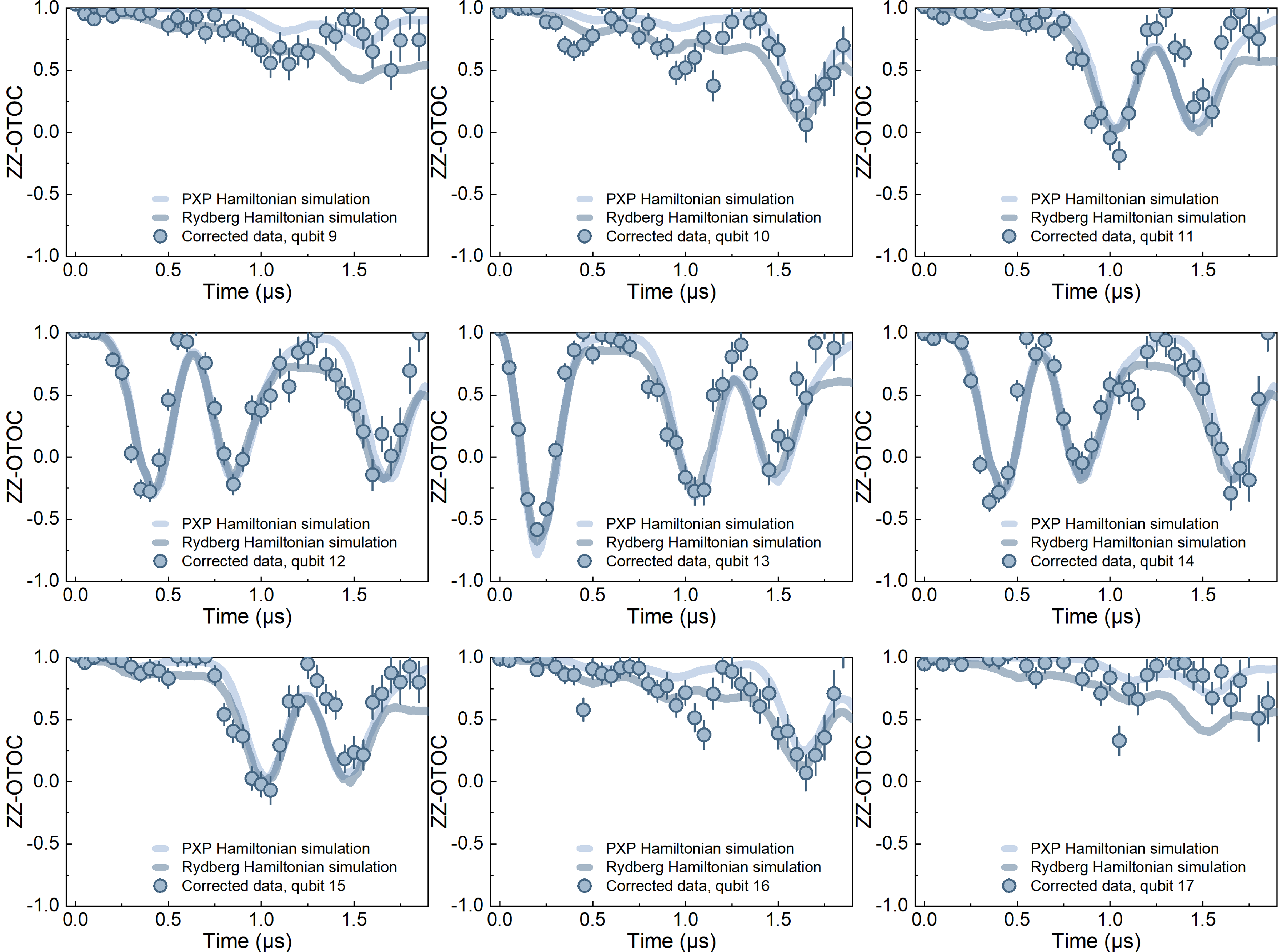}
\caption{\textbf{Mitigated ZZ-OTOC results for $\ket{\mathbb{Z}_2}$ state.} 
Blue points represent mitigated experimental ZZ-OTOC data for the $\ket{\mathbb{Z}_2}$ state, corrected based on IZ-OTOC results.
Light blue curves represent the simulated ZZ-OTOC dynamics under the PXP Hamiltonian (equation~(\ref{Eq:PXP})), accounting for local perturbation imperfections.
Dark blue curves show the simulated ZZ-OTOC dynamics under the Rydberg Hamiltonian (equation~(\ref{Rydberg Hamiltonian})), also incorporating local perturbation imperfections.
The imperfections in the local perturbation make the OTOC patterns less distinct, which cannot be mitigated using IZ-OTOCs. 
Therefore, these imperfections are incorporated in the numerical simulations using Monte Carlo methods (see Supplementary Information section 4).
}
\label{Mitigated_data}
\end{figure*}

\nolinenumbers

\vspace*{1\baselineskip}

\noindent
\textbf{\large Data availability}

\noindent
Data that support the figures within this paper are available from the corresponding author upon reasonable request.

\vspace*{1\baselineskip}

\noindent
\textbf{\large Code availability}

\noindent
The code used in this study is available from the corresponding author upon reasonable request. 

\vspace*{1\baselineskip}

\noindent
\textbf{\large Acknowledgments}

\noindent
The authors thank Dong-Ling Deng for his valuable inputs to this work. The authors acknowledge insightful discussions with Weibin Li, Yijia Zhou, Yue Chang, Tao Shi, Xiong-Jun Liu, Long Zhang, and Yiqiu Ma. The authors thank Ren Liao, Zerui Song, Jianhao Yu, and Hui-Feng Xie for experimental assistance. D.-S.X. thanks Pei-Yuan Cai for the helpful advice on numerical simulation speedup. The simulations in this work are completed in the HPC Platform of Huazhong University of Science and Technology. This work was supported by the National Key Research and Development Program of China (Grant No.~2021YFA1402003), the National Natural Science Foundation of China (Grant Nos.~12374329 and U21A6006), and the National Science and Technology Major Project of the Ministry of Science and Technology of China (Grant No.~2023ZD0300900). D.Y. acknowledges support from the National Natural Science Foundation of China (Grant No.~123B2072).

\vspace*{1\baselineskip}

\noindent
\textbf{\large Author contributions}

\noindent
All authors contributed substantially to this work.

\vspace*{1\baselineskip}

\noindent
\textbf{\large Competing interests}

\noindent
The authors declare no competing interests.

\end{document}


\title{Supplementary Information for: Observation of quantum information collapse-and-revival in a strongly-interacting Rydberg atom array}
\author
{De-Sheng Xiang$^{1,\ast}$, Yao-Wen Zhang$^{1,\ast}$, Hao-Xiang Liu$^{1,\ast}$, Peng Zhou$^{1,\ast}$, Dong Yuan$^{2,\ast}$, Kuan Zhang$^{1,\ast}$, Shun-Yao Zhang$^{3}$, Biao Xu$^{1}$, Lu Liu$^{1}$, Yitong Li$^{1}$, and Lin Li$^{1,\dagger}$\\
\normalsize{$^{1}$\textit{MOE Key Laboratory of Fundamental Physical Quantities Measurement,}}
\normalsize{\textit{Hubei Key Laboratory of Gravitation and Quantum Physics, PGMF,}}\\
\normalsize{\textit{Institute for Quantum Science and Engineering, School of Physics,}}\\
\normalsize{\textit{Huazhong University of Science and Technology, Wuhan 430074, China}}\\
\normalsize{$^{2}$\textit{Center for Quantum Information, IIIS, Tsinghua University, Beijing 100084, China}}\\
\normalsize{$^{3}$\textit{Lingang Laboratory, Shanghai 200031, China}}\\
\normalsize{$^\ast$ These authors contributed equally to this work.}\\
\normalsize{$^\dagger$ E-mail: li\_lin@hust.edu.cn}
}

\noindent

\begin{minipage}{0.95\textwidth}
\maketitle
\tableofcontents
\end{minipage}

\newpage
\nolinenumbers

\section{Experimental details}
\label{section:experiment}
\LL{
Programmable atomic arrays have emerged as a novel platform for quantum simulation and computation, offering exceptional control and scalability for exploring quantum many-body phenomena.
In the past decade, significant progress has been made in this field, including the preparation of atomic array~\cite{piotrowicz2013two,endres2016atom,barredo2016atom,lee2017defect,barredo2018synthetic,kumar2018sorting,sheng2022defect,barnes2022assembly,okuno2022high,liu2023realization,pause2024supercharged}, efforts towards quantum computation~\cite{isenhower2010demonstration,jau2016entangling,omran2019generation,bluvstein2022quantum,graham2022multi,fu2022high,schine2022long,singh2023mid,evered2023high,scholl2023erasure,ma2023high,zhao2023floquet,bluvstein2024logical,gregory2024second,shaw2024benchmarking}, quantum optimization~\cite{graham2022multi,ebadi2022quantum,kim2022rydberg,bluvstein2024logical}, quantum simulations~\cite{bernien2017probing,keesling2019quantum,ebadi2021quantum,scholl2021quantum,semeghini2021probing,fang2024probing,DeLeseleuc2019,scholl2022microwave,chen2023continuous,kim2024realization}, and quantum metrology~\cite{bornet2023scalable,eckner2023realizing}.
This section presents the details of our experiment, including the experimental apparatus, timing sequence, and $\ket{\mathbb{Z}_2}$ state preparation and characterization.
}

\subsection{Experimental setup}
\label{section:setup}

\LL{
Our experimental platform is built around a dual-chamber vacuum system, consisting of a 2D magneto-optical trap (MOT) chamber and a science chamber. In the 2D-MOT chamber, an $^{87}$Rb atom source (ampule) produces a diffuse atomic vapor. These atoms are cooled and confined by the magnetic fields and 780-\SI{}{\nano\meter} lasers, 
$2\pi\times \SI{30}{\MHz}$ red-detuned from the $\ket{5S_{1/2}, F=2} \rightarrow \ket{5P_{3/2}, F=3}$ cycling transition. The pre-cooled atomic ensemble is then transferred through a differential pumping aperture into the science chamber by a 780-\SI{}{\nano\meter} pushing beam.
The science chamber is a custom-designed rectangular glass cell (Japan Cell) with large optical access. Ultra-high vacuum conditions inside the chamber are maintained by a non-evaporable getter pump (NEXTorr D 200-5, SAES), achieving a pressure well below \(10^{-11}\) mbar. This low background pressure initially allowed for single-atom trapping lifetimes exceeding 10 minutes. However, due to the malfunction of an ion pump (SP-4, JJJvac) in the 2D MOT chamber, the lifetime has since been reduced to approximately \SI{90} seconds. Atoms are captured and cooled in the science chamber by a three-dimensional magneto-optical trap (3D MOT) using three pairs of counter-propagating 780-\SI{}{\nano\meter} laser beams, with a magnetic field gradient of \SI{15}{\gauss\per\cm}. Each beam contains a cooling light red-detuned by $2\pi\times \SI{24}{\MHz}$ from the \(\ket{5S_{1/2}, F=2} \rightarrow \ket{5P_{3/2}, F=3}\) cycling transition, and a repumping light resonant with the \(\ket{5S_{1/2}, F=1} \rightarrow \ket{5P_{3/2}, F=2}\) transition.
}

\LL{
Single atoms are trapped in a static two-dimensional optical tweezer array generated by an 808-nm laser (TA pro, Toptica) operating in free-running mode. The laser beam illuminates a phase-control spatial light modulator (SLM, HED 6010-NIR-080-C, Holoeye) loaded with a phase hologram generated via the weighted Gerchberg-Saxton (WGS) algorithm. Additionally, a system of atom-shuttling tweezers, utilizing the same 808-nm laser source as the static array but with orthogonal polarization, allows for precise atom rearrangement. The atom-shuttling tweezers are controlled by a pair of orthogonally oriented acousto-optic deflectors (AODs, DTSX-400-800.850, AA Opto-Electronic), driven by radio frequencies with independent arbitrary waveforms generated by a dual-channel arbitrary waveform generator (AWG, M4i.6631-x8, Spectrum).
}

\LL{A high numerical aperture objective (G Plan Apo 50$\times$, Mitutoyo, NA = 0.5) focuses both the static and movable tweezers while also collecting atom fluorescence, which is directed to an electron-multiplying CCD (EMCCD, iXon Ultra 888, Andor) camera for detection. Additionally, a 480-\SI{}{\nano\meter} beam for local Rydberg control, utilizing a similar SLM-based approach, is focused through the same objective (see Extended Data Fig.~1a). This shared configuration for trapping, addressing, and fluorescence detection enhances the stability of the system.}
\LL{
Opposite the Mitutoyo objective is a home-made objective with a numerical aperture of 0.4. A CCD camera images the static tweezer array, consisting of 36 \(\times\) 2 tweezers with \SI{7}{\um} spacing and a beam waist of approximately \SI{0.9}{\um}. Through iterative feedback and adjustments, the intensity variation across the entire array is kept well below 1\%. The 795-\SI{}{\nano\meter} addressing laser beams, essential for \(\ket{\mathbb{Z}_2}\) state preparation and local perturbation, are focused through the home-made objective. Two counter-propagating laser beams---one at \SI{780}{\nm} and the other at \SI{480}{\nm}---enable global ground-Rydberg coherent manipulation. A microwave antenna, positioned near the science chamber, generates microwave pulses for Rydberg state manipulation and detection.
}

\LL{
The experimental setup incorporates multiple laser systems for state preparation, qubit control, and detection. The 780-\SI{}{\nano\meter} laser system, a tapered amplifier laser (TA Pro, Toptica), is used for both the MOT cooling beams and Raman light in the Rydberg excitation scheme. Its frequency is stabilized to an ultra-low expansion reference cavity (SLS) with a finesse of 26,000. An AWG drives a single-pass acousto-optic modulator (AOM, SGT200-780-0.5TA-B), optimized for modulation bandwidth, to dynamically adjust pulse frequency and intensity.
For coupling the \(\ket{e} \rightarrow \ket{r}\) transition, a 480-\SI{}{\nano\meter} laser system (frequency-doubled TA-SHG Pro, Toptica) is employed, with the 960-nm seed laser locked to the same reference cavity as the 780-\SI{}{\nano\meter} laser. An AOM driven by a direct digital synthesis (DDS, AD9910) controls the 480-\SI{}{\nano\meter} laser beam, which is intensity-stabilized and remains on throughout the entire Rydberg operation sequence. Fast rising and falling edges of the Rydberg excitation lasers are achieved using electro-optic modulators (EOMs) for precise on-and-off switching.
}

\begin{figure}
  \centering
  \includegraphics[width=\textwidth]{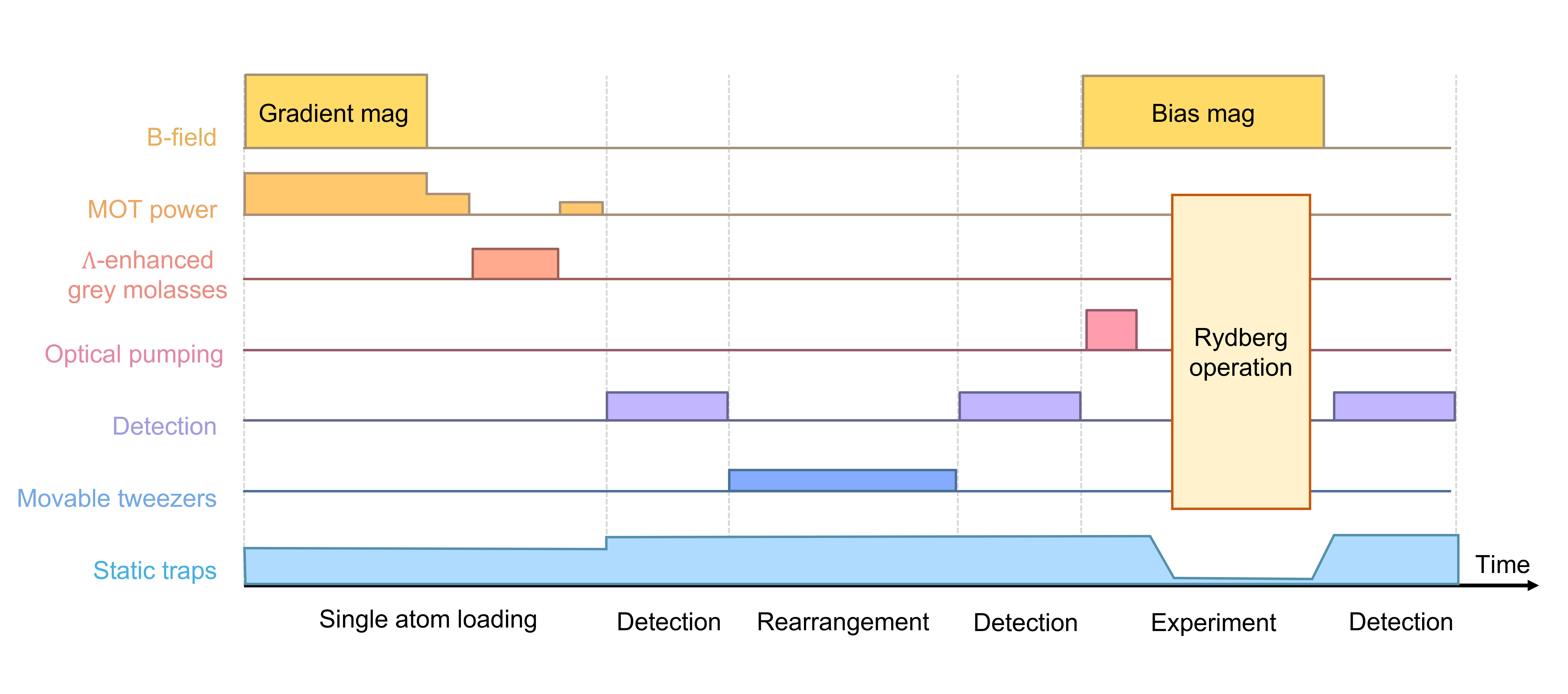}  
  \caption{
    \textbf{Experimental timing sequence.} 
    Summary of a typical experimental sequence as described in section~\hyperref[section:timing sequence]{1.2}. Detailed Rydberg experimental sequences for OTOC and Holevo information measurements are provided in Figs.~\ref{SI_OTOC_sequence},\ref{SI_HI_sequence}.
    }
  \label{Fig:S1}
\end{figure}

\subsection{Experiment timing sequence} 
\label{section:timing sequence}

\LL{
The experimental sequence begins with the preparation of a cold atomic ensemble in the science chamber. A 3D MOT is loaded for \SI{200}{\ms}, producing an atomic cloud with a diameter of \SI{500}{\um} and a temperature of \SI{150}{\micro\kelvin}, measured via time-of-flight (TOF) expansion. To further reduce the temperature, polarization gradient cooling (PGC) is applied. This process involves quickly extinguishing the magnetic field gradient (within \SI{500}{\us}) while increasing the cooling light detuning to approximately $2\pi\times\SI{90}{\MHz}$ and reducing the intensity of both the cooling and repumping lasers. As a result, the atoms are cooled further to approximately \SI{40}{\micro\kelvin}.
}
\LL{
The laser-cooled atomic ensemble serves as a reservoir for stochastic loading of the programmable atomic array. For stochastic loading, \(\Lambda\)-enhanced grey molasses (\(\Lambda\)GM) is implemented using two counter-propagating 795-\SI{}{\nano\meter} laser beams.
An additional stage of polarization gradient cooling (PGC), optimized for in-trap cooling, is then applied. This combined cooling approach results in a single-atom loading efficiency of approximately 80\%, with an average atom temperature of \SI{15}{\micro\kelvin} in traps with a depth of \SI{1}{\milli\kelvin}, as measured using the release-and-recapture (R\&R) method. Atom fluorescence detection utilizes the same 795-\SI{}{\nano\meter} beams used for \(\Lambda\)GM, multiplexed after increasing the optical trap depth to \SI{1.3}{\milli\kelvin}. By carefully balancing heating and cooling rates, a detection fidelity exceeding 99.9\% is achieved with a \SI{30}{\ms} exposure time, while maintaining an average atom loss per detection below 1\%. The use of 795-\SI{}{\nano\meter} fluorescence imaging minimizes crosstalk from the strong 780-\SI{}{\nano\meter} beams, ensuring accurate atom detection.
}

\LL{
An atom rearrangement procedure generates a defect-free, one-dimensional atomic chain. The process begins by linearly ramping up the intensity of the movable tweezers from zero to three times the static trap depth, transferring atoms from the static SLM-generated traps to the movable tweezers. The AODs are then driven with time-dependent waveforms to transport the atoms at an average speed of \SI{100}{\um\per\ms}, following a sinusoidal velocity profile to ensure smooth acceleration and deceleration. The rearrangement sequence is determined using a modified Hungarian algorithm \cite{lee2017defect}, optimizing atom movement column by column. To minimize unnecessary atom loss and heating from moving tweezers sweeping through static traps, the tweezer paths include additional segments to bypass intervening traps. All AOD waveforms for the rearrangement process are pre-computed, allowing for rapid execution of the sequence. Post-rearrangement measurements show that atom temperatures increase to approximately \SI{50}{\micro\kelvin} in the \SI{1}{\milli\kelvin} deep traps.
}

\begin{figure}[t]
  \centering
  \includegraphics[width=0.9\textwidth]{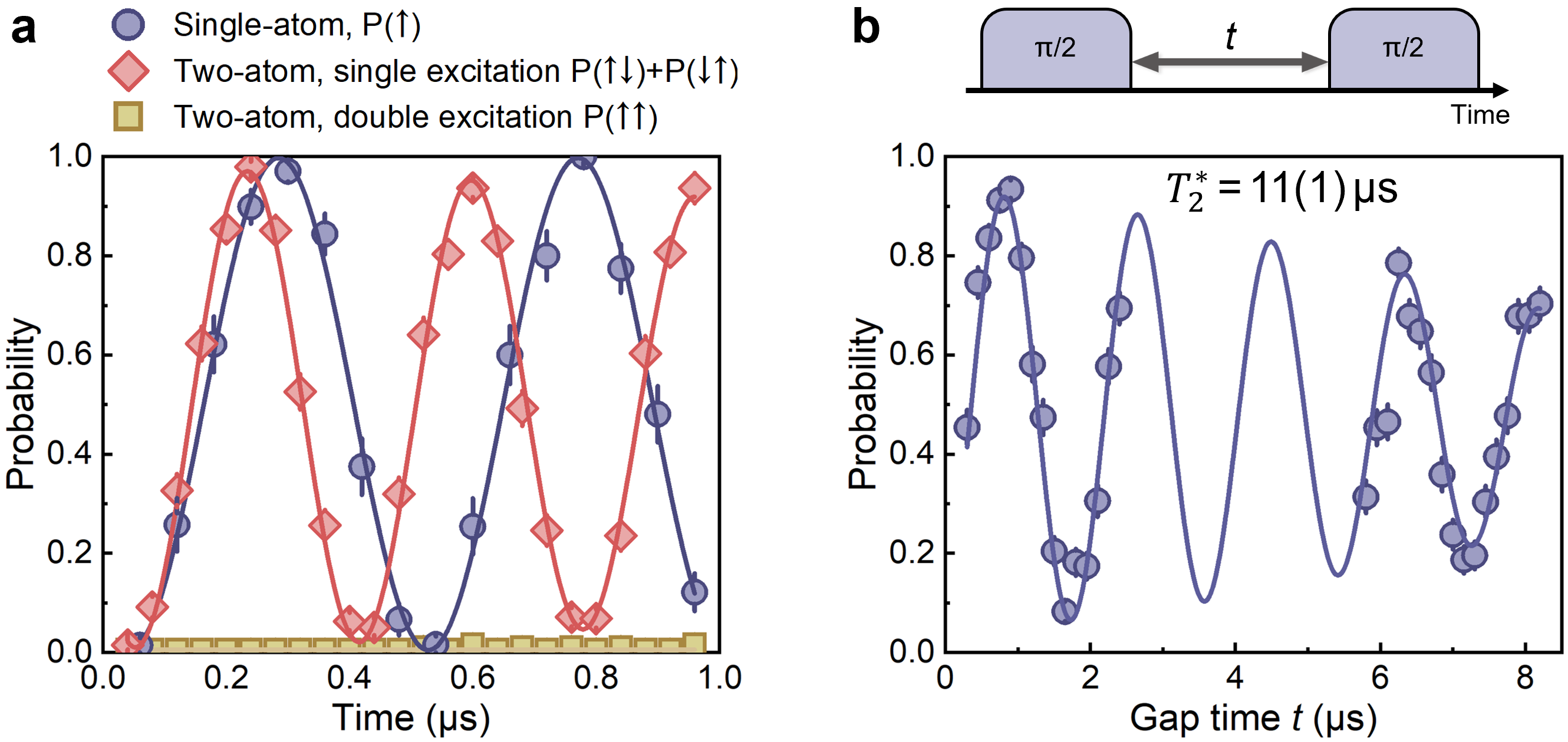}  
  \caption{
\textbf{Ground-Rydberg state Rabi oscillation, Rydberg blockade, and Ramsey coherence.} 
\LL{
\textbf{a}, High-contrast ground-Rydberg state Rabi oscillation (blue circles) is observed when a single atom is driven by 480-\SI{}{\nano\meter} and 780-\SI{}{\nano\meter} Raman lasers. For a pair of atoms within the blockade radius, double Rydberg excitation (yellow squares) is strongly suppressed, and the Rabi frequency of the single excitation (red diamonds) is enhanced by a factor of \(\sqrt{2}\) due to the Rydberg blockade effect. curves are damped sinusoidal fits.
\textbf{b}, Ramsey oscillation showing Gaussian-type inhomogeneous dephasing with a coherence time of \(T^*_2 = \SI{11(1)}{\micro\second}\). The measured probability represents \(\frac{1}{2}(\langle \sigma^y \rangle + 1)\) after the gap time.
}
}
  \label{SI_Block_Rabi}
\end{figure}

\LL{
For coherent Rydberg excitation, a two-photon Raman scheme is employed. A red-detuned 780-\SI{}{\nano\meter} laser with \(\sigma^+\) polarization couples the ground state to the intermediate state \(\ket{5P_{3/2}, F=3, m_F=3}\) (see Extended Data Fig.~1b). The collimated 780-\SI{}{\nano\meter} laser is directed onto the atoms with a beam waist of approximately \SI{300}{\um} and a maximum Rabi frequency of \(\sim 2\pi\times\SI{100}{\MHz}\). Simultaneously, a blue-detuned 480-\SI{}{\nano\meter} laser, also with \(\sigma^+\) polarization, connects the intermediate state to the Rydberg state \(\ket{\uparrow} = \ket{68D_{5/2}, m_J=5/2}\). The 480-\SI{}{\nano\meter} laser is focused onto the atoms with a beam waist of \SI{13}{\um}, with a maximum Rabi frequency of \(\sim 2\pi\times\SI{70}{\MHz}\). Both lasers are detuned from the intermediate state by \(\Delta = 2\pi\times\SI{1.16}{\GHz}\).
A bias magnetic field of \SI{30}{\gauss} is applied throughout the experiment.
Figure~\ref{SI_Block_Rabi}a (blue circles) illustrates the high-contrast ground-Rydberg state Rabi oscillation for a single atom, driven by 780-\SI{}{\nano\meter} and 480-\SI{}{\nano\meter} lasers. When two atoms are placed within the Rydberg blockade region, $\Omega \ll V$, double Rydberg excitations are strongly suppressed, as indicated by the yellow squares in Fig.~\ref{SI_Block_Rabi}a. Red diamonds in Fig.~\ref{SI_Block_Rabi}a shows that the Rabi oscillation between the Bell state $\frac{1}{\sqrt{2}}(\ket{\downarrow}\ket{\uparrow}+\ket{\uparrow}\ket{\downarrow})$ and the ground state $\ket{\downarrow}\ket{\downarrow}$ is enhanced by a factor of \(\sqrt{2}\). Figure~\ref{SI_Block_Rabi}b presents the ground-Rydberg coherence for a single atom, where we perform two \(\pi/2\) ground-Rydberg rotations, separated by a variable gap, to implement a Ramsey sequence. The decay of the Ramsey oscillation fringe reveals a ground-Rydberg coherence time of $T^*_2=\SI{11(1)}{\us}$. The high-contrast Rabi oscillations and long coherence times achieved in our system lay a solid foundation for high-fidelity operations in subsequent experiments.
}

\LL{
To perform local operations on the ground and Rydberg states, we employ 795-\SI{}{\nano\meter} and 480-\SI{}{\nano\meter} addressing laser beams generated by SLMs. The 795-\SI{}{\nano\meter} laser beams are blue-detuned by \(2\pi \times \SI{15}{\GHz}\) from the \(\ket{5S_{1/2}, F=2} \rightarrow \ket{5P_{1/2}}\) transition, providing a light shift of \(2\pi \times \SI{12.2(3)}{\MHz}\). This enables the creation of an alternating pattern of excitable and non-excitable atoms, allowing the system to be prepared in a $\mathbb{Z}_2$-ordered configuration.
Simultaneous imaging of the 795-\SI{}{\nano\meter} atomic fluorescence and the 795-\SI{}{\nano\meter} addressing laser beams on the EMCCD ensures precise spatial overlap between the lasers and the atoms. Measurements indicate that crosstalk to neighbouring atoms in the chain is suppressed to less than \(2\pi \times \SI{1}{\kHz}\).
The 480-\SI{}{\nano\meter} laser beams, resonant with the $\ket{r}-\ket{e}$ transition, are used to enable selective Rydberg-to-ground state transfer and create EIT conditions in nearest-neighbour (NN) and next-nearest-neighbour (NNN) sites. The 480-\SI{}{\nano\meter} laser beam alignment is optimized by maximizing Rydberg-to-ground state transfer efficiency on the target site while minimizing crosstalk on the neighbouring sites. After optimization, more than 96\% of the Rydberg state population in the target site is quickly transferred to the ground state, while the Rydberg state population in the neighbouring sites is reduced by less than 1\%.
}

\LL{
Finally, a state detection scheme is implemented to distinguish between ground and Rydberg states. Within \SI{1}{\us} after the Rydberg operation, the trap depth is rapidly increased to \SI{1.3}{\milli\kelvin} to expel Rydberg atoms from the trapping region while recapturing ground-state atoms. Following this, a strong \SI{2.4}{\GHz} microwave pulse is applied to deplete the remaining Rydberg populations. This comprehensive detection scheme ensures that the probability of misidentifying a Rydberg atom as being in the ground state is less than 1\%.
}

\subsection{$\ket{\mathbb{Z}_2}$ state preparation and characterization}
\label{section:Z2}

\begin{figure}
  \centering
  \includegraphics[width=\textwidth]{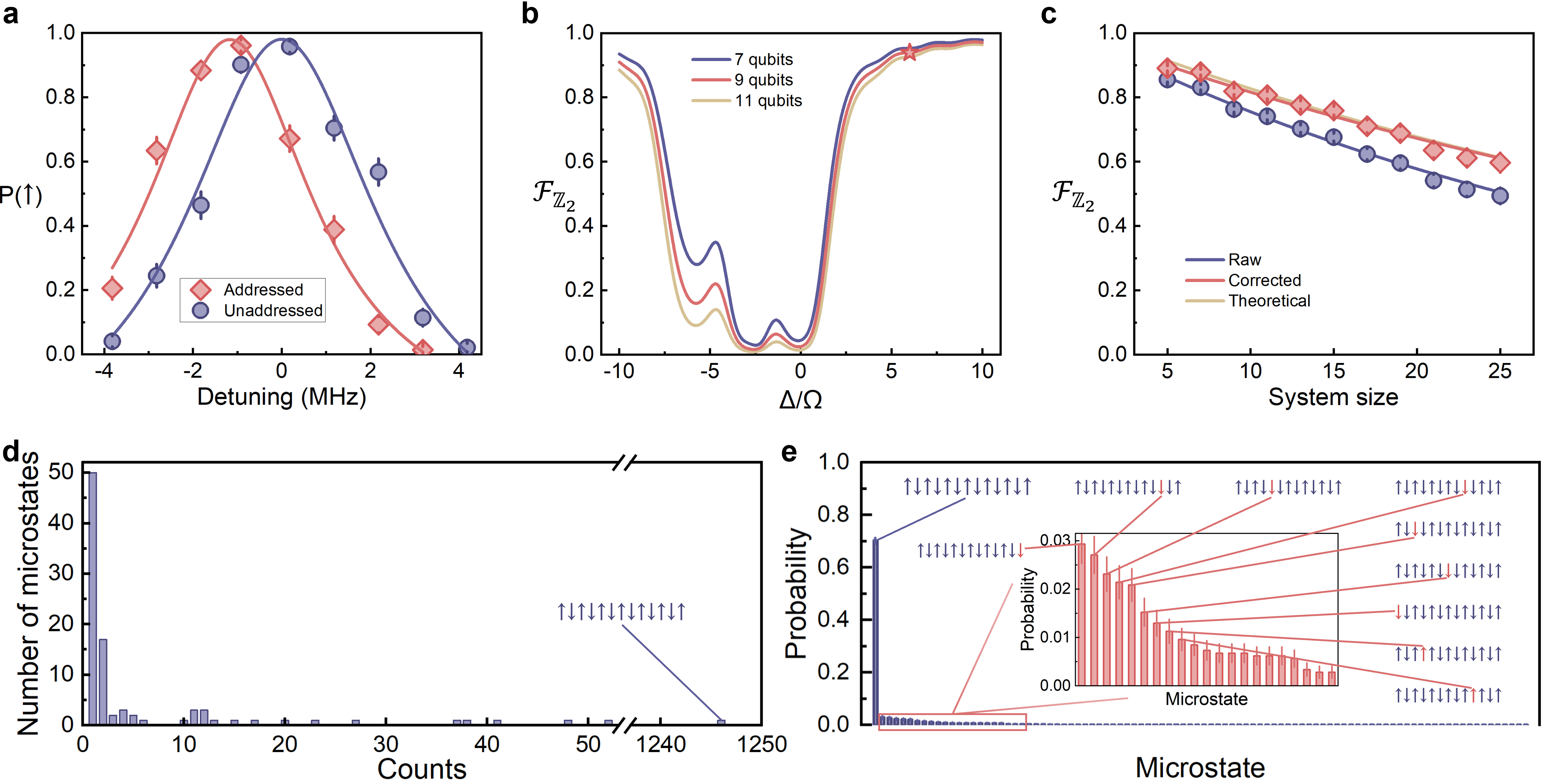}  
  \caption{
    \textbf{$\ket{\mathbb{Z}_2}$ state preparation details.} 
    \textbf{a}, \LL{Rydberg excitation spectra of addressed (red) and unaddressed (blue) atoms. The Rydberg population is shown as a function of Raman excitation laser detuning.}
    \textbf{b}, \LL{Impact of the anti-blockade effect and addressing laser-induced light shift on state preparation fidelity. Simulated $\ket{\mathbb{Z}_2}$ state preparation fidelity as a function of 795-\SI{}{\nano\meter} addressing laser light shift (in units of Rydberg Raman excitation Rabi frequency $\Omega$) for various system sizes, with the nearest-neighbour Rydberg interaction strength $V_{i,i+1}=3\Omega$. The red star marks our experimental condition.}
    \textbf{c}, \LL{$\ket{\mathbb{Z}_2}$ state preparation fidelity as a function of system size. Red and blue lines: the exponential fit for corrected and uncorrected $\ket{\mathbb{Z}_2}$ state fidelity. Yellow line: theoretical upper limit for current experimental addressing approach.}
    \textbf{d}, \LL{Number of 13-qubit microstates as a function of occurrences from 1,774 experiments. Successfully prepared $\ket{\mathbb{Z}_2}$ state: 1,246 counts (70\% of events).}
    \textbf{e}, \LL{Measured 13-qubit microstate distribution. Inset: dominant error states.}
  }
  \label{Fig:S2}
\end{figure}

\LL{
Rapid thermalization is ubiquitous in ergodic quantum many-body systems, leading to extensive research aimed at uncovering systems with fundamentally different dynamics~\cite{deutsch1991quantum,srednicki1994chaos,rigol2008thermalization,deutsch2018eigenstate,lewis2019dynamics}.
Recently, studies involving Rydberg atom arrays have identified many-body scar states~\cite{bernien2017probing,turner2018weak,choi2019emergent,ho2019periodic,bluvstein2021controlling,turner2021correspondence,serbyn2021quantum,maskara2021discrete,zhang2023many,dong2023disorder}, which show weak breaking of ergodicity and maintain a degree of coherence over extended timescales.
A previous theoretical work~\cite{yuan2022quantum} has predicted that when the $\mathbb{Z}_2$-ordered scar states are used as initial states for studying the spatial-temporal evolution of out-of-time-ordered correlators and Holevo information, one may observe persistent information backflow and an unusual breakdown of quantum chaos. These results suggest novel avenues for investigating unique dynamics of quantum information scrambling in kinetically constrained many-body systems.
Here, we investigate quantum information scrambling using two initial states: (1) the $\mathbb{Z}_2$-ordered state \(\ket{\mathbb{Z}_2} = \ket{\uparrow \downarrow \uparrow \downarrow \uparrow...}\) and (2) the trivial product state \(\ket{\mathbf{0}} = \ket{\downarrow \downarrow \downarrow \downarrow \downarrow...}\). While \(\ket{\mathbf{0}}\) can be readily prepared using optical pumping, achieving high-fidelity preparation of \(\ket{\mathbb{Z}_2}\) state in large-scale systems remains challenging.
Previous studies have successfully prepared \(\ket{\mathbb{Z}_2}\) states in one-dimensional and two-dimensional atom arrays through adiabatic state transfer techniques~\cite{bernien2017probing,omran2019generation,bluvstein2021controlling}, where the ground state of an engineered Hamiltonian is adiabatically transformed from \(\ket{\mathbf{0}}\) to \(\ket{\mathbb{Z}_2}\). However, as the system size increases, the exponential growth of the Hilbert space results in diminishing energy gaps, which causes a rapid decline of the state preparation fidelity.
}

\LL{
Our experiment employs a scalable state preparation approach that combines global Rydberg excitation with site-selective laser addressing. A SLM is used to generate a customized light shift pattern across the atom array, selectively detuning certain sites from the ground-Rydberg transition (Fig.~\ref{Fig:S2}a), thereby creating an alternating arrangement of excitable and non-excitable atoms.
The 795-\SI{}{\nano\meter} addressing beams, $2\pi \times \SI{15}{\GHz}$ detuned from the D1 line resonance, induce an energy shift of $2\pi \times \SI{12.2(3)}{MHz}$ on the ground state. The scattering rate caused by the 795-\SI{}{\nano\meter} addressing beams is approximately $2\pi \times \SI{5}{kHz}$, significantly lower than that caused by the Raman laser beams. Numerical simulations (Fig.~\ref{Fig:S2}b) indicate that to maximize the fidelity of the \(\ket{\mathbb{Z}_2}\) state, the sign of the light shift must be opposite to that of the Rydberg interaction, avoiding anti-blockade effects. Under our experimental conditions, numerical simulations yield a preparation infidelity of $\sim 0.012$ per qubit.
}

\LL{
Our method remains highly effective as the system size scales up (Fig.~\ref{Fig:S2}c). The preparation infidelity is primarily attributed to uncorrelated single-qubit flip errors: 0.8\% per qubit from Rydberg excitation inefficiency, 1.2\% per qubit due to finite energy shifts, and 1\% ($\ket{\downarrow}\rightarrow\ket{\uparrow}$) and 0.5\% ($\ket{\uparrow}\rightarrow\ket{\downarrow}$) from state detection errors. This decomposition of many-body scar state preparation into single-qubit state preparations enhances the scalability of our approach compared to adiabatic transfer protocols.
In an array of up to 25 atoms, we achieved the target crystalline state with a measured fidelity of 49(3)\%, which is corrected to 60(3)\% after considering the detection errors. This high-fidelity preparation, in a Hilbert space of dimension $2^{25}$, underscores the robustness and scalability of our protocol.
}

\LL{
Microstate distribution analysis of the experimentally prepared $\ket{\mathbb{Z}_2}$ state reveals non-Poissonian error occurrences (Fig.~\ref{Fig:S2}d).
Figure~\ref{Fig:S2}e shows the measured microstates distribution, highlighting that the dominant errors are single-qubit flips from $\ket{\uparrow}$ to $\ket{\downarrow}$, further confirming that our approach employs single-qubit operations to prepare a many-body state. This predictable error distribution facilitates error mitigation in the experimentally measured data for quantum information scrambling in $\ket{\mathbb{Z}_2}$ state, as discussed in section~\ref{section:error}.
}

\begin{figure}
  \centering  \includegraphics[width=0.75\textwidth]{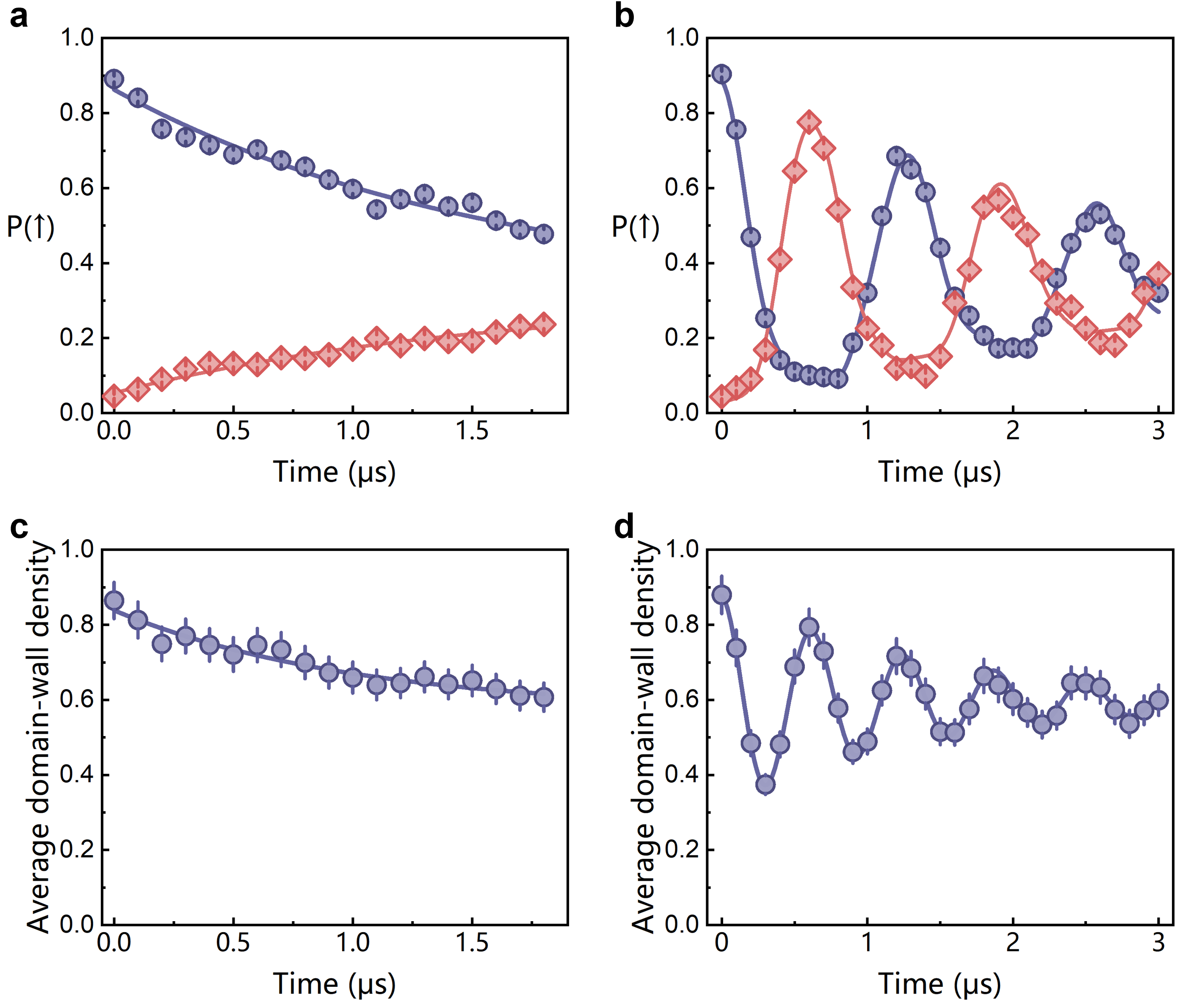}  
  \caption{
\textbf{$\ket{\mathbb{Z}_2}$ state dynamics under PXP Hamiltonian evolution.} 
\LL{
\textbf{a} and \textbf{b}, Population of \(\ket{\uparrow}\) (blue) and \(\ket{\downarrow}\) (red) initialized qubits during \textbf{a}, forward-and-backward evolution \((e^{-iHt}\) followed by \(e^{iHt})\) and \textbf{b}, forward-only \((e^{-iHt})\) evolution. Curves in \textbf{a} represent exponential fits,
while \textbf{b} is fitted with damped Fourier series up to the 5th order.
\textbf{c} and \textbf{d}, Average domain-wall density of the central 13 qubits during \textbf{c}, forward-and-backward evolution, and \textbf{d}, forward-only evolution. The blue curve in \textbf{c} is an exponential fit,
while \textbf{d} is fitted with a damped sinusoidal function.
}
}
  \label{Fig:S3}
\end{figure}

\LL{Next, the evolution and lifetime of the $\ket{\mathbb{Z}_2}$ state were investigated. The dynamics of the $\ket{\mathbb{Z}_2}$ state were measured under both forward-and-backward evolution (\(e^{-iHt}\) followed by \(e^{iHt}\)) and forward-only evolution (\(e^{-iHt}\)) (Fig.~\ref{Fig:S3}) using the PXP Hamiltonian \(H = \sum_i P_i \sigma^x_{i+1} P_{i+2}\)~\cite{lesanovsky2012interacting,turner2018weak,cheng2024emergent}. To characterize the evolution, the Rydberg state population \(P(\uparrow)\) (Fig.~\ref{Fig:S3}a,b) and the average domain-wall density (Fig.~\ref{Fig:S3}c,d) are measured.
Exponential fits to the data in Fig.~\ref{Fig:S3}a,c show the decay rate of the $\ket{\mathbb{Z}_2}$ state under forward-reverse evolution, with a 1/e lifetime of approximately \SI{1.6(1)}{\micro\second} for population, and \SI{1.0(3)}{\micro\second} for average domain-wall density. This decay limits the contrast in the raw ZZ-OTOC data presented in Fig. 3f of the main text. Forward-only evolution of the average domain-wall density (Fig.~\ref{Fig:S3}d) is fitted to a damped sinusoidal function, yielding a $\ket{\mathbb{Z}_2}$ state lifetime of approximately \SI{1.5(1)}{\micro\second}. 
\LL{
The Rydberg population dynamics (Fig.~\ref{Fig:S3}b) are fitted using a damped Fourier series, yielding a 1/e decay time of approximately \SI{2.8(2)}{\us}.
}
Additionally, in section~\ref{section:error}, the data from Fig.~\ref{Fig:S3}b are fitted to the error model to characterize the noise in the driving fields. The finite lifetime of the $\ket{\mathbb{Z}_2}$ state results in an overall loss during transport, as reflected in the global decay of the measured Holevo information in Fig. 4b of the main text.
The breakdown of the defect-free $\mathbb{Z}_2$-ordered system into subsystems, indicated by the decay of domain-wall density (Fig.~\ref{Fig:S3}c) and the reduction in oscillation contrast (Fig.~\ref{Fig:S3}d), smears the dynamics and impedes quantum information propagation.
While the state preparation errors can be easily corrected, the mitigation of errors during the evolution is far more complex. The above-detailed characterization of the prepared $\ket{\mathbb{Z}_2}$ state, particularly the decay rates during driven evolution, provides insights into the underlying noise sources (see section~\ref{section:error} for details on the noise model) and informs subsequent error mitigation strategies.}

\LL{
In summary, a scalable approach for preparing $\ket{\mathbb{Z}_2}$ states in large atomic arrays has been demonstrated. By combining global Rydberg excitation with site-selective addressing, high-fidelity state preparation was achieved in systems of up to 25 qubits. Detailed characterization of the state evolution provides valuable understanding of quantum scar dynamics and systematic noise sources. These findings lay the groundwork for further exploration of quantum information scrambling in scarred systems.
}

\section{Theoretical modeling and numerical simulation}

\subsection{Effective Hamiltonian and numerical simulation methods}

\LL{Our experimental setup consists of a linear array of 25 individual atoms trapped in optical tweezers. The system dynamics are governed by the microscopic Hamiltonian:
\begin{equation}
H = \sum_i \left[\frac{\Omega}{2} \sigma^x_i - \Delta n_i \right] + \sum_{i,j} V_{ij} n_i n_j,
\label{eq:rydberg_hamiltonian}
\end{equation}
where $\Omega$ is the Rabi frequency, $\Delta$ is the detuning, and $n_j = (1 + \sigma^z_j)/2$ is the projector onto $\ket{\uparrow}$ at site $j$, indicating whether the atom is in the Rydberg state. The interaction term $V_{ij} = C_6/R^6_{ij}$ represents the van der Waals interaction between atoms in $\ket{\uparrow}$ state at sites $i$ and $j$, with $C_6$ being the van der Waals coefficient and $R_{ij}$ the distance between atoms.}

\LL{In the regime of strong nearest-neighbour interactions ($\Omega \ll V_{i,i+1}$), neglecting longer-range interactions ($V_{i,j>i+1}$), and setting the detuning $\Delta = 0$, the Hamiltonian simplifies to the effective PXP Hamiltonian via the Schrieffer-Wolff transformation:
\begin{equation}
H_\text{PXP} = \sum_i P_i \sigma^x_{i+1} P_{i+2},
\label{eq:pxp_hamiltonian}
\end{equation}
where $P_i = (1 - \sigma^z_i)/2$ is the projector onto the $\ket{\downarrow}$ state at site $i$, and $\sigma^{x,y,z}_i$ are Pauli matrices for the $i$-th qubit.
The local three-body terms $P_i \sigma^x_{i+1} P_{i+2}$ impose kinetic constraints, allowing a Rydberg atom state to flip only if both neighbouring atoms are in the spin-down $\ket{\downarrow}$ state. This constraint effectively rules out configurations $\ket{\cdots \uparrow_i\uparrow_{i+1}\cdots}$ from the computational basis, as adjacent Rydberg excitations are forbidden by the blockade effect.
The low-energy subspace, spanned by configurations without adjacent excited states, can be described by the projector:
\begin{equation}
\mathcal{P} = \prod_j \left( 1 - n_j n_{j+1}\right).
\label{eq:projector}
\end{equation}
This constrained subspace forms the effective Hilbert space of reduced dimensionality, governing the system's constrained dynamics. Remarkably, the dimension of this effective Hilbert space grows according to the Fibonacci sequence, scaling as $\phi^N$, where $\phi = (1+\sqrt{5})/2$ is the golden ratio and $N$ is the system size. 
This reduced dimensionality reflects the exclusion of certain configurations due to the kinetic constraints, which significantly simplifies the dynamics.}

\LL{The PXP model exhibits three notable symmetries:
(1) Discrete spatial inversion symmetry $\mathcal{I}$, mapping $j \rightarrow N-j+1$.
(2) Translational symmetry, applicable under periodic boundary conditions.
(3) Particle-hole symmetry, represented by $\mathcal{C} = \prod_j \sigma_j^z$, resulting in $\mathcal{C} H_\text{PXP}\mathcal{C} = -H_\text{PXP}$.
The particle-hole symmetry plays a crucial role in reversing the Hamiltonian evolution $\text{exp}(-iH t)$ in our experiment. However, this symmetry is only present in the PXP Hamiltonian $H_\text{PXP}$, not in the full Rydberg Hamiltonian $H$ governing the experiment, which causes imperfections in time reversal.}

\LL{When initialized in the $\ket{\mathbb{Z}_2}$ state, the Rydberg atom system exhibits wavefunction oscillations with a slow decay. This decay is partially due to state losses and decoherence during the experimental evolution, but also arises from imperfections in the many-body scars and a small overlap with thermal states in the underlying PXP model.
To better understand the intrinsic dynamics' contribution to the observed decay, we performed numerical simulations of the revival behaviour, initializing the system of size $N=25$ in the $\ket{\mathbb{Z}_2}$ state. The results, shown in Fig.~\ref{Fig:SI_toy}c, provide evidence that a significant portion of the observed decay can be attributed to intrinsic features of the PXP model.}

\begin{figure}[t]
\centering
\includegraphics[width=0.7\textwidth]{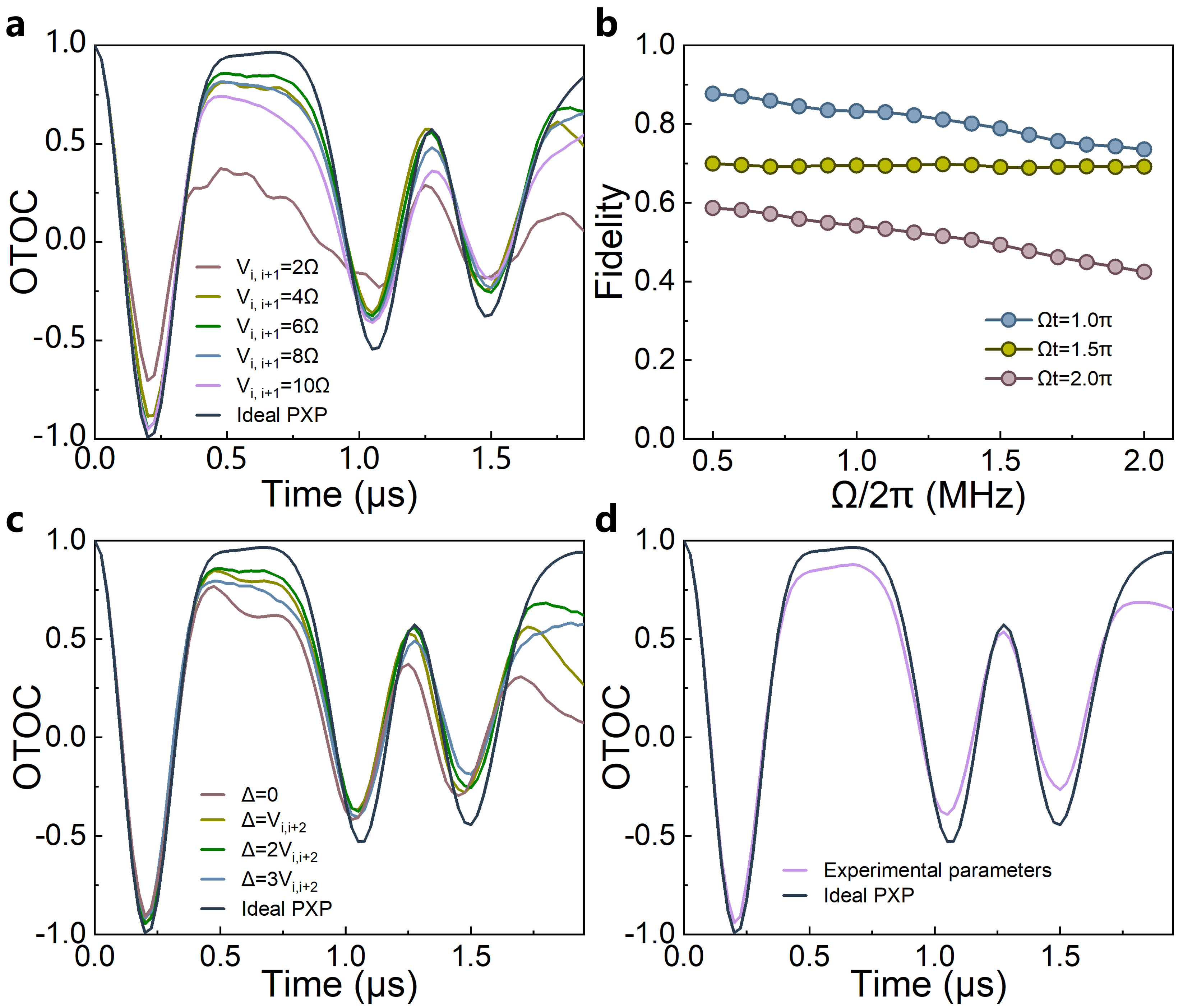}
\caption{\textbf{Optimization of experimental parameters for closely approximating the ideal PXP Hamiltonian dynamics.} 
\LL{Numerical simulations of OTOC and time-reversal fidelity of the central qubit in a 10-qubit chain with periodic boundary conditions. \textbf{a}, Simulated OTOC dynamics for distinct nearest-neighbour (NN) Rydberg interactions $V_{i,i+1}$, in comparison to those of the ideal PXP model. The interaction strengths vary from $2\Omega$ to $10\Omega$ in steps of $2\Omega$. The optimal choice, $V_{i,i+1} = 6\Omega$, best matches the ideal case.
\textbf{b}, Time-reversal fidelity as a function of Rabi frequency $\Omega$, with $V_{i,i+1}$ fixed at $6\Omega$. Different time evolutions are shown for $\Omega t = 1.0\pi$ (blue symbols), $\Omega t = 1.5\pi$ (yellow symbols), and $\Omega t = 2.0\pi$ (red symbols). As $\Omega$ increases, the fidelity decreases.
\textbf{c}, OTOC dynamics for different detuning values $\Delta$, ranging from $\Delta = 0$ to $3V_{i,i+2}$, where $V_{i,i+2}$ is the next-nearest-neighbour (NNN) interaction. The black curve represents the ideal PXP model. The optimal detuning, $\Delta = 2V_{i,i+2}$, best matches the ideal dynamics.
\textbf{d}, A comparison of OTOC dynamics using the optimized experimental parameters (violet curve) and the ideal PXP model (black curve), showing close agreement. The optimized parameters effectively reproduce the key oscillations observed in the OTOC.}
} 
\label{FIG:experimental parameters}
\end{figure}

\LL{For all numerical simulations, we employ the Rydberg Hamiltonian (\ref{eq:rydberg_hamiltonian}) to closely approximate the experimental physical conditions. For atomic chains of length $\leq$ 13, we utilize exact diagonalization to efficiently compute the full time evolution. However, for chains exceeding 13 atoms, exact diagonalization becomes computationally prohibitive due to exponentially increasing memory requirements and computation time, as well as limitations in available computational resources. Consequently, we implement the Matrix Product Operator (MPO) method to accelerate our numerical calculations in these cases. This approach allows us to extend our simulations to longer atomic chains while maintaining computational feasibility and numerical accuracy. Unless otherwise specified, all simulations employ parameters identical to those in the experimental setup (detailed in Section \ref{section:experiment}). To account for experimental uncertainties, we implement a Monte Carlo method: accounting for fluctuations in Rabi frequency and detuning, laser noise, uncertainties in atomic positions and other relevant experimental parameters, all variables are randomly sampled from their respective probability distributions. Typically, we perform 200 runs and average the results. This comprehensive approach provides a robust representation of the system's behaviour under realistic experimental conditions.}

\subsection{Parameter tuning for optimal quantum dynamics}

\label{section:experimental parameter}

\LL{The approximation of the effective PXP Hamiltonian relies on two key assumptions:
(1) Neglecting longer-range interactions $V_{i,j>i+1}$.
(2) Ensuring that the nearest-neighbour interactions dominate over the Rabi frequency ($V_{i,i+1} \gg \Omega$).}

\LL{However, in practice, the next-nearest-neighbour interactions $V_{i,i+2}$ cannot be ignored. To approximate the PXP model closely, the system must operate in the regime where $V_{i,i+2} \ll \Omega \ll V_{i,i+1}$. This poses a challenge because, in our 1D geometry with equally spaced atoms, the ratio $V_{i,i+1}/V_{i,i+2}$ is fixed at approximately 64. For example, setting $V_{i,i+1} \sim 16\Omega$ to meet the second assumption results in $\Omega \sim 4V_{i,i+2}$, making it difficult to fully suppress the effects of $V_{i,i+2}$. This creates a trade-off between the two key assumptions.}

\LL{To find an optimal ratio of $V_{i,i+1}/\Omega$, we performed numerical simulations of the ZZ-OTOC with varying parameters. We identified the optimal ratio that minimizes the \textit{collapse-and-revival} decay, as shown in Fig.~\ref{FIG:experimental parameters}a. Based on these results, we set the experimental ratio of $V_{i,i+1}/\Omega$ to 6, which differs from the theoretical intermediate value of $\Omega=\sqrt{V_{i,i+1} V_{i,i+2}}$.}

\LL{The selection of the Rabi frequency involves balancing two competing factors: (1) Time-reversal fidelity: Our simulations (Fig.~\ref{FIG:experimental parameters}b) show that increasing the Rabi frequency reduces time-reversal fidelity when $V_{i,i+1}/\Omega$ is fixed at 6 and the interval between forward and backward evolution is fixed at \SI{200}{\nano\second}. (2) Single-atom coherence: Higher Rabi frequencies improve single-atom coherence, which is mainly limited by laser phase noise.}

\LL{After carefully weighing these factors, we chose a Rabi frequency of $\Omega = 2\pi \times \SI{1.21(1)}{MHz}$. This value strikes a balance between maintaining adequate time-reversal fidelity and ensuring sufficient single-atom coherence for our experimental needs.}

\LL{In addition to optimizing $V_{i,i+1}$ and $\Omega$, we introduced a small detuning $\Delta$ to counteract the residual next-nearest-neighbour interactions $V_{i,i+2}$. Our simulations indicate that setting $\Delta = 2V_{i,i+2}$ best captures the OTOC \textit{collapse-and-revival} phenomenon (Fig.~\ref{FIG:experimental parameters}c).
}

\LL{
In our OTOC measurement, a gap is inserted between the forward and backward evolutions, for the implementation of local and global single-qubit $\sigma^z$ operations.
To minimize the gap time to $\sim$ \SI{200}{ns}, we execute the local perturbations and global $\sigma^z$ rotations in parallel, exploiting their commutative properties. We numerically investigated the impact of this finite gap on the \textit{collapse-and-revival} behaviour of information observed in the experiment, as shown in Fig.~\ref{FIG:experimental parameters}d. Our analysis reveals that the \SI{200}{ns} single-qubit operation time used in the experiment does not significantly affect the \textit{collapse-and-revival} phenomenon.
}

\LL{
These parameter optimizations allow us to closely approximate the PXP model in our experimental setup (Fig.~\ref{FIG:experimental parameters}d), despite the inherent constraints of our system.
These results highlight the importance of precise parameter tuning for accurately implementing the target Hamiltonian and maximizing the fidelity of quantum many-body scars' evolution in Rydberg atom systems. The optimized parameters facilitate the experimental realization of coherent spin rotation dynamics under kinetic constraints, closely approximating the ideal behaviour of the PXP model.}

\subsection{Boundary and finite-size effects}
\label{section:boundary}

\begin{figure}[t]
\centering
\includegraphics[width=1.0\textwidth]{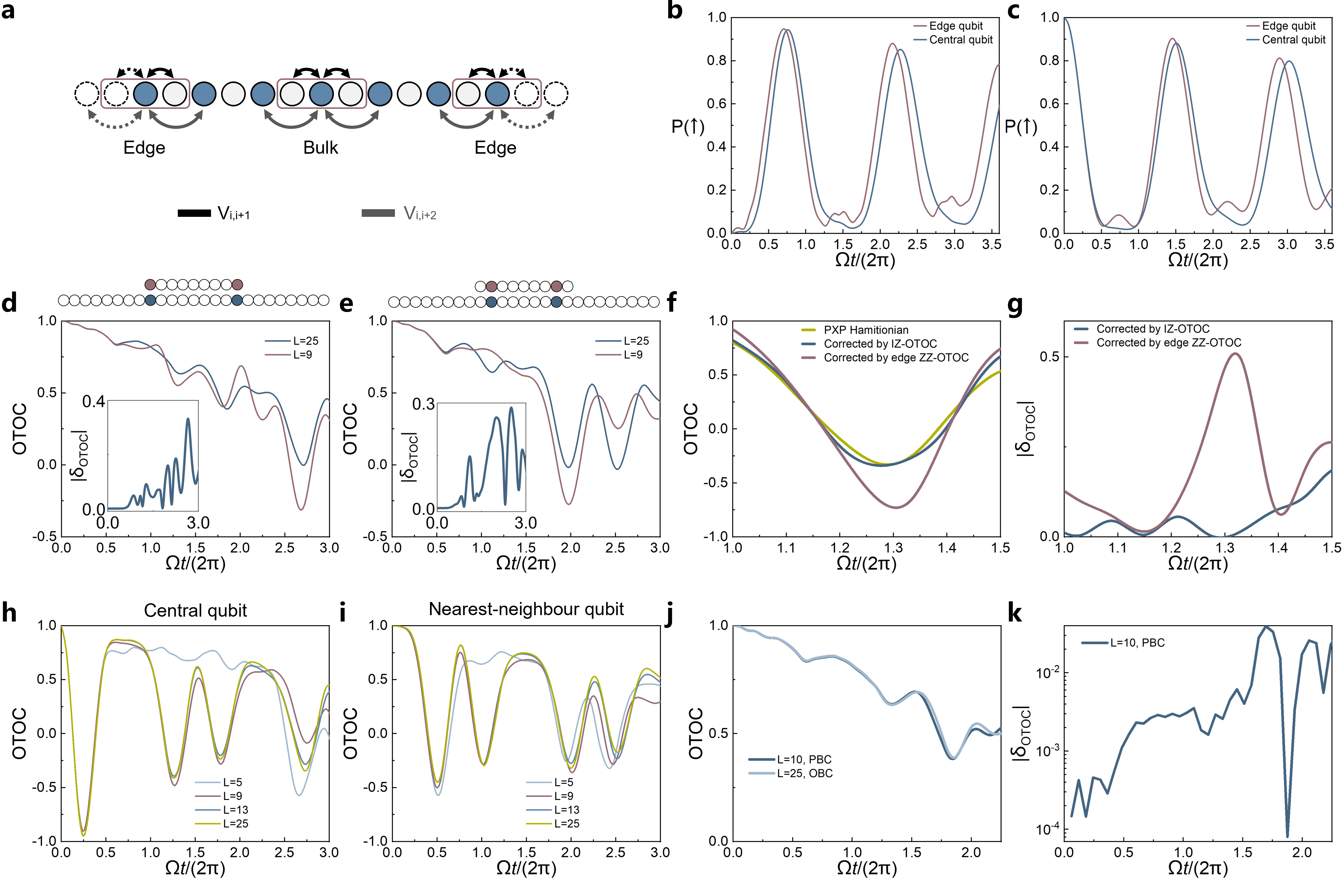}
\caption{\LL{\textbf{Boundary and finite-size effects.}
Simulated dynamics of OTOCs and time evolution under the ideal Rydberg Hamiltonian for various system sizes and boundary conditions, initialized in the $\ket{\mathbb{Z}_2}$ state.
\textbf{a}, Schematic illustration of boundary effects in a 13-qubit chain. Boundary qubits lack nearest and next-nearest neighbours, breaking the translational symmetry.
\textbf{b} and \textbf{c}, Numerical results of $\left\langle n_j(t) \right\rangle$ for edge (red) and central (blue) qubits in a 13-qubit chain, starting from the $\ket{\downarrow}$ state (\textbf{b}) and the $\ket{\uparrow}$ state (\textbf{c}). Accelerated oscillations of edge qubits reflect boundary effects.
\textbf{d} and \textbf{e}, \ZP{OTOC dynamics for \textbf{d}, the edge qubit and \textbf{e}, its neighbouring qubit in 9-qubit chain (red curves), compared to the corresponding qubits in 25-qubit chain (blue curves). The schematics above each plot, marked with coloured circles, illustrate the specific qubit positions within their respective chains. Insets show deviations $|\delta_\text{OTOC}|$  between the two dynamics, indicating significant boundary effects that propagate into the interior of the chain.}
\textbf{f}, OTOC evolution under the ideal PXP Hamiltonian (yellow), and under the noisy Rydberg Hamiltonian normalized using IZ-OTOC (blue) and the edge qubit's ZZ-OTOC (red). 
\textbf{g}, Deviations of normalized OTOCs from the ideal PXP case (red: edge ZZ-OTOC, blue: IZ-OTOC). Significant deviations observed in the case normalized using ZZ-OTOC from the edge qubit.
\textbf{h} and \textbf{i}, Simulated OTOC dynamics for the central qubit (\textbf{h}) and nearest-neighbour (NN) qubit (\textbf{i}) in chains of various lengths (5, 9, 13, and 25 qubits) with open boundary conditions. Distinct variations of OTOC dynamics in smaller systems show the finite-size effects.
\textbf{j} and \textbf{k}, Comparison of OTOC dynamics for qubits furthest from the perturbation in a 10-qubit chain (periodic boundary conditions, PBC, dark blue) and edge qubits in a 25-qubit chain (open boundary conditions, OBC, light blue). The negligible deviation $|\delta_\text{OTOC}|$ validates the use of a 10-qubit PBC system to approximate bulk qubit behaviour in a larger 25-qubit OBC system, matching our experimental conditions.}}
\label{Fig:Boundary and finite-size effect}
\end{figure}

\LL{In finite-sized qubit chains, the existance of boundary qubits breaks the translational symmetry of the PXP Hamiltonian for certain initial states such as $\ket{\mathbb{Z}_2} = \ket{...\uparrow \downarrow\uparrow\downarrow\uparrow...}$ and $\ket{\mathbf{0}} = \ket{...\downarrow \downarrow\downarrow\downarrow\downarrow...}$. This symmetry breaking results in different interaction strengths between edge and bulk qubits (Fig.~\ref{Fig:Boundary and finite-size effect}a), introducing significant boundary effects. Additionally, the finite number of particles leads to variations in the dynamical evolution of bulk atoms across different system sizes, introducing finite-size effects. For the Rydberg Hamiltonian, which includes long-range interactions, these boundary effects are further amplified due to differences in the residual interactions at the edges compared to the bulk.}

\LL{To quantitatively analyse the boundary effect, we numerically simulate the evolution dynamics of both the central and the edge qubit in a 13-qubit chain, starting from the $\ket{\mathbb{Z}_2}$ state under the ideal Rydberg Hamiltonian. As shown in Fig.~\ref{Fig:Boundary and finite-size effect}b,c, boundary effects cause the edge qubits to exhibit accelerated periodic oscillations. This acceleration arises from the reduced constraints on edge qubits, resulting in a stronger effective driving strength.}

\LL{Moreover, both numerical simulations and experimental results reveal that boundary effects gradually alter the spin rotations from the outer edge inward, causing the initially uniform propagating wavefront to bend during the evolution of the $\ket{\mathbb{Z}_2}$ initial state, underscoring the critical role of boundary effects in shaping the system's dynamics.}

\LL{Additionally, we investigate the impact of boundary effects on OTOC measurements by \ZP{comparing the OTOC of the edge qubit and its neighbouring qubit in a 9-qubit chain with the corresponding qubits in a 25-qubit chain.} As shown in the inset of Fig.~\ref{Fig:Boundary and finite-size effect}d,e, significant differences are observed, highlighting the substantial influence of boundary effects on the OTOC dynamics.}
\ZYW{To ensure the accuracy of the observed OTOC dynamics over extended evolution times, a larger system size is required.}

\LL{Notably, as discussed in section~\hyperref[subsection:Error Mitigate]{4.4}, a reference OTOC measurement is required to serve as the normalization denominator for mitigating experimental imperfections. Theoretical analysis suggests that the IZ-OTOC is an appropriate reference for this purpose~\cite{swingle2018resilience,mi2021information}. In our experiment, qubits near the boundary in the long chain remain unaffected by the local perturbation for a certain period, making them potentially suitable for approximating the IZ-OTOC dynamics for central qubits. To evaluate the effectiveness of different normalization schemes, we compared the edge qubit’s ZZ-OTOC with the central qubits’ IZ-OTOC as the normalization denominator in a noisy environment. Our numerical simulations indicate that using the edge qubit’s ZZ-OTOC as the normalization denominator introduces significant over-corrections (as large as 50\%) and temporal misalignment, as shown in Fig.~\ref{Fig:Boundary and finite-size effect}f,g. These findings underscore the importance of employing the IZ-OTOC for accurate normalization and reliable mitigation.}

\LL{
We further study finite-size effects by comparing the OTOC dynamics of the central two qubits in chains of varying lengths (5-, 9-, 13-, and 25-qubit chains with open boundary conditions). Numerical simulations reveal that in smaller atomic systems, finite-size effects are very pronounced, as shown in Fig.~\ref{Fig:Boundary and finite-size effect}h,i.} 
\XDS{This suggests that we need to extend the chain length to ensure that our experimental observations accurately reflect the scrambling of quantum information. However, simulating 25 atoms requires substantial computational resources, even when using the MPO method. To address this challenge, we simulate both a 25-qubit chain with open boundary conditions (OBC) and a 10-qubit chain with periodic boundary conditions (PBC), shown in Fig.~\ref{Fig:Boundary and finite-size effect}j,k. We compare the OTOC dynamics of the atom furthest from the perturbation.  The results reveal negligible differences within the experimental timescale, as the atom remains unaffected by the equivalent local perturbations at both ends during this period.}

\XDS{Based on this analysis, when simulating the dynamics of OTOC, we employ simulations of a 10-qubit chain with PBC. For experimental studies of quantum information scrambling within a constrained Hilbert space, we employ a 25-qubit chain to shield the central 13 atoms, thereby avoiding boundary effects and finite-size effects.}

\section{Probing quantum information scrambling and transport dynamics}

\LL{
Quantum information dynamics, the study of how local quantum information propagates in complex many-body systems, plays a crucial role in the understanding of many fundamental questions.  It can be employed to study the limits on the speed of information propagation in quantum systems~\cite{
lieb1972finite,hastings2006spectral,sekino2008fast,hauke2013spread,eisert2013breakdown,foss2015nearly,maldacena2016bound,von2018operator,nahum2018operator,nachtergaele2019quasi,else2020improved,gong2022bounds,chen2023speed,
cheneau2012light,langen2013local,richerme2014non,jurcevic2014quasiparticle}.
It is also deeply linked to quantum chaos and quantum thermodynamics~\cite{hosur2016chaos,kukuljan2017weak,lewis2019dynamics,yuan2022quantum,kaufman2016quantum,brydges2019probing,zhu2022observation}, providing insights into dynamics in thermal and non-ergodic systems ~\cite{
nandkishore2015many,schreiber2015observation,choi2016exploring,abanin2019colloquium,lukin2019probing,rispoli2019quantum,
huang2017out,chen2017out,deng2017logarithmic,he2017characterizing,swingle2017slow,fan2017out,banuls2017dynamics}.
Moreover, in black hole physics, quantum information scrambling is related to the information paradox~\cite{hayden2007black,almheiri2013black,shenker2014black,shenker2015stringy,qi2018does}.
Additionally, quantum information dynamics has broad potential applications. In quantum computing, understanding information spreading is vital for developing noise-resistant systems and enhancing quantum error correction~\cite{landsman2019verified,mi2021information}; in quantum metrology, it could inspire novel precision measurement protocols~\cite{li2023improving}.
In this section, we present the details of our study on quantum information scrambling and transport using a Rydberg atom array, including the measurements of out-of-time-ordered correlators, and Holevo information.
}

\subsection{OTOC measurements details}
\label{Sec:OTOC}
\begin{figure}[t]
  \centering
  \includegraphics[width=0.9\textwidth]{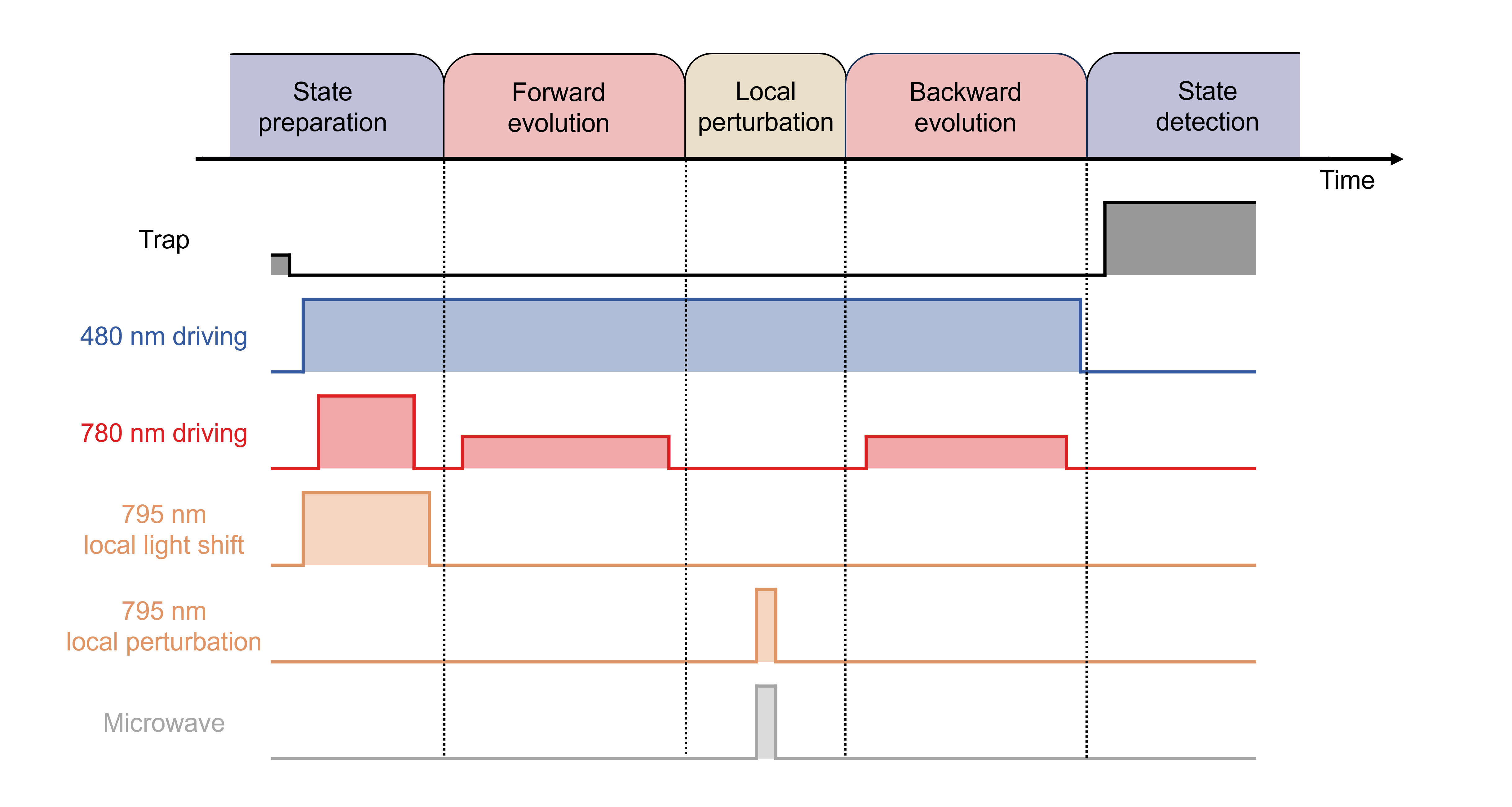}  
  \caption{
\textbf{Pulse sequence for OTOC measurements.} 
}
  \label{SI_OTOC_sequence}
\end{figure}

\LL{
Out-of-time-ordered correlators (OTOCs) have become a powerful tool for investigating quantum information scrambling, revealing how local perturbations propagate through quantum systems~\cite{Larkin1969,swingle2018unscrambling,xu2020accessing,xu2024scrambling,huang2017out,fan2017out,chen2016universal,chen2017out,he2017characterizing,swingle2017slow,hashimoto2017out,lewis2019dynamics}. 
Experimental demonstrations of OTOC measurements have been achieved in several quantum platforms, including
superconducting circuits~\cite{
mi2021information,blok2021quantum,braumuller2022probing,zhao2022probing,wang2022information},
trapped ions~\cite{
landsman2019verified,garttner2017measuring,joshi2020quantum,green2022experimental},
nuclear magnetic resonance (NMR)~\cite{
li2017measuring,wei2018exploring,wei2019emergent,sanchez2020perturbation,niknam2020sensitivity,nie2020experimental},
NV centres~\cite{chen2020detecting},
degenerate Fermi gases~\cite{pegahan2021energy} and
cavity quantum electrodynamics (cavity QED) system~\cite{li2023improving}.
In this section, we provide a detailed description of the OTOC measurements.
}

\LL{
Figure~\ref{SI_OTOC_sequence} illustrates the detailed pulse sequence for measuring the ZZ-OTOC. The measurement protocol begins with the preparation of two distinct initial states, \(\ket{\mathbb{Z}_2}\) and \(\ket{\mathbf{0}}\) (see section~\hyperref[section:Z2]{1.3} for details on state preparations). The system then undergoes forward time evolution under the Hamiltonian \(H\) for a duration \(t\).
Next, a local perturbation \(\sigma_j^z\) is selectively applied to the central (13th) atom by inducing a \(\pi\)-phase shift with a 795-\SI{}{\nano\meter} addressing laser. At the same time, a global \(\prod_i\sigma^z_i\) rotation is performed on all qubits via a microwave field.
This global rotation, combined with the subsequent Hamiltonian evolution, effectively implements the time-reversed Hamiltonian $-H$ for an equal duration $t$. 
This carefully designed sequence realizes the desired OTOC measurement, \(F_{ij}(t) = \bra{\psi}W_i^\dagger(t)V_j^\dagger W_i(t)V_j\ket{\psi}\).
}

\LL{
To mitigate the differential AC-Stark shifts induced by the tweezer traps on the ground and Rydberg states, the traps are switched off before Rydberg excitation and turned back on after state evolution. Given the estimated atomic temperature of approximately \SI{10}{\micro\kelvin} and the release and recapture time of \SI{10}{\micro\second}, the resulting atomic loss is estimated to be around 1\%.
}

\begin{figure}[t]
  \centering
  \includegraphics[width=0.4\textwidth]{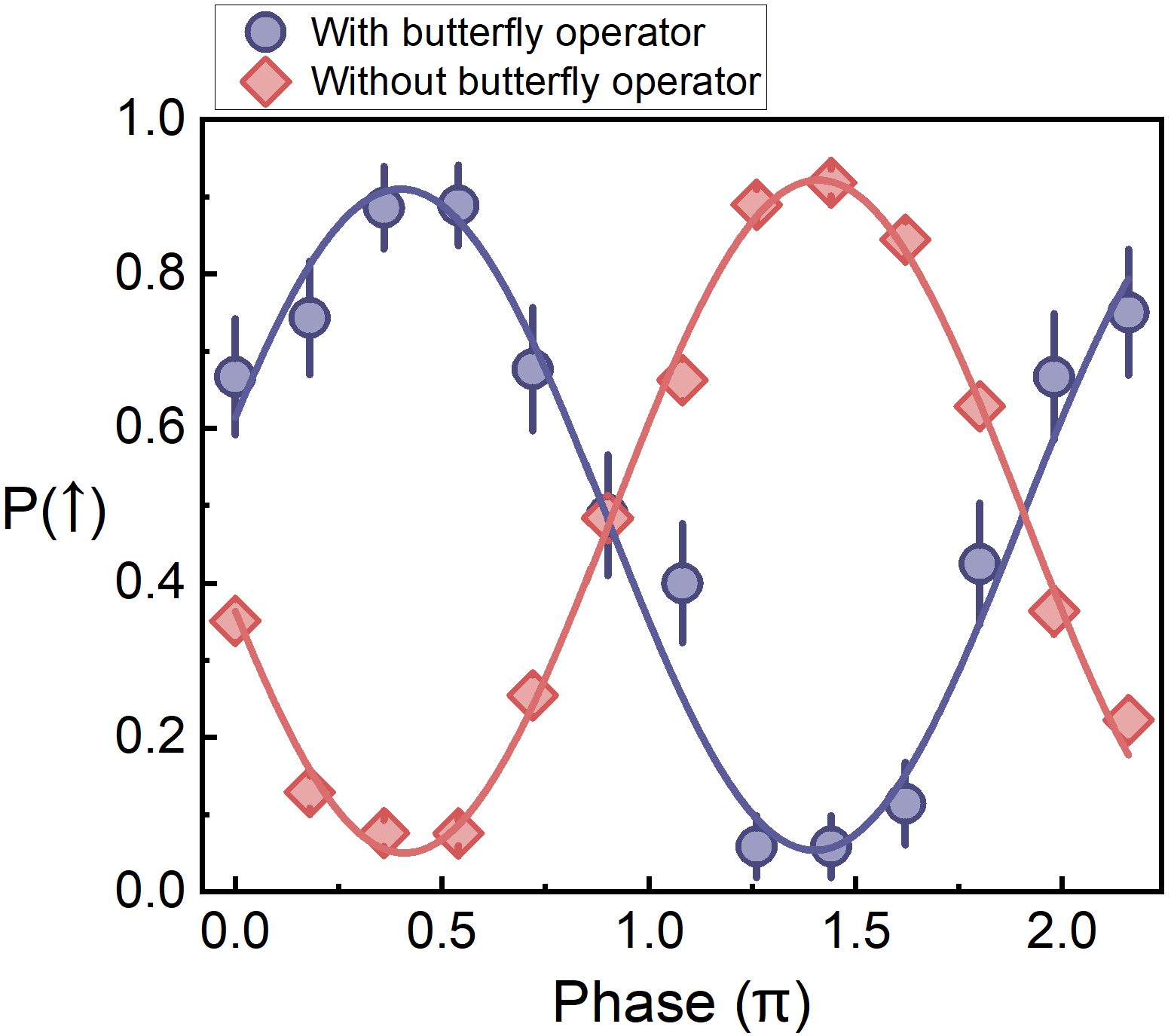}  
  \caption{
\textbf{Local perturbation.} 
\LL{Phase-dependent Ramsey oscillation of the central qubit, with (blue) and without (red) the local butterfly operator $\sigma_c^z$. }
}
  \label{SI_OTOC_details}
\end{figure}

\LL{
In the regime where \(V_{i,i+2} \ll \Omega \ll V_{i,i+1}\), the Rydberg blockade effect introduces kinetic constraints, excluding configurations with adjacent qubits in the Rydberg state, \(\ket{\cdots \uparrow_i \uparrow_{i+1} \cdots}\), from the computational basis. This kinetically constrained system is well approximated by the PXP model. However, due to the nature of van der Waals interactions, the ratio \(V_{i,i+1} / V_{i,i+2}\) is fixed at approximately 64, making it difficult to fully separate these energy scales. The experimental parameters are carefully optimized, with \(\Omega\) set to approximately \(V_{i,i+1}/6\), \ZP{as detailed in section~\hyperref[section:experimental parameter]{2.2}.}
Additionally, a small non-zero detuning \(\Delta\) between the Raman excitation lasers and ground-Rydberg transition is introduced to mitigate the residual next-nearest-neighbour interactions. Numerical simulations (Fig.~\ref{SI_OTOC_details}a) indicate that a detuning of $2 V_{i,i+2} \sim 2\pi \times \SI{0.2}{MHz}$ better preserves OTOC oscillations, closely approximating the expected PXP dynamics.
}

\LL{
The 795-\SI{}{\nano\meter} addressing laser used for local perturbation \(\sigma_j^z\) and the laser array for generating the alternating light shifts in \(\ket{\mathbb{Z}_2}\) state preparation are both directed onto different regions of the same SLM. These two independent regions display distinct holograms, allowing for different addressing laser patterns. This spatial multiplexing enables sub-microsecond switching of 795-\SI{}{\nano\meter} addressing laser patterns between \(\ket{\mathbb{Z}_2}\) state preparation and local perturbation, without being limited by the refresh rates of devices such as AODs (microsecond scale) and SLMs (millisecond scale).
The duration of the local perturbation pulse is \(\sim\)\SI{110}{\nano\second}, resulting in a \(\pi\)-phase shift on the ground states. The effectiveness of this local perturbation is experimentally verified by applying it to a single atom between two \(\pi/2\) pulses separated by a fixed interval. By varying the phase of the second pulse, Ramsey-type oscillations of both addressed and unaddressed atoms are observed (Fig.~\ref{SI_OTOC_details}b). The oscillations of the addressed atom are shifted by \(1.01(2)\pi\) relative to those of the unaddressed atoms, confirming a controlled phase shift on individual atoms.
}

\LL{
One of the key challenges in measuring OTOCs is implementing the inverse Hamiltonian evolution \(\exp(-iHt)\) in a many-body system. In the Rydberg PXP model, we overcome this difficulty by exploiting particle-hole symmetry, represented by \(\mathcal{C} = \prod_j \sigma_j^z\), which leads to the relation \(\mathcal{C} H_\text{PXP} \mathcal{C} = -H_\text{PXP}\), effectively reversing the sign of the Hamiltonian. This symmetry allows us to implement the time-reversed Hamiltonian \((\prod_i \sigma_i^z)H(\prod_i \sigma_i^z) = -H\). Experimentally, this is achieved by applying a global \(\sigma^z\) gate~\cite{feldmeier2024quantum} to all qubits using a $\SI{\sim180}{ns}$-long far-detuned microwave (MW) field. The MW field off-resonantly couples the Rydberg states $\ket{\uparrow}=\ket{r} = \ket{68D_{5/2}, m_J=5/2}$ and $\ket{r'} = \ket{69P_{3/2}}$, and induces a \(\pi\)-phase shift on the state $\ket{\uparrow}$ via the AC Stark effect. This method enables the forward-and-backward evolution of the \(\ket{\mathbb{Z}_2}\) state, with the measured results presented in Fig. 3c of the main text. Our digital-analogue approach offers an efficient and elegant way to implement time reversal in programmable Rydberg atom arrays, enabling precise measurements of quantum information scrambling via OTOCs.
}
\LL{For the ZZ-OTOC measurement, we apply a local butterfly operator $W_i = \sigma^z_c$ to perturb the central qubit (the 13th qubit), while the measurement operator $V_j = \sigma^z_j$ acts on the $j$-th qubit. The ZZ-OTOC, denoted as $F_{ij}(t)$, where $i = c$ for the central qubit, can be expressed as:
\begin{equation}
F_{ij}(t) = F_{cj}(t) = \langle \Psi_0 | \sigma_c^z(t) \sigma_j^z \sigma_c^z(t) \sigma_j^z | \Psi_0 \rangle = \langle \Psi_0 | \sigma_j^z | \Psi_0 \rangle \langle \Psi_c(t) | \sigma^z_j | \Psi_c(t) \rangle,
\end{equation}
where $\ket{\Psi_0}$ is the initial state, and $\ket{\Psi_c(t)} = e^{iHt} \sigma^z_c e^{-iHt} \ket{\Psi_0}$ represents the time-evolved state after forward-and-backward Hamiltonian evolution over time $t$.
For consistency, we fix $\langle \Psi_0 | \sigma_j^z | \Psi_0 \rangle=1$.
The expectation value $\langle \Psi_c(t) | \sigma^z_j | \Psi_c(t) \rangle$, which represents the correlation between the central qubit and the $j$-th qubit, can be directly obtained from site-resolved measurements of the Rydberg state population. Specifically, it is given by:
\begin{equation}
\langle \Psi_c(t) | \sigma^z_j | \Psi_c(t) \rangle = 2P_j(\uparrow) - 1,
\end{equation}
where $P_j(\uparrow)$ is the measured population of the $\ket{\uparrow}$ state for the $j$-th qubit in the array. Consequently, the ZZ-OTOC can be written as:
\begin{equation}
F_{cj}(t) = 2P_j(\uparrow) - 1.
\end{equation}
In the experiment, the Rydberg state population $P(\uparrow)$ is measured for each qubit after time evolution, allowing us to extract the full spatio-temporal dynamics of the OTOC across the atomic array.}

\LL{
To ensure consistency when measuring qubits in the $\ket{\mathbb{Z}_2}$ state, the measured $j$-th qubit is always initialized in the $\ket{\uparrow}$ state. Consequently, the indexing of the initial $\ket{\mathbb{Z}_2}$ state must be adjusted based on whether the measured qubit index $j$ is odd or even. For the central 13 qubits, the initial $\ket{\mathbb{Z}_2}$ state is defined as $\ket{\mathbb{Z}_2} = \ket{...\downarrow_{j-1}\uparrow_j \downarrow_{j+1}...}$, where the measured qubit is always initialized as $\ket{\uparrow}$, with the central qubit labeled as qubit 13.
The data acquisition and processing procedure depends on the odd or even nature of the measured qubit index.
For odd indices, $\ket{\uparrow}$-initialized qubit 13 is perturbed, and the OTOC data is collected from qubits 7, 9, ..., 19.
For even indices, $\ket{\downarrow}$-initialized qubit 12 is perturbed, OTOC data is collected from qubits 7, 9, ..., 17, and aligned with qubit indices 8, 10, ..., 18 for consistency in indexing.
Throughout the measurement process, the initial 25-qubit $\ket{\mathbb{Z}_2}$ state remains unchanged to maintain consistent initial state preparation fidelity.
}

\LL{
To mitigate boundary effects inherent in finite-size systems (Fig.~\ref{Fig:Boundary and finite-size effect}), the OTOC measurements focus on the central 13 qubits of the prepared 25-qubit $\ket{\mathbb{Z}_2}$ state. Boundary qubits experience fewer neighbouring interactions, which can lead to deviations in their dynamics. By concentrating on the central region, we limit the influence of these boundary effects and ensure that the measured dynamics more accurately represent the bulk behaviour of the PXP model.
This strategy allows us to observe the intrinsic quantum information scrambling and \textit{collapse-and-revival} phenomena with higher fidelity, as the central qubits are less affected by edge-induced artifacts. Thus, the measured OTOCs from these qubits provide a better approximation of the expected PXP behaviour.
}

\subsection{Holevo information measurements details}

\LL{Holevo information, introduced by Alexander Holevo in 1973~\cite{holevo1973bounds}, is a fundamental concept in quantum information theory that sets an upper bound on the amount of information that can be reliably transmitted through a quantum channel~\cite{holevo2012information,holevo2012quantum}. It is formally defined as the difference between the von Neumann entropy of the average output state and the average of the von Neumann entropies of the individual output states. Mathematically, if $\rho_\text{X} = \sum_i p_i \rho_i$ is the average output state corresponding to the quantum channel's output ensemble $\{p_i, \rho_i\}$, the Holevo information $\mathbb{X}$ is given by:
\[
\mathbb{X} = S(\rho_\text{X}) - \sum_i p_i S(\rho_i),
\]
where $S(\rho)$ denotes the von Neumann entropy, defined as $S(\rho) = -\text{Tr}(\rho \log_2 \rho)$. 
Importantly, Holevo information represents the upper bound on the accessible information that can be shared between two parties using a quantum channel, reflecting the best-case scenario for the amount of information that can be transmitted, regardless of the specific measurement strategy employed by the receiver.}

\LL{To illustrate this, we consider an ensemble with two equally probable quantum states $\rho$ and $\rho'$. Holevo information in this case can be interpreted as a measure of the distinguishability between the two states. Suppose Alice selects either $\rho$ or $\rho'$ with a probability of $1/2$ and sends it to Bob through a quantum channel. Bob then performs a measurement to gain as much information as possible about which state Alice has sent. The information Bob retrieves from his measurement is upper bounded by the Holevo information $\mathbb{X}$. For instance, if $\rho$ and $\rho'$ are completely indistinguishable (i.e., $\rho = \rho'$), Bob cannot obtain any information about Alice’s choice, and the Holevo information is zero ($\mathbb{X} = 0$). In this case, regardless of Bob's measurement, the outcome gives no clue about whether $\rho$ or $\rho'$ was sent. \LL{On the other hand, if $\rho$ and $\rho'$ are orthogonal, e.g. $\rho = \ket{\uparrow}\bra{\uparrow}$ and $\rho' = \ket{\downarrow}\bra{\downarrow}$, Bob can fully determine Alice’s choice using an appropriate measurement, such as a $\sigma^z$ measurement.} In this case, the Holevo information reaches its maximum value of $\mathbb{X} = 1$, meaning Bob retrieves all the information about Alice’s selection.}

\LL{Holevo information has also been proposed as a powerful tool for studying scrambling and transport dynamics in many-body quantum systems~\cite{yuan2022quantum,zhuang2022phase,zhuang2023dynamical}.}
\R{Compared to OTOCs, which measure the scrambling of local perturbations in a system, Holevo information provides a different perspective on quantum information dynamics, as it does not rely on specific perturbations that may be ineffective under certain conditions.}

\begin{figure}[t]
  \centering
  \includegraphics[width=0.9\textwidth]{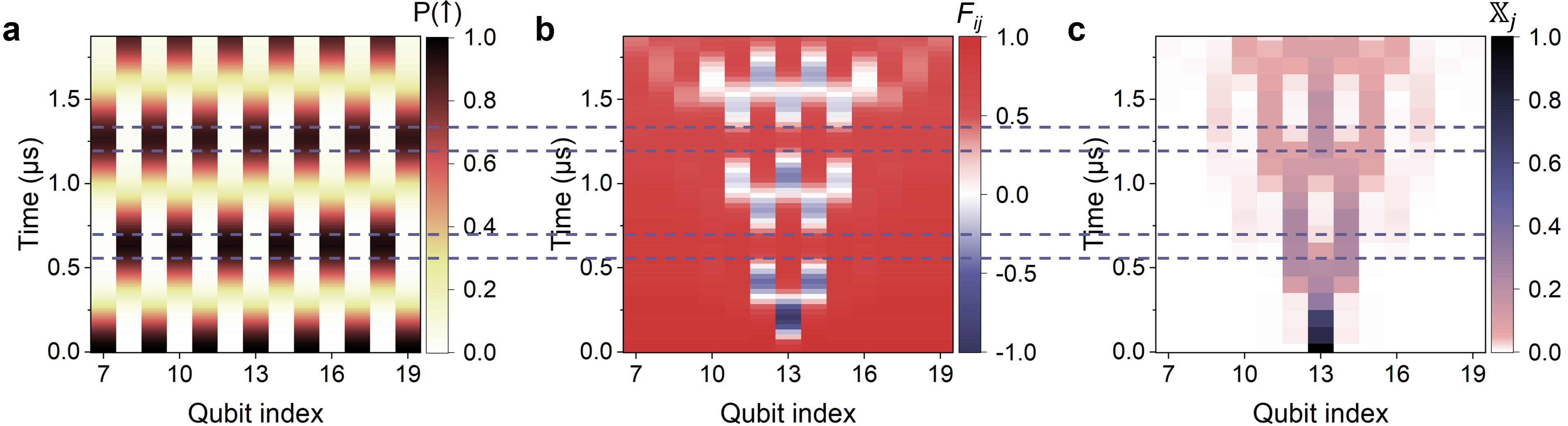}  
  \caption{\R{
\textbf{Illustration of ineffective local perturbations.}
Numerical simulations of \textbf{a}, $\ket{\mathbb{Z}_2}$ state evolution, \textbf{b}, ZZ-OTOC dynamics and \textbf{c}, Holevo information dynamics with the same time scale, mirroring Figs. 2e, 3l and 4c in the main text. Time intervals when all the qubits approach the $\ket{\uparrow}$ or $\ket{\downarrow}$ pure states are marked with two pairs of dashed lines (\SI{0.6}{\us}--\SI{0.7}{\us} and \SI{1.2}{\us}--\SI{1.3}{\us}). During these intervals, the ZZ-OTOCs ($F_{ij}$) for all qubits approach 1 since the $\sigma^z$ butterfly operator has no effect on the perturbed qubit.
In contrast, Holevo information remains effective throughout, enabling uninterrupted tracking of quantum information dynamics.
}}
  \label{Fig:SI_fake_collapse}
\end{figure}

\R{In systems exhibiting quantum many-body scars, ZZ-OTOC measurements may occasionally become less informative due to the periodic oscillation of the scar state wavefunction. 
For example, during certain time intervals (such as \SI{0.6}{\us}--\SI{0.7}{\us} and \SI{1.2}{\us}--\SI{1.3}{\us} in Fig.~3j,l and Fig.~\ref{Fig:SI_fake_collapse}), the ZZ-OTOC values for all qubits approach 1.
The reason is that during these intervals, the wavefunction of the scar state $\ket{\mathbb{Z}_2}$ partially revives, causing the perturbed qubit to be mostly in either the $\ket{\uparrow}$ or $\ket{\downarrow}$ pure state, where the butterfly operator $\sigma^z$ becomes ineffective. Consequently, no effective perturbation occurs, and no measurable scrambling is observed. As a result, ZZ-OTOC values of all qubits approach 1, due to the lack of an effective perturbation.}
\R{In contrast, Holevo information can continuously capture quantum information dynamics, even during periods when the butterfly operator in ZZ-OTOCs become less effective (Fig.~\ref{Fig:SI_fake_collapse}c).}

\LL{This distinction is further emphasized when comparing Holevo information to classical Shannon information. While Shannon information only accounts for classical probability distributions, ignoring quantum phases, Holevo information captures both classical and quantum aspects of information, including coherence and entanglement. For example, between $\ket{\leftarrow}$ and $\ket{\rightarrow}$ states, Shannon information may be minimized to zero, while Holevo information can still be maximized, reflecting the quantum coherence between these states.
In the specific context of the PXP model with the initial state $\ket{\mathbb{Z}_2}$, Holevo information reveals the retarded spin dynamics within the light-cone structure (see Fig.~\ref{Fig:S0}). This analysis highlights how kinetic constraints lead to a persistent phase delay in the propagation of spin rotations, influencing the overall information dynamics. Within this light cone, spins resume their constrained rotations; however, the distinguishability of spin states periodically collapses and revives, as captured by Holevo information. This behaviour reflects the constrained dynamics of the PXP model, where the Rydberg blockade effect induces delayed spin rotations near the central flipped spin, with this delay propagating outwards in a light-cone-like wavefront.}

\LL{To understand how Holevo information applies to our experimental system, we consider the scenario where Alice and Bob transmit information through the $\ket{\mathbb{Z}_2}$ state. Based on the $\ket{\mathbb{Z}_2}$ state, Alice at the central site chooses whether to flip her qubit at $t=0$, and Bob at site $j$ measures his qubit at time $t$ to infer Alice's choice. Outside the light cone, Bob retrieves no information ($\mathbb{X}_j(t) = 0$) due to the finite speed of information propagation. Inside the light cone, Bob periodically gains and loses information, as the distinguishability of spin states collapses and revives, leading to corresponding oscillations in Holevo information. 
This behaviour reflects the constrained dynamics of the PXP model, where the Rydberg blockade effect induces delayed spin rotations near the central flipped spin, with this delay propagating outwards. Notably, even when Bob measures the same qubit perturbed by Alice, the phenomenon of collapse and revival can still occur due to the constrained dynamics in the system. Additionally, information can be retrieved from other sites, as the quantum information propagates across the system, even when Bob’s qubit does not provide it.}

\LL{Furthermore, Holevo information provides a robust measure of non-Markovianity, capturing the information backflow from surrounding spins to the central spin. Each spin exhibits periodic increases in Holevo information, signaling the backflow of quantum information, which is a clear signature of positive non-Markovianity~\cite{fanchini2014non,megier2021entropic,smirne2022holevo}.
This unique \textit{collapse-and-revival} behaviour of Holevo information is distinct from the quantum scar state oscillations observed in similar systems, as it incorporates thermal eigenstates, thereby providing a broader picture of quantum dynamics beyond the scarred subspace.}

\LL{To the best of our knowledge, this work presents the first experimental investigation of many-body dynamics in Rydberg atom array using Holevo information. In the rest of this section, we provide details regarding the experimental sequence and parameters utilized in our measurements of Holevo information.}

\begin{figure}[t]
  \centering
  \includegraphics[width=0.9\textwidth]{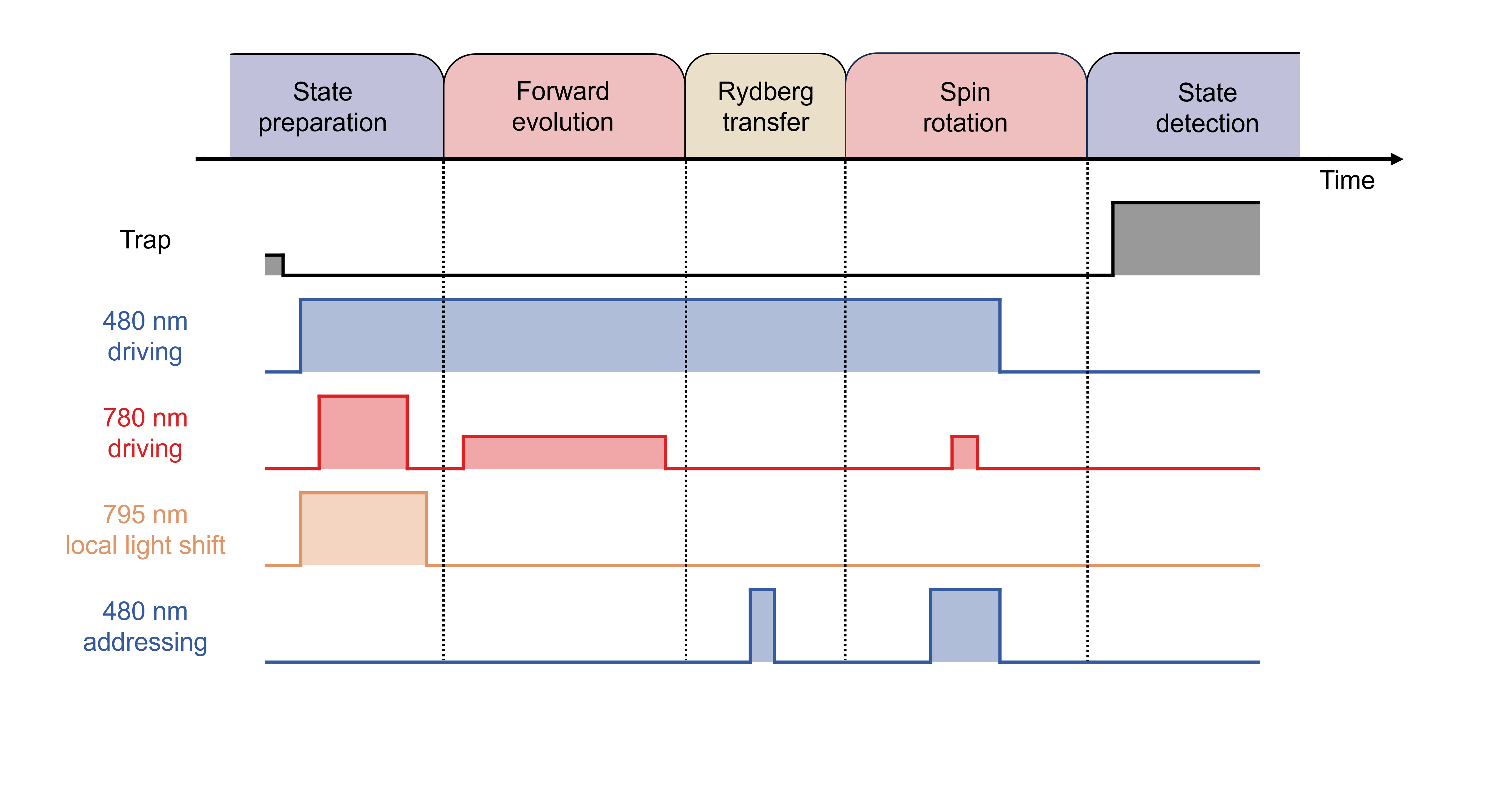}  
  \caption{
\textbf{Pulse sequence for Holevo information measurements.} 
}
  \label{SI_HI_sequence}
\end{figure}

\LL{
Figure~\ref{SI_HI_sequence} illustrates the pulse sequence used to measure Holevo information dynamics, corresponding to Fig. 4a in the main text. The protocol begins with the preparation of two distinct initial states: \(\ket{\mathbb{Z}_2}\) and \(\sigma^x_c \ket{\mathbb{Z}_2}\), where \(\sigma^x_c\) acts on the central spin in the chain. These initial states evolve under the Rydberg Hamiltonian, after which quantum state tomography is performed to reconstruct the density matrix \(\rho_j\) for each qubit, enabling a detailed investigation of quantum information transport dynamics under kinetic constraints.
}

\LL{
The Holevo information is extracted from the reconstructed density matrix, which includes both diagonal and off-diagonal elements. The diagonal elements of \(\rho_j(t)\) and \(\rho'_j(t)\) are obtained through projective measurement of \(\sigma^z_j\) on the \(j\)-th qubit, which can be accessed via Rydberg population measurement. However, measuring the off-diagonal elements requires a single-qubit \(\pi/2\)-rotation with a variable phase. This process is particularly challenging in strongly interacting Rydberg atom systems due to the constraints imposed by the PXP model, which requires nearest-neighbouring Rydberg atoms to be in the excitation blockade regime. The strong interactions between neighbouring Rydberg atoms create significant obstacles for performing spin rotations on any given qubit. If a nearest-neighbouring atom of the target qubit is in the Rydberg state, the blockade effect will prevent the target qubit from undergoing rotation. Even if only a next-nearest-neighbouring atom is in the Rydberg state, though not causing a full blockade, the residual Rydberg interaction will still affect the phase during quantum state tomography. Moreover, even when both nearest-neighbouring and next-nearest-neighbouring atoms are in the ground state, performing a spin rotation on the target qubit still remains difficult, as surrounding atoms can become entangled with the target qubit due to their involvement in the Rydberg excitation, resulting in a complex many-body quantum state. Therefore, to carry out the spin rotations required for quantum state tomography on the target qubit, it is necessary for the nearest-neighbouring and next-nearest-neighbouring atoms to neither be in the Rydberg state nor resonant with the ground-Rydberg transition.
}

\begin{figure}
  \centering
  \includegraphics[width=0.9\textwidth]{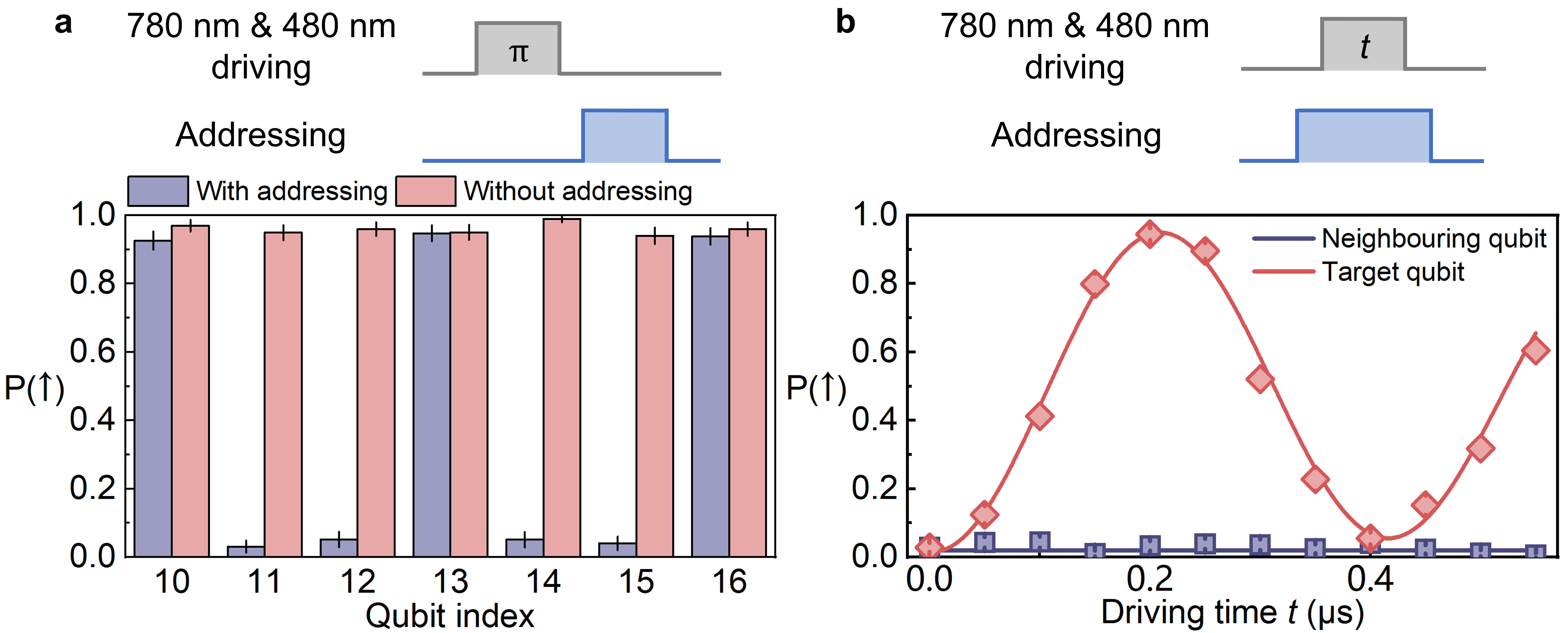}
  \caption{
\textbf{Holevo information measurement details.}
\textbf{a}, \LL{Rydberg population of the central 7 qubits, with (blue) and without (red) the Rydberg population transfer process. The data is obtained via Rydberg population measurement of even atoms and odd atoms, respectively.}
\textbf{b}, \LL{Under the driving of global 480-\SI{}{\nano\meter} and 780-\SI{}{\nano\meter} lasers, the target qubit (red diamonds) exhibits Rabi oscillation between the ground and Rydberg states, while the neighbouring qubits (blue squares), due to the EIT condition created by the 480-\SI{}{\nano\meter} addressing lasers, do not participate in the spin rotation process.}
}
  \label{SI_HI_dissipative}
\end{figure}

\LL{
Experimentally, we implement a novel approach for density matrix reconstruction using global rotations combined with 480-\SI{}{\nano\meter} addressing lasers. The addressing beams, resonant with the transition from the excited state $\ket{e}$ to the Rydberg state $\ket{r}$, are selectively applied to the four neighbouring qubits, transferring their Rydberg populations to the ground state via spontaneous emission from the intermediate state \(\ket{e}\). 
The decay rate of the Rydberg population during the transfer process, $\Gamma_r$, can be estimated by:
\begin{equation}
\Gamma_r = \Gamma_e\frac{\Omega_{480}^2/4}{\delta^2 + \Gamma^2/4 + \Omega_{480}^2/2}.
\end{equation}
Here, $\Gamma_e = 2\pi \times \SI{6.06}{\MHz}$ is the natural linewidth of the excited state $\ket{5P_{3/2}}$. The terms $\Omega_{480}$ and $\delta$ denote the Rabi frequency and detuning of the addressing beams, respectively.
This process lasts for approximately \SI{200}{\nano\second}, which is sufficient to deplete the Rydberg population at neighbouring sites (Fig.~\ref{SI_HI_dissipative}a), while being short enough to avoid the unwanted effects on the target qubit due to crosstalk from the 480-\SI{}{\nano\meter} addressing lasers. 
After the state transfer process, the Rydberg population in the neighbouring qubits is reduced by 96(1)\%, with negligible crosstalk-induced reduction in the Rydberg population of the target qubit.
}

\LL{
Next, the 480-\SI{}{\nano\meter} addressing lasers split the bare Rydberg state \(\ket{r}\) into two dressed states: \(\ket{+} = \frac{1}{\sqrt{2}}(\ket{r} + \ket{e})\) and \(\ket{-} = \frac{1}{\sqrt{2}}(\ket{r} - \ket{e})\), separated by \(\hbar\Omega_{480}\), where $\Omega_{480} \sim 2\pi \times \SI{20}{MHz}$ is the Rabi frequency of the 480-\SI{}{\nano\meter} addressing laser.
The dressed states are significantly detuned from the ground-Rydberg transition.
The off-resonant excitation from the ground state \(\ket{g}\) to the dressed states \(\ket{+}\) and \(\ket{-}\) causes destructive interference, preventing population transfer to the Rydberg state. This creates an electromagnetically induced transparency (EIT) condition, ensuring neighbouring qubits do not participate in the ground-Rydberg coherent driving implemented by the global excitation lasers.
}

\LL{
Finally, single-qubit rotation on the target qubit is implemented using global excitation lasers.
Figure~\ref{SI_HI_dissipative}b demonstrates that the target qubit can undergo single-qubit rotation (Rabi oscillation between the ground and Rydberg states) in the presence of neighbouring qubits addressed by the 480-\SI{}{\nano\meter} laser. The addressed neighbouring qubits remain ineligible for the ground-Rydberg excitation process due to the EIT condition induced by the 480-\SI{}{\nano\meter} laser.
}

\LL{In an ideal scenario with periodic boundary conditions, the PXP model would exhibit a stationary expectation value of the Pauli-X operator $\langle \sigma^x \rangle$ throughout the evolution. For initial states like $\ket{\mathbb{Z}_2}$ and $\sigma_c^x \ket{\mathbb{Z}_2}$, this implies that $\langle \sigma^x \rangle = 0$. However, during the gap time between the evolution and the projection measurement, the residual van der Waals interaction between the atoms could lead to the accumulation of phases, causing $\langle \sigma^x \rangle$ to become non-zero. This introduces challenges in accurately measuring the desired off-diagonal elements of the system's density matrix. 
\LL{Due to the difference in the initial states ($\ket{\mathbb{Z}_2}$ and $\sigma_c^x \ket{\mathbb{Z}_2}$), even after the same PXP evolution, the two output states may accumulate different residual phases. This phase difference introduces additional distinguishability when directly measuring the off-diagonal elements, increasing the difference between the two output states $\rho_j(t)$ and $\rho'_j(t)$. As a result, this extra distinguishability reduces the accuracy of the Holevo information.}
To address this, we measure the full parity oscillation curve by scanning the phase shift between two $\pi/2$ pulses. From this curve, we extract the oscillation amplitude $A$ by fitting it to a sinusoidal function.}

\LL{With the measured values of $P(\uparrow)$ and $A$ for each qubit, we reconstruct the density matrix $\rho(t)$ at time $t$ using the following expression:
\begin{equation}
\rho(t) = \frac{1}{2}(\mathbb{I} + (2P(\uparrow) - 1)\sigma^z) + \epsilon(t)A\sigma^y,
\end{equation}
where $\epsilon(t)$ represents the sign of $\langle \sigma^y \rangle$, with $\epsilon(t) = 1$ when $P(\uparrow)$ is expected to be increasing, and $\epsilon(t) = -1$ otherwise. Here, $\mathbb{I}$ is the identity matrix, and $\sigma^y$ and $\sigma^z$ are the Pauli-Y and Pauli-Z matrices, respectively.}

\LL{The Holevo information for qubit $j$ at time $t$ can be obtained:
\begin{equation}
\mathbb{X}_j(t) = S\left(\frac{\rho_j(t) + \rho'_j(t)}{2}\right) - \frac{S(\rho_j(t)) + S(\rho'_j(t))}{2},
\label{Eq:Holevo information}
\end{equation}
where $\rho_j(t)$ and $\rho'_j(t)$ are the density matrices of the $j$-th qubit evolved from distinct initial states $\ket{\mathbb{Z}_2}$ and $\sigma^x_c \ket{\mathbb{Z}_2}$.
The von Neumann entropy $S(\rho) = -\text{Tr}(\rho \log_2 \rho)$
is used to quantify the information content of the density matrices.}

\LL{The measurement of off-diagonal elements is what distinguishes quantum information from classical Shannon information. Shannon information, which only considers the diagonal elements of the density matrix, is a simple measure of classical probability distributions. In contrast, quantum information involves off-diagonal elements, which capture quantum effects like coherence and entanglement---features not present in classical systems.
We emphasize that von Neumann entropy goes beyond Shannon entropy in quantum information science. While Shannon entropy only reflects the uncertainty in classical probability distributions, von Neumann entropy plays a central role in quantum systems. It is essential for quantifying information in quantum states and determining the capacities of quantum channels. More importantly, it captures quantum phenomena, such as entanglement, which are critical for understanding quantum systems. This is why we have made significant efforts to measure the off-diagonal elements, as they provide deeper insights into the unique aspects of quantum information.}

\subsection{Non-Markovian quantum information dynamics in a strongly-interacting Rydberg atom array}

\LL{Non-Markovian quantum dynamics are often characterized by memory effects, where a system's evolution depends on its past interactions with the environment. Unlike Markovian dynamics, where information is irreversibly lost to the environment, non-Markovian systems can experience information backflow, allowing for the recovery of previously lost information~\cite{rivas2014quantum,breuer2016colloquium, deVega2017Dynamics,li2018concepts}.
In many-body quantum systems, strong interactions could lead to non-Markovian dynamics, significantly influencing the spread and preservation of quantum information
}

\LL{In this work, we observe non-Markovian dynamics in a strongly interacting Rydberg atom array, focusing on the unusual behaviour of information backflow. We designate the central spin in an atomic array as the ``system'' and the surrounding spins as the ``environment''. The strong spin interactions allow information to transfer between the system and environment in complex ways, making it possible to observe non-Markovian effects.}

\LL{Our experimental setup allows for precise quantum state tomography of each spin, enabling real-time tracking of information flow across the system. This capability provides detailed insight into non-Markovian behaviour by directly measuring how information, initially lost to the environment, returns to the spins that originally held it.}

\begin{figure}[t]
\centering
\includegraphics[width=0.45\textwidth]{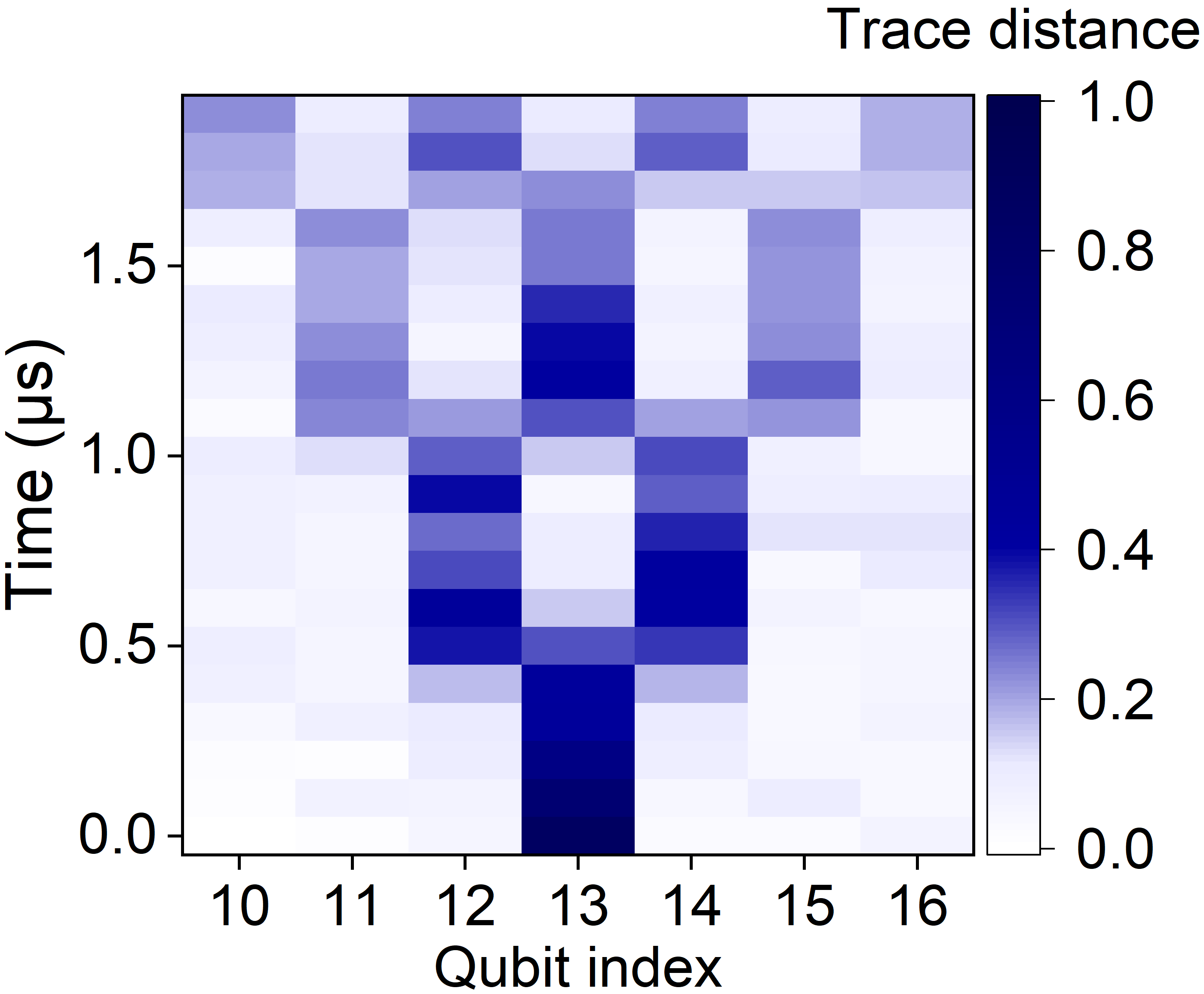}
\caption{
\XDS{\textbf{Quantify non-Markovian dynamics via trace distance measurements.} 
Spatio-temporal dynamics of trace distance between $\ket{\mathbb{Z}_2}$ and $\sigma_c^x\ket{\mathbb{Z}_2}$ states under Rydberg Hamiltonian. Density matrices reconstructed from Holevo information measurements. The measurements reveal a linear light cone and \textit{collapse-and-revival} pattern, demonstrating periodic increases in state distinguishability. This non-monotonic behaviour provides evidence of quantum information backflow, indicating non-Markovian dynamics in the Rydberg atom array.
}}
\label{SI_Fig_trace_distance}
\end{figure}

\LL{To quantify non-Markovianity, we assess the degree of information backflow using metrics like
trace distance~\cite{breuer2009measure,fan2023collapse} and
Holevo information~\cite{fanchini2014non,megier2021entropic,smirne2022holevo},
which track the evolution of quantum state distinguishability over time. The constrained spin rotations within the system drive this backflow, causing periodic collapses and revivals in information distinguishability. This allows us to capture key signatures of non-Markovian dynamics as information spreads, collapses, and recovers between the system and environment. These findings offer valuable insights into quantum information dynamics and hold promise for advancing quantum memory technologies.
}

\LL{
The trace distance is defined as 
\begin{equation}
D(\rho_1,\rho_2) = \frac{1}{2}\mathrm{Tr}{|\rho_1-\rho_2|},
\end{equation}
where $\rho_1$ and $\rho_2$ are density matrices of $\ket{\mathbb{Z}_2}$ and $\sigma_c^x\ket{\mathbb{Z}_2}$, respectively. 
Figure~\ref{SI_Fig_trace_distance} shows the experimentally measured spatio-temporal dynamics of the trace distance between the $\ket{\mathbb{Z}_2}$ state and $\sigma_c^x\ket{\mathbb{Z}_2}$ under the Rydberg Hamiltonian. Density matrices are reconstructed from Holevo information measurements. The plot reveals a clear linear light cone structure and a spatial-temporal \textit{collapse-and-revival} pattern, mirroring observations in the Holevo information data presented in the main text. Trace distance, a widely used metric for quantifying non-Markovianity in quantum systems, tracks the distinguishability between two quantum states over time. While Markovian processes exhibit monotonic decay of trace distance, indicating irreversible information loss to the environment, our data shows periodic increases in trace distance. This observed pattern provides compelling evidence for quantum information backflow and the non-Markovian nature of the system's dynamics. These dynamics, marked by changes in state distinguishability, align well with theoretical expectations from the PXP model and display features of information backflow, where quantum information is periodically exchanged between the system and its environment rather than being permanently lost. Our findings are consistent with previous studies on non-Markovian behaviour in quantum systems~\cite{breuer2009measure}.}

\section{Error Analysis and Mitigation}
\label{section:error}

\LL{In our Rydberg atom quantum simulator, imperfect Hamiltonian evolution and finite qubit coherence lead to the accumulation of both coherent and incoherent errors in the dynamics of OTOCs and Holevo information. This section analyses the major error sources, develops an error model to identify and characterize primary error mechanisms, and implements error mitigation techniques for ZZ-OTOC, enhancing the performance of the quantum simulator.}

\subsection{Initial state preparation error}
\label{Z2_F}

\LL{As the number of atomic qubits increases, the exponential growth in the dimension of the density matrix amplifies the complexity of quantum state dynamics and the impact of initial state preparation errors. 
In studying ZZ-OTOC or Holevo information dynamics for $\ket{\mathbb{Z}_2}$ initial states within the constrained Hilbert space, rapid thermalization of the error states could affect the behaviour of quantum information scrambling. Therefore, the fidelity of initial $\ket{\mathbb{Z}_2}$
state preparation is crucial for accurately probing quantum information dynamics in our system.}

\vspace*{0.5\baselineskip}

\noindent

\textbf{1. Influence on OTOCs}

\vspace*{0.5\baselineskip}

\noindent

\LL{
As mentioned in section~\hyperref[section:Z2]{1.3}, the $\ket{\mathbb{Z}_2}$ state preparation under global coherent Rabi excitation with site-selective addressing technique introduces errors primarily from infidelity in single-qubit operations. Our measurements of the microstate distribution revealed that errors in $\ket{\mathbb{Z}_2}$ state preparation are predominantly attributable to single qubit $\ket{\uparrow}\rightarrow\ket{\downarrow}$ flips. These error states account for 86(2)\% of all occurrences in conjunction with the $\ket{\mathbb{Z}_2}$ state (Fig.~\ref{Fig:S2}d,e). The remaining error contribution primarily stems from detection errors (approximately 1\% per qubit). Consequently, the prepared density matrix can be expressed as a weighted sum of the target $\ket{\mathbb{Z}_2}$ state and the major error states. Given the $\ket{\mathbb{Z}_2}$ state preparation fidelity of $\mathcal{F}_{\mathbb{Z}_2}$, the measurement results of ZZ-OTOC $F_{ij}^{exp}(t)$ can be therefore expressed as:
}

\begin{equation}
F_{ij}^{exp}(t) =\mathcal{F}_{\mathbb{Z}_2} \cdot F_{ij}^{\mathbb{Z}_2}(t) + \sum_{\alpha} \rho_{\alpha\alpha}(0) F_{ij}^{\alpha}(t)
\end{equation}

\begin{figure}
\centering
\includegraphics[width=0.7\textwidth]{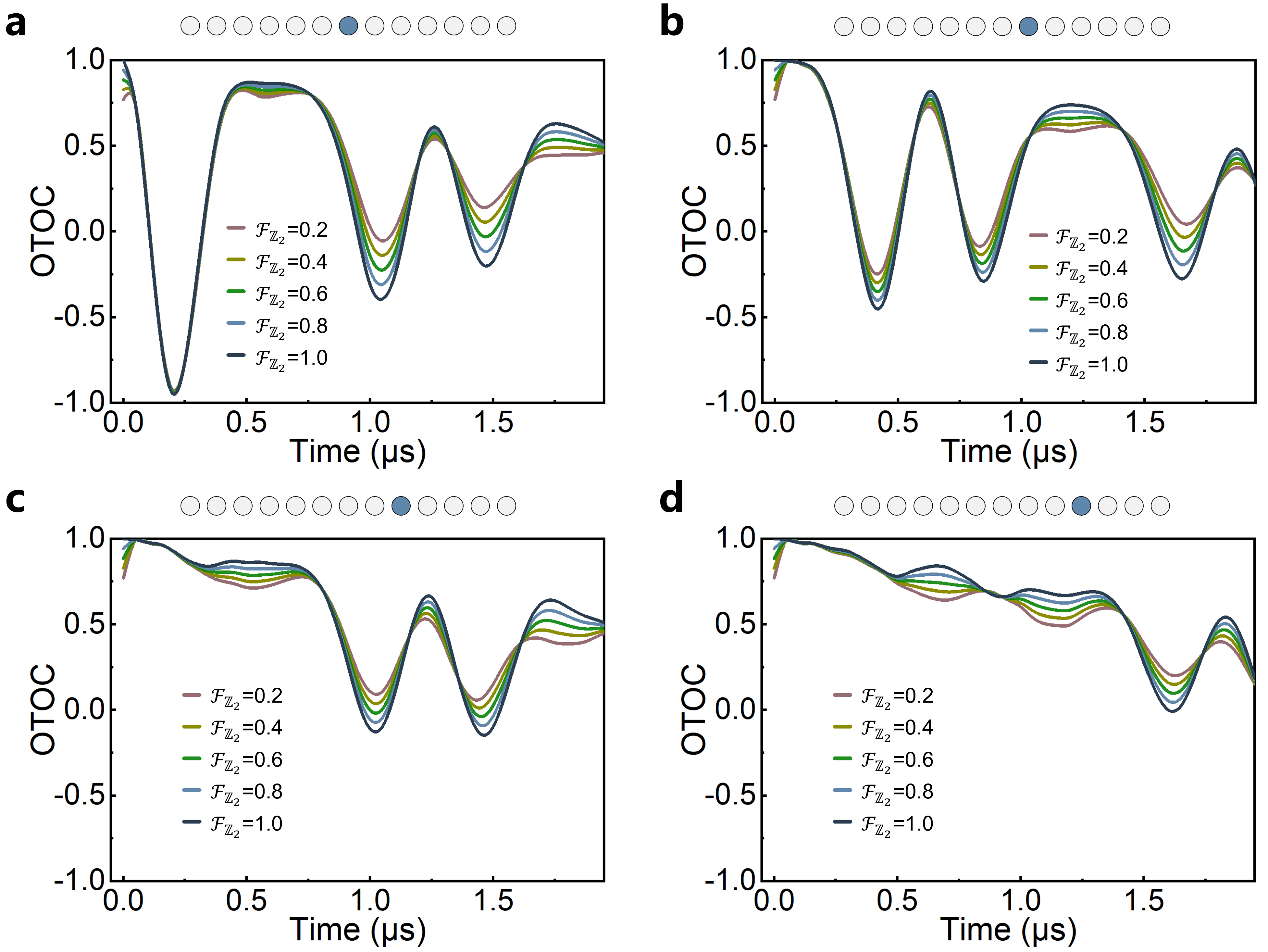}
\caption{\textbf{Impact of $\ket{\mathbb{Z}_2}$ state preparation fidelity on ZZ-OTOC dynamics.} 
\LL{
\textbf{a}--\textbf{d} show ZZ-OTOC evolution for four qubits exhibiting non-trivial dynamics. The coloured curves represents increasing fidelity values (0.2--1.0). The blue circle in the qubit chain diagram (top) indicates the qubit in each plot.
}}
\label{Z2_Fidelity}
\end{figure}

\begin{figure}[t]
\centering
\includegraphics[width=0.95\textwidth]{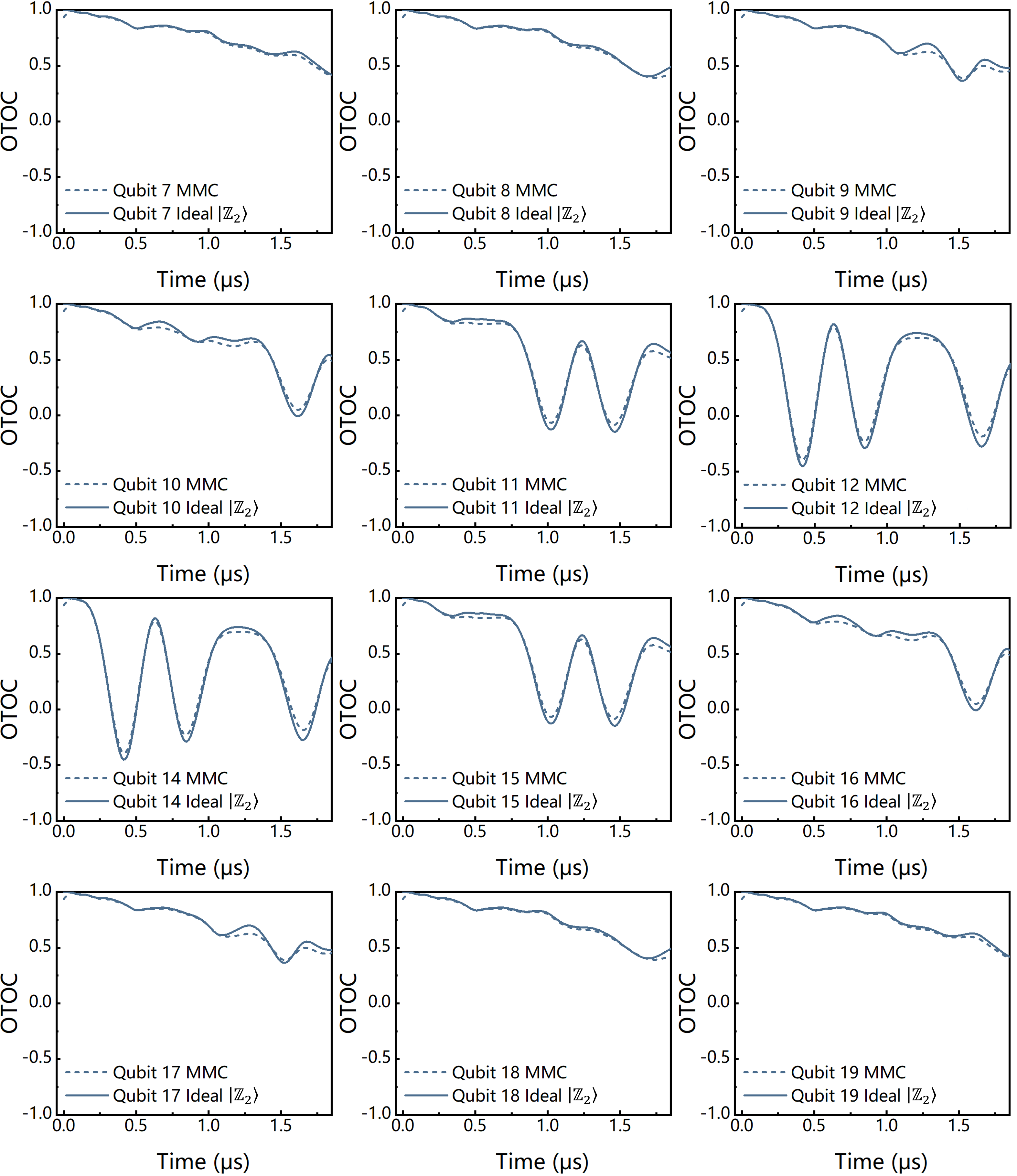}
\caption{
\LL{\textbf{Impact of experimental $\ket{\mathbb{Z}_2}$ state preparation errors on ZZ-OTOC dynamics.} Solid and dashed lines represent numerical simulations of OTOC evolution with the experimentally measured microstate combination (MMC) and perfect $\ket{\mathbb{Z}_2}$ state as initial states, respectively, showing a negligible difference for each qubit in the 13-atom array. The results for the central (13th) qubit are shown in Figure 3b of the main text.} 
}
\label{Z2_Fidelity_70}
\end{figure}

\LL{Here, $F_{ij}^{\mathbb{Z}_2}(t)$ represents the ZZ-OTOC dynamics with the ideal initial $\ket{\mathbb{Z}_2}$ state, and $F_{ij}^{\alpha}(t)$ denotes the dynamics for the $\alpha$-th error state in the prepared density matrix $\rho(0) = \mathcal{F}_{\mathbb{Z}_2}\ket{\mathbb{Z}_2}\bra{\mathbb{Z}_2}+\rho_{\alpha\alpha}(0)\ket{\alpha}\bra{\alpha}$. To investigate the impact of $\ket{\mathbb{Z}_2}$ state preparation fidelity on ZZ-OTOCs, we simulated the ZZ-OTOC dynamics with various $\ket{\mathbb{Z}_2}$ state preparation fidelity (ranging from 0.2 to 1.0), as shown in Fig.~\ref{Z2_Fidelity}. The atomic array in our experiment exhibits symmetry around the central qubit, which means that the dynamics of qubits at symmetrically equivalent positions are exactly the same. Taking advantage of this symmetry, we can fully characterize the system's behaviour by examining a subset of qubits. We present simulation results for four qubits that exhibit non-trivial dynamics and represent the distinct behaviours observed in the array. For these error states, we considered uniformly distributed $\ket{\uparrow}\rightarrow\ket{\downarrow}$ errors. This error state distribution aligns with the typical experimental conditions as characterized in section~\hyperref[section:Z2]{1.3}. The simulations reveal that the $\ket{\mathbb{Z}_2}$ state preparation fidelity significantly affects the contrast of the \textit{collapse-and-revival} pattern within the light cone.
Moreover, given our high $\ket{\mathbb{Z}_2}$ state preparation fidelity (78(1)\% for 13-qubit chain, after detection error correction.) and the measured microstate distribution, numerical simulation results (Fig.~3b of the main text for qubit 13, and Fig.~\ref{Z2_Fidelity_70} for other qubits) demonstrate that the characteristic \textit{collapse-and-revival} pattern remains clearly observable.}

\vspace*{0.5\baselineskip}

\noindent

\textbf{2. Influence on Holevo information}

\vspace*{0.5\baselineskip}

\noindent

\LL{For Holevo Information, we consider the evolution of both the $\ket{\mathbb{Z}_2}$ and the $\sigma_c^x \ket{\mathbb{Z}_2}$ states, accounting for imperfections in state preparation. 
With the prepared density matrices $\rho(0) = \mathcal{F}_{\mathbb{Z}_2}\cdot\ket{\mathbb{Z}_2}\bra{\mathbb{Z}_2} + \sum_{\alpha} \rho_{\alpha\alpha}(0) \ket{\alpha} \bra{\alpha}$ and $\rho'(0) = \mathcal{F}_{\mathbb{Z}_2^x}\sigma_c^x\ket{\mathbb{Z}_2}\bra{\mathbb{Z}_2}\sigma_c^x+\sum_{\beta}\rho'_{\beta\beta}(0)\ket{\beta}\bra{\beta}$ for $\ket{\mathbb{Z}_2}$ and $\sigma_c^x\ket{\mathbb{Z}_2}$ respectively, the final density matrices after evolution are given by:
\begin{equation}
\rho(t) =\mathcal{F}_{\mathbb{Z}_2}\cdot\ket{\mathbb{Z}_2(t)}\bra{\mathbb{Z}_2(t)} + \sum_{\alpha} \rho_{\alpha\alpha}(0) \ket{\alpha(t)} \bra{\alpha(t)}
\end{equation}
for imperfect $\ket{\mathbb{Z}_2}$ and:
\begin{equation}
\rho'(t) =\mathcal{F}_{\mathbb{Z}_2^x}\cdot\ket{\mathbb{Z}^x_2(t)}\bra{\mathbb{Z}^x_2(t)} + \sum_{\beta} \rho_{\beta\beta}(0) \ket{\beta(t)} \bra{\beta(t)}
\end{equation}
for imperfect $\sigma_c^x\ket{\mathbb{Z}_2}$. Here, $\ket{\mathbb{Z}_2(t)} = e^{-iHt} \ket{\mathbb{Z}_2} $ and $\ket{\mathbb{Z}^x_2(t)} = e^{-iHt} \sigma_c^x \ket{\mathbb{Z}_2}$ represent the final states after evolution for the ideal $\ket{\mathbb{Z}_2}$ and $\sigma_c^x\ket{\mathbb{Z}_2}$ initial states under the Rydberg Hamiltonian $H$, respectively. $\mathcal{F}_{\mathbb{Z}_2}$ and $\mathcal{F}_{\mathbb{Z}_2^x}$ denote the preparation fidelity for $\ket{\mathbb{Z}_2}$ and $\sigma_c^x\ket{\mathbb{Z}_2}$, respectively. The summation terms account for contributions from the evolution of the various initial error states $\ket{\alpha}$ and $\ket{\beta}$ : $\ket{\alpha(t)} = e^{-iHt}\ket{\alpha}$ and $\ket{\beta(t)} = e^{-iHt}\ket{\beta}$, with weights $\rho_{\alpha\alpha}(0)$ and $\rho_{\beta\beta}(0)$ representing their respective probabilities in the initial density matrices.}

\LL{Using these final states, we calculate the reduced density matrices for each qubit:
\begin{equation}
\rho_j(t) = \mathrm{Tr}_{i\neq j}\rho(t)
\end{equation}
\begin{equation}
\rho'_j(t) = \mathrm{Tr}_{i\neq j}\rho'(t)
\end{equation}
where $\mathrm{Tr}_{i\neq j}$ denotes the partial trace over all qubits except the $j$-th qubit. And the Holevo information is then calculated following the equation (\ref{Eq:Holevo information}).}

\begin{figure}[t]
  \centering
  \includegraphics[width=0.9\textwidth]{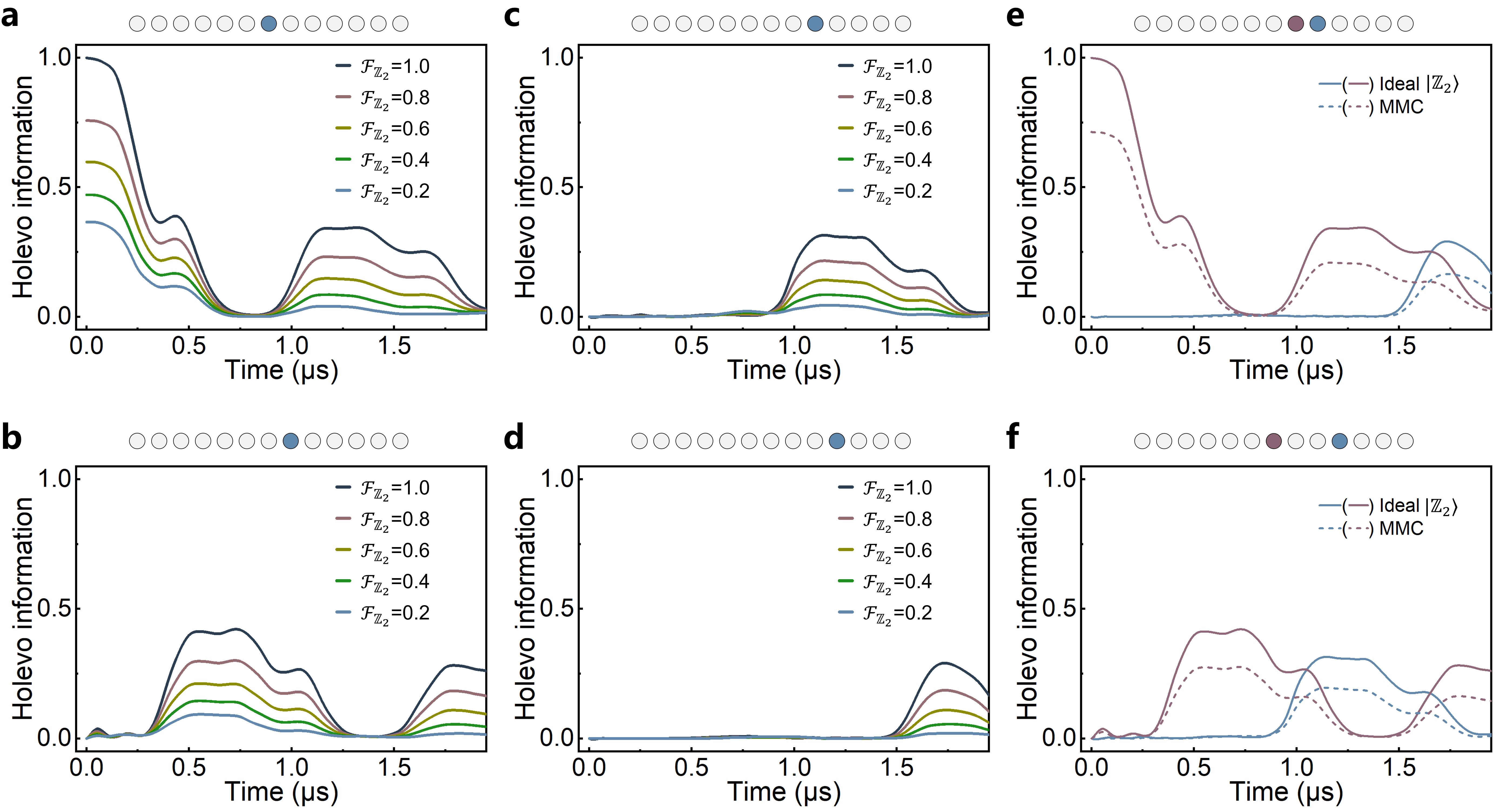}  
  \caption{
\textbf{Impact of initial state preparation fidelity on Holevo information dynamics.} 
\LL{
\textbf{a}--\textbf{d}, Simulated Holevo information dynamics with initial state preparation fidelities varying from 0.2 to 1. 
\textbf{e} and \textbf{f},
Solid lines represent dynamics with perfect initial states, while dashed lines show dynamics with experimentally prepared initial state fidelities.} The qubit chain above each plot indicates the positions of the qubits under consideration (coloured circles).
}
\label{Fig:Holevo_Z2}
\end{figure}

\LL{To assess the impact of imperfect initial state preparation on the Holevo information dynamics, we performed numerical simulations with different initial state preparation fidelities. Figure~\ref{Fig:Holevo_Z2}a--d demonstrates the Holevo information dynamics for initial state preparation fidelity ranging from 0.2 to 1.0. 
The results suggest that when the preparation fidelity decreases to 0.4 or lower, the \textit{collapse-and-revival} phenomenon after $\SI{1}{\micro\second}$ (driving Rabi frequency $\sim 2\pi \times\SI{1.2}{MHz}$) becomes significantly weakened and more challenging to observe experimentally, particularly for edge qubits.
These results further emphasize the necessity of high $\ket{\mathbb{Z}_2}$ state preparation fidelity for observing the \textit{collapse-and-revival} of quantum information. Based on the measured microstates combinations of the experimental prepared initial states, equal preparation fidelities for both $\ket{\mathbb{Z}_2}$ and $\sigma_c^x\ket{\mathbb{Z}_2}$ states were assumed in the numerical simulations, with error states uniformly distributed across all $\ket{\uparrow}$-initialized qubits. To this end, Fig.~\ref{Fig:Holevo_Z2}e,f compare the Holevo information dynamics from the experimentally prepared $\ket{\mathbb{Z}_2}$ and $\sigma_c^x\ket{\mathbb{Z}_2}$ states (dashed lines) with perfectly initial states (solid lines). The results suggest that, given the experimentally achieved high initial state preparation fidelity, the distinctive features of Holevo information dynamics persist and remain readily discernible.}

\subsection{Detection and evolution error}
\LL{Imperfections in quantum state evolution and detection introduce errors into experimental results, which degrade the performance of the Rydberg quantum simulator. This section identifies and analyzes two main categories of error: detection errors and evolution errors.}

\vspace*{0.5\baselineskip}

\noindent

\textbf{1. Detection error}

\vspace*{0.5\baselineskip}

\noindent

\LL{
The gap time (as described in section~\hyperref[Sec:OTOC]{3.1}) between the end of evolution and the start of detection results in Rydberg-state atoms decaying to the ground state due to their finite lifetime, contributing to detection errors.
Additionally, measurements of both Rydberg and ground states are subject to inherent detection errors (quantum state discrimination error).
}

\LL{To quantify these errors, we introduce two parameters: $\varepsilon$ to represent the detection error for the Rydberg state and $\eta$ to account for atomic loss in the ground state, both of which arise from the factors mentioned above. The experimentally measured ground state population, $P(\downarrow)$, can then be expressed as:}

\begin{equation}
P(\downarrow) = \varepsilon(1-\eta) P'(\uparrow) + (1-\eta) P'(\downarrow)
\end{equation}

\LL{Here, $P'(\uparrow)$ and $P'(\downarrow)$ represent the actual Rydberg and ground state population after experimental evolution, respectively.}

\vspace*{0.5\baselineskip}

\noindent

\textbf{2. Evolution error}

\vspace*{0.5\baselineskip}

\noindent

\LL{While we consider only two states (ground state $\ket{\downarrow}$ and Rydberg state $\ket{\uparrow}$) in numerical simulations, a third state (intermediate state $\ket{e} = \ket{5P_{3/2}}$, with a linewidth of $\Gamma_e$ $\approx$ 2$\pi \times \SI{6.06}{MHz}$) is involved in the evolution driven by the Raman lasers, and introduces incoherent errors in the dynamics of OTOC and Holevo information. 
The 480-\SI{}{\nano\meter} (780-\SI{}{\nano\meter}) Raman laser couples the $\ket{\uparrow}$ ($\ket{\downarrow}$) state to $\ket{e}$, leading to unwanted scattering and depolarization between $\ket{\downarrow}$ and $\ket{\uparrow}$. 
Furthermore, the radiative lifetime of the Rydberg state also contributes to the depolarization during the evolution process. These effects can be summed up and characterized by one parameter,
the depolarization time $T_1$,
accounting for the amplitude damping of both OTOC and Holevo information oscillations.}

\LL{Another major error source is coherent evolution error, also referred to as evolution noises. During the evolution, two dominant noise sources emerge: fluctuations in the relative phase between $\ket{\downarrow}$ and $\ket{\uparrow}$, and variations in the Rabi frequency. These sources contribute to non-unitarity in the forward-and-backward Hamiltonian evolution and decoherence in Rabi oscillations. 
The time-dependent noisy Rydberg Hamiltonian is modeled as:}

\begin{equation}
H(t) = \sum_i \left[\frac{\Omega(t) e^{-\mathrm{i}\phi(t)}}{2} \sigma^x_i - \Delta(t) n_i\right] + \sum_{i<j} V_{ij}(t) n_i n_j
\label{eq:Noisy-Hamiltonian}
\end{equation}

\LL{Here, $\phi(t)$ and $\Omega(t)$ represent the time-dependent phase and Rabi frequency, respectively. $\Delta(t)$ accounts for time-dependent laser frequency detuning. These time-dependent fluctuations contribute to the single-atom decoherence time $T_2^*$, arising from various sources including laser noises and Doppler effects. Additionally, $V_{ij}(t)$ denotes the time-dependent Rydberg-Rydberg interaction strength between atoms $i$ and $j$ in the many-body system, whose uncertainty is introduced by atomic motion and the disorder in the initial atomic distance.}

\LL{In our experiment, as described in section~\hyperref[section:setup]{1.1}, we employ the Pound-Drever-Hall (PDH) technique to frequency-stabilize the Rydberg excitation laser to a ULE cavity. This method effectively suppresses laser frequency noise below the cavity linewidth. }
\LL{We treat the high-frequency noises above the linewidth in two components: one attributed to servo bumps \cite{levine2018high, de2018analysis}, and the remaining noise, which can be modeled as spectrally uniform (white) phase noise \cite{jiang2023sensitivity}.}
\LL{As a result, $\Delta(t)$ in equation (\ref{eq:Noisy-Hamiltonian}) can be approximated by a Gaussian distribution with a root-mean-square (RMS) amplitude of $\delta\Delta$. Moreover, variations in laser power and spatial inhomogeneities induce Gaussian-type perturbations in the Rabi frequency, characterized by an RMS amplitude of $\delta\Omega$. Additionally, the parameter $\delta\phi$ is introduced to represent the uncertainty in $\phi(t)$, to account for the fluctuations in the relative phase between the Rydberg state and the ground state during the evolution process, typically arising from the servo bump of the excitation lasers.}

\begin{figure}[t]
\centering
\includegraphics[width=0.7\textwidth]{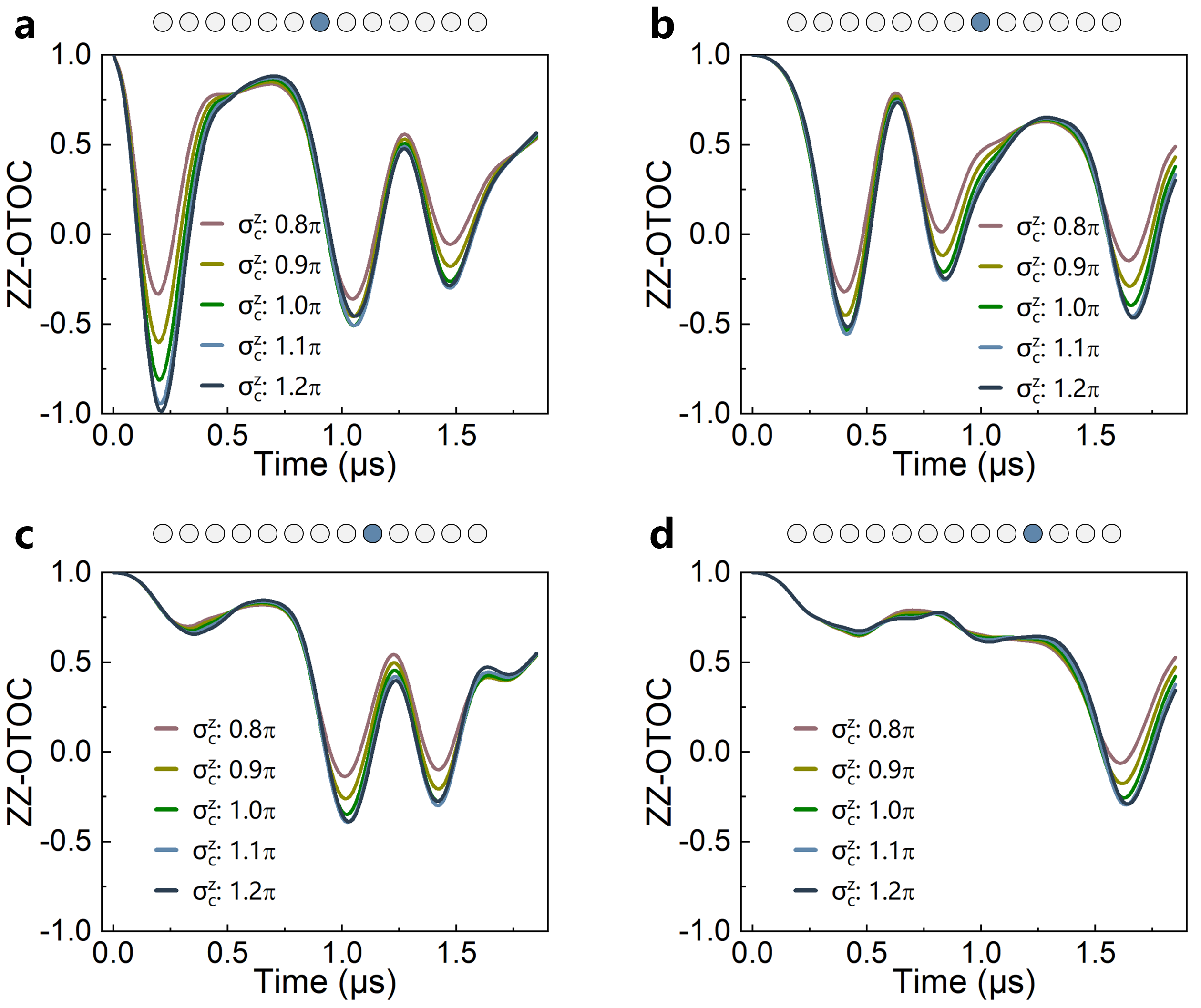}
\caption{\LL{
\textbf{Effect of local perturbation fidelity on ZZ-OTOC measurements.} Simulated ZZ-OTOC dynamics for four representative qubits in a 13-qubit chain, demonstrating the impact of $\sigma_i^z$ fidelity on OTOC oscillations. The $\sigma_c^z$ fidelity is represented by the accumulated phase, ranging from 0.8$\pi$ to 1.2$\pi$. \textbf{a}-\textbf{d}, ZZ-OTOC evolution for different qubits. The qubit configuration is shown above each plot, with the orange circle indicating the qubit under consideration.
}}
\label{SigmaZ}
\end{figure}

\LL{The fidelity of the local perturbation $\sigma_c^z$ also significantly impacts the experimentally measured ZZ-OTOC values. Numerical simulations show that the fidelity of $\sigma_c^z$ operations directly affects the contrast of OTOC oscillations within the light cone. This relationship is illustrated in Fig.~\ref{SigmaZ}, presenting simulation results for a 13-qubit array. Since the $\sigma_c^z$ gate is accomplished with a relative $\pi$ phase shift between the Rydberg state and ground state induced by far-detuned 795-\SI{}{\nano\meter} addressing laser beams, the infidelity mainly stems from uncertainty in the accumulated phase.}

\LL{This comprehensive error model captures the primary sources of noise in our system. By identifying and formalizing these errors, we establish a framework for accurately interpreting experimental results. This approach provides a solid foundation for error benchmarking and mitigation protocols, which are crucial for enhancing the accuracy of OTOC and Holevo information measurements in probing quantum information collapse and revival.}

\subsection{Error Characterization}
\label{subsection:error model}

\LL{To quantify and mitigate errors in our quantum simulator, we conducted a series of calibration experiments to characterize the error sources identified in our model.}

\vspace*{0.5\baselineskip}

\noindent

\textbf{1. Detection error}

\vspace*{0.5\baselineskip}

\noindent

\LL{
We systematically characterize detection errors in our system. For atoms in the ground state, the raw detection error $\eta$ is approximately 1\%. For Rydberg states, the detection error is more complex, consisting of a raw error $\varepsilon' \approx 1\%$ along with an additional time-dependent component arising from the finite Rydberg state lifetime.
}

\LL{To quantify this time dependence, we measure the lifetime $T_R$ of the Rydberg state used in our experiment. Atoms are first prepared in the Rydberg state using a global Raman $\pi$-pulse, after which we vary the time interval between the $\pi$-pulse and the population measurement. The population measurement is performed by turning on the optical tweezers to recapture ground state atoms while repelling the Rydberg atoms. An exponential fit to the data yields a 1/e time constant of $T_R = \SI{140(15)}{\us}$. Given the interval time $t_i$, which depends on the evolution time $t$ in our sequence, the Rydberg state detection error accumulates over time. This error is expressed as $\varepsilon'(t) = 1 - e^{-t_i/T_R}$, representing the probability of Rydberg atoms decaying to the ground state during the interval.}

\vspace*{0.5\baselineskip}

\noindent

\textbf{2. Evolution error}

\vspace*{0.5\baselineskip}

\noindent

\begin{figure}[t]
  \centering
  \includegraphics[width=0.8\textwidth]{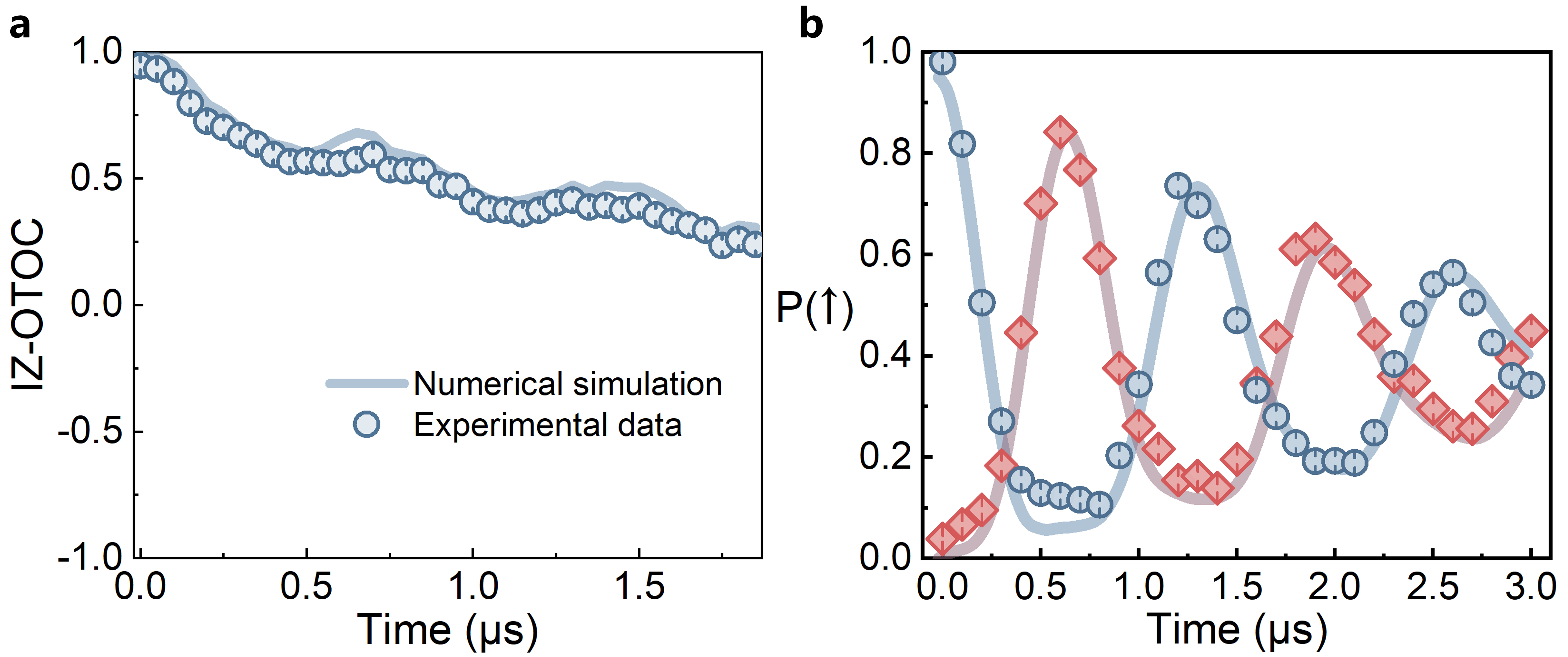}  
  \caption{
\textbf{Investigating the decay mechanisms in OTOC and Holevo information dynamics.} 
\LL{\textbf{a}, Comparison of experimental data (blue points, corrected the detection errors and the incoherent errors introduced by the depolarization effect) with numerical simulations (solid lines) for IZ-OTOC with initial $\ket{\mathbb{Z}_2}$ state. \textbf{b}, Rydberg Hamiltonian evolution dynamics of initial $\ket{\mathbb{Z}_2}$ state. Blue circles and red diamonds represent detection errors corrected experimental results for $\ket{\uparrow}$- and $\ket{\downarrow}$- initialized state, respectively, while solid lines show corresponding numerical simulations. The excellent agreement confirms that our understanding of the decay mechanisms in OTOC and Holevo information dynamics is accurate.}
}
  \label{IZ-OTOC}
\end{figure}

\LL{Evolution errors in our system arise from three main sources: the depolarization caused by spontaneous emission, the finite temperature of the atoms, and laser noise.}

\B{Depolarization effects}. \LL{We accounted for depolarization time $T_1$ due to spontaneous emission via the intermediate state and Rydberg state radiative decay. 
The error probability of a Rydberg state decaying to the ground state during evolution time $t$ is approximated as $\eta = \gamma t$ with $\gamma t \ll 1$. Here, $\gamma = 1/T_1$ is treated as a free parameter due to the complexity of many-body evolution. This complexity arises from two factors: (1) during evolution, different initial states (e.g., $\ket{\mathbb{Z}_2}$ and $\ket{\mathbf{0}}$) lead to variations in the average Rydberg population, resulting in different effective decay rates; and (2) the exponentially growing Hilbert space and complex interactions in many-body state evolution cause the effective decay rate to differ from the more easily measured decay rate in single-atom evolution.}

\LL{\B{Finite-temperature effects}. The thermal motion of atoms leads to two effects: fluctuations in atomic positions and Doppler shifts. Position fluctuations affect the Rydberg-Rydberg interactions, which scale as $1/R^6$, where R is the inter-atomic distance. We estimated the standard deviation of position fluctuations to be about \SI{0.3}{\um}, directly impacting the strength of Rydberg-Rydberg interactions. Doppler shifts, on the other hand, introduce frequency detuning in the Rydberg excitation. We characterized these thermal effects by measuring the average temperature of the atoms at the beginning of evolution using the release and recapture method, finding it to be approximately \SI{10}{\micro K}. From this, we calculated the standard deviation of the atomic velocity distribution as $\sigma_v = \sqrt{k_B T/M}$, where $k_B$ is the Boltzmann constant, $T$ is the temperature, and $M$ is the atomic mass. 
For our counter-propagating two-photon excitation scheme with a 480-\SI{}{\nano\meter} $\sigma^+$-polarized and a 780-\SI{}{\nano\meter} $\sigma^+$-polarized light, 
this corresponds to a Doppler broadening with a standard deviation of $\delta{\Delta_1}=k \sigma_v \approx 2 \pi \times \SI{25}{kHz}$, where $k\approx \SI{1.25}{\per \um}$ is the effective two-photon wave vector.}

\LL{\B{Laser noise}. We characterized both intensity and phase noise of our laser sources. The Rabi frequency fluctuation $\delta \Omega/\Omega$ is related to the laser intensity fluctuation $\delta I/I$ by $\delta \Omega/\Omega \approx \delta I/(2I)$. High-bandwidth measurements of 780-\SI{}{\nano\meter} and 480-\SI{}{\nano\meter} Raman laser power variations using fast photodiodes revealed an RMS amplitude noise of $\delta \Omega/\Omega \approx 0.01$. For laser frequency noise, we analysed the in-loop PDH error signal at Fourier frequencies above the cavity linewidth ($\gamma_\text{cav} \approx 2\pi \times \SI{110}{kHz}$ at \SI{960}{nm} and $2\pi \times \SI{60}{kHz}$ at \SI{780}{nm}). We estimated the laser frequency noise by integrating the noise spectral density $S_{\nu}(f)$:}
\begin{equation}
\delta{\Delta_2} = \sqrt{\int_{\gamma_\text{cav}}^{f_h} S_{\nu}(f) df},
\end{equation}
\LL{where $f_h \approx 1/\delta t$. This yields an RMS frequency noise of $\delta{\Delta_2} \approx 2\pi \times\SI{5}{kHz}$ for the combined 780-\SI{}{\nano\meter} and 480-\SI{}{\nano\meter} laser contributions. Since this frequency noise contributes to the uncertainty in the relative phase between the qubit and the driving field, $\delta\phi$, in the same way as the Doppler effect, we consider only $\delta\phi$ instead of the combination of $\delta{\Delta_1}$ and $\delta{\Delta_2}$ in the following analysis.}

\begin{figure}[t]
  \centering
  \includegraphics[width=0.95\textwidth]{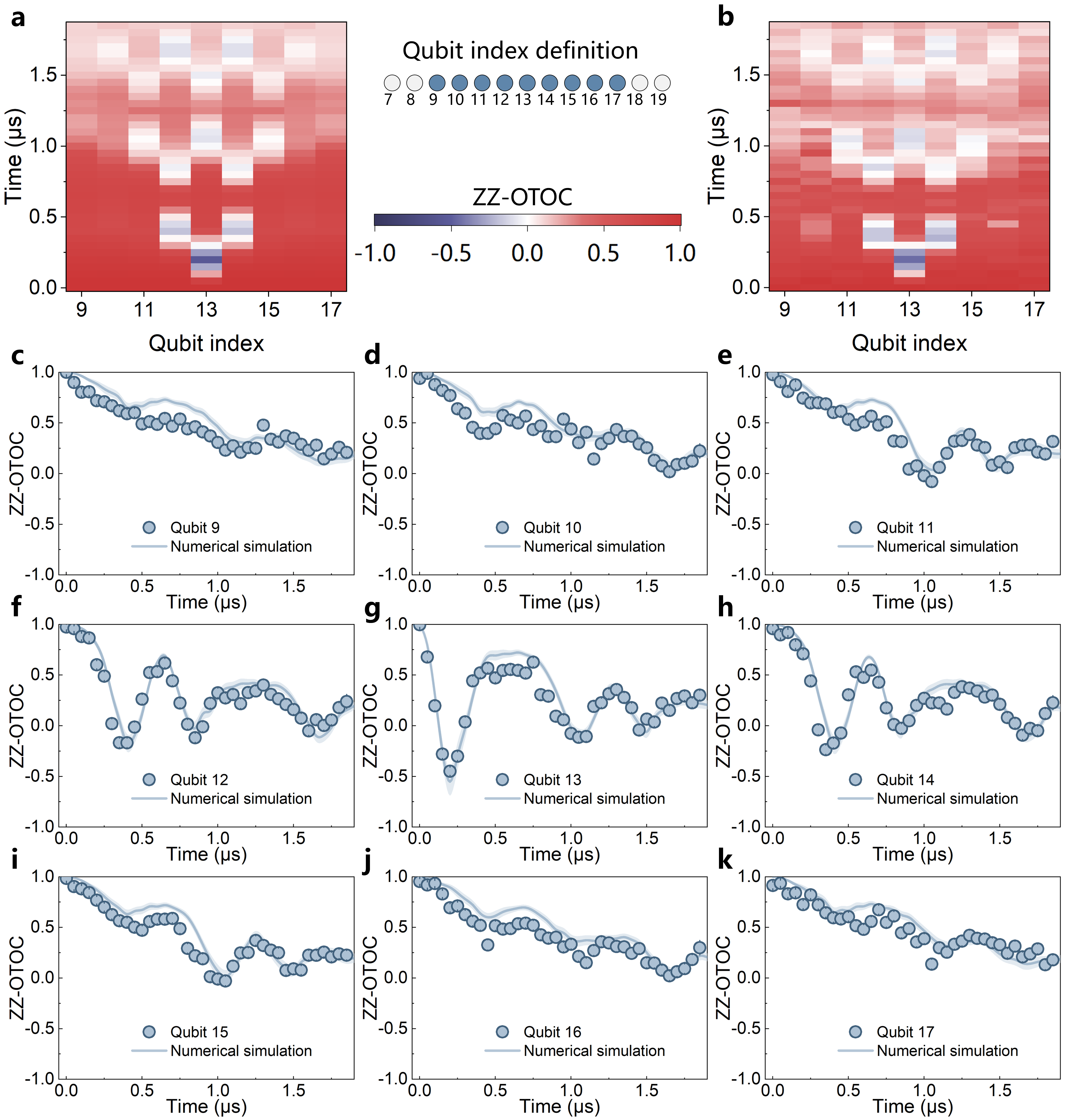}  
  \caption{
\textbf{Dynamics of ZZ-OTOC for $\ket{\mathbb{Z}_2}$  considering experimental imperfections.} \LL{
\textbf{a}, Simulated spatio-temporal evolution of ZZ-OTOC for $\ket{\mathbb{Z}_2}$ state under the time-dependent noisy Hamiltonian (\ref{eq:Noisy-Hamiltonian}). \textbf{b}, Experimental ZZ-OTOC data for $\ket{\mathbb{Z}_2}$ state, corrected for detection errors and incoherent errors arise from the depolarization effect. Inset, The qubit index definition of the exhibited qubits (highlighted) in 13-qubit chain (top).
\textbf{c}--\textbf{k}, Detailed dynamics plots of the corrected experimental data (blue points) and the simulation results (solid curve). The shaded areas around the curves represent the error bar from the numerical simulation. The corresponding qubit index of plots is respectively marked; good agreement between experimental data and numerical results is found for all qubits. 
      }}
  \label{Noise_model}
\end{figure}

\LL{To further calibrate $\delta\phi$ and $\gamma$, we measured the IZ-OTOC (main text), which follows the same sequence as our ZZ-OTOC experiments but without the local perturbation. Numerical simulations treating $\delta\phi$ as a free parameter yield excellent agreement with corrected experimental data for $\delta\phi = 0.08\pi$ and $\gamma = \SI{0.035}{\per \micro\second}$ (Fig.~\ref{IZ-OTOC}a). For Holevo information evolution, we measured Rydberg Hamiltonian evolution dynamics of the initial $\ket{\mathbb{Z}_2}$ state, finding good agreement when $\delta\phi = 0.08\pi$ (Fig.~\ref{IZ-OTOC}b).}

\LL{
The experimental results shown in Fig.~\ref{IZ-OTOC}a and Fig.~\ref{Noise_model} have been corrected for detection errors and partially corrected for evolution errors. Specifically, while detection errors were fully accounted for, only the evolution errors related to Rydberg state decay to the ground state ($\gamma t$) were addressed. First, detection error correction was applied. Then, we subtracted the accumulating population of Rydberg states decaying to the ground state ($\gamma t$) from the experimentally measured ground state population to compensate for the incoherent errors introduced by the intermediate state during the experiment.
}

\LL{
For ZZ-OTOC measurements, an extra error source is the $\sigma^z$ local perturbation infidelity. In order to characterize the uncertainty in local perturbation, we conducted a comparative analysis of the results obtained from Ramsey experiments with and without 795-\SI{}{\nano\meter} addressing employed for local $\sigma_i^z$. The findings indicate that uncertainty is approximately 0.09$\pi$. This uncertainty contributes to the overall evolution noise and affects the fidelity of our local perturbation.
}

\begin{figure}[t]
\centering
\includegraphics[width=1\textwidth]{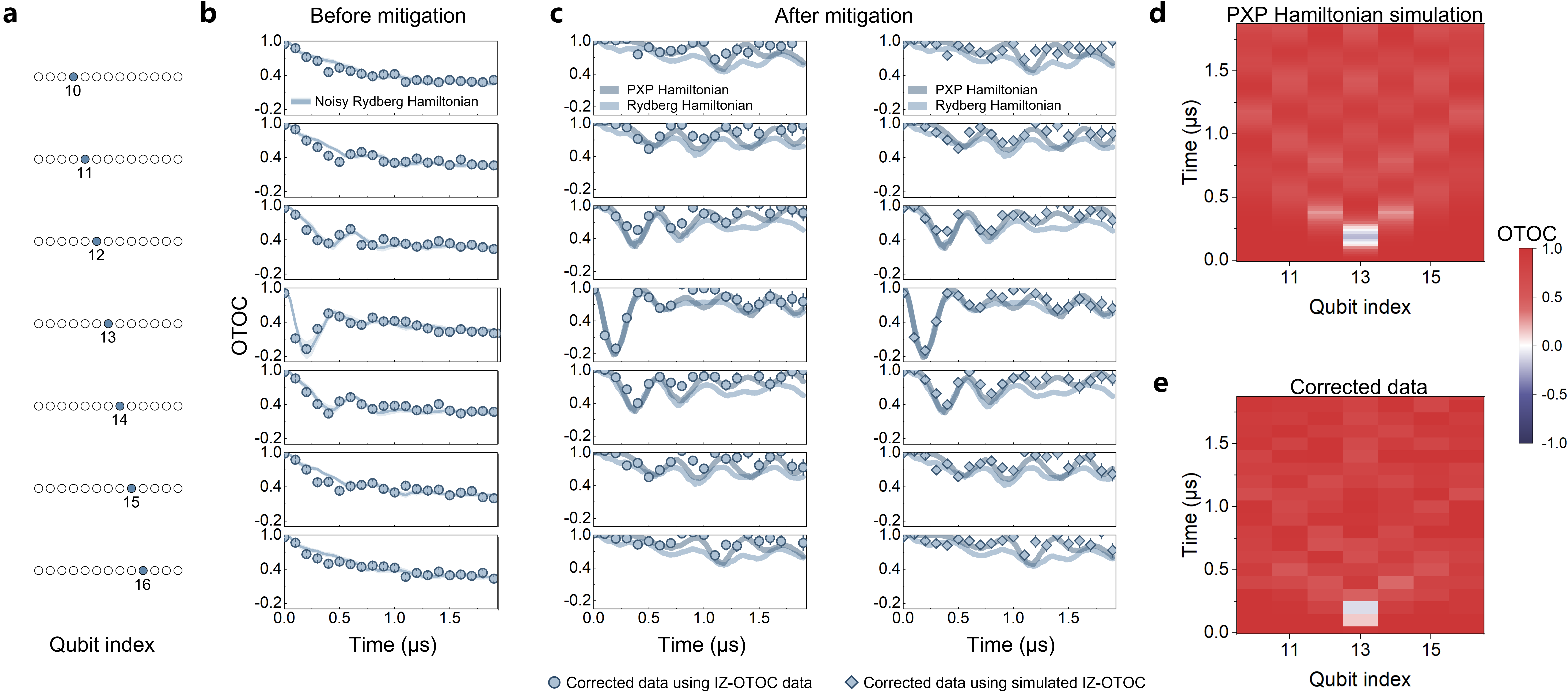}
\caption{\LL{\textbf{Error mitigation for OTOC dynamics of the initial $\ket{\mathbf{0}}$ state.} \textbf{a}, Qubit index definition for each row in panels b and c (blue filled circle). The central 7 qubits from the spin chain are used for analysis.
\textbf{b}, OTOC dynamics before error mitigation. Blue curves with shaded error bars represent simulations using the modeled noisy Rydberg Hamiltonian. Circles are the experimental data, demonstrating excellent agreement with the simulations.
\textbf{c}, OTOC dynamics after error mitigation. Left: Experimental data corrected using measured IZ-OTOC (circles). Right: Experimental data corrected using simulated IZ-OTOC (diamonds). In both cases, light blue curves represent simulations with the ideal Rydberg Hamiltonian, while dark blue curves show the ideal PXP Hamiltonian dynamics. 
\textbf{d--e}, Spatio-temporal OTOC dynamics. \textbf{d}, Simulated OTOC dynamics for the central 7 qubits using the PXP Hamiltonian, incorporating imperfections from local perturbations. \textbf{e}, Experimental data corresponding to the left panel data in \textbf{c}. The colour bar corresponds to the values of the OTOC.
The excellent agreement between corrected data and simulations demonstrates the effectiveness of the error mitigation scheme.
}}

\label{FIG:Ground_Mitigated_data}
\end{figure}

\LL{Using the noise parameters mentioned above, we simulated the dynamics of the ZZ-OTOC with initial $\ket{\mathbb{Z}_2}$ state under the noisy Rydberg Hamiltonian (\ref{eq:Noisy-Hamiltonian}) based on the Monte Carlo method. We compared the simulation results with the experimental results for the central 9 qubits (the qubit index definition is shown in Fig.~\ref{Noise_model} inset). For the OTOC dynamics of the initial state $\ket{\mathbf{0}}$, we employed the same error model. Figure~\ref{FIG:Ground_Mitigated_data}b presents the comparison between the simulation results and the experimental data for the central 7 qubits. The excellent agreement between these simulations and our experimental data (Fig.~\ref{Noise_model} for $\ket{\mathbb{Z}_2}$ state and Fig.~\ref{FIG:Ground_Mitigated_data}b for $\ket{\mathbf{0}}$ state) provides solid validation for our error model, which accounts for both detection and evolution errors, and enhances our understanding of the complex many-body dynamics.
}

\subsection{Error mitigation for ZZ-OTOC}
\label{subsection:Error Mitigate}

\begin{figure}
  \centering
  \includegraphics[width=0.8\textwidth]{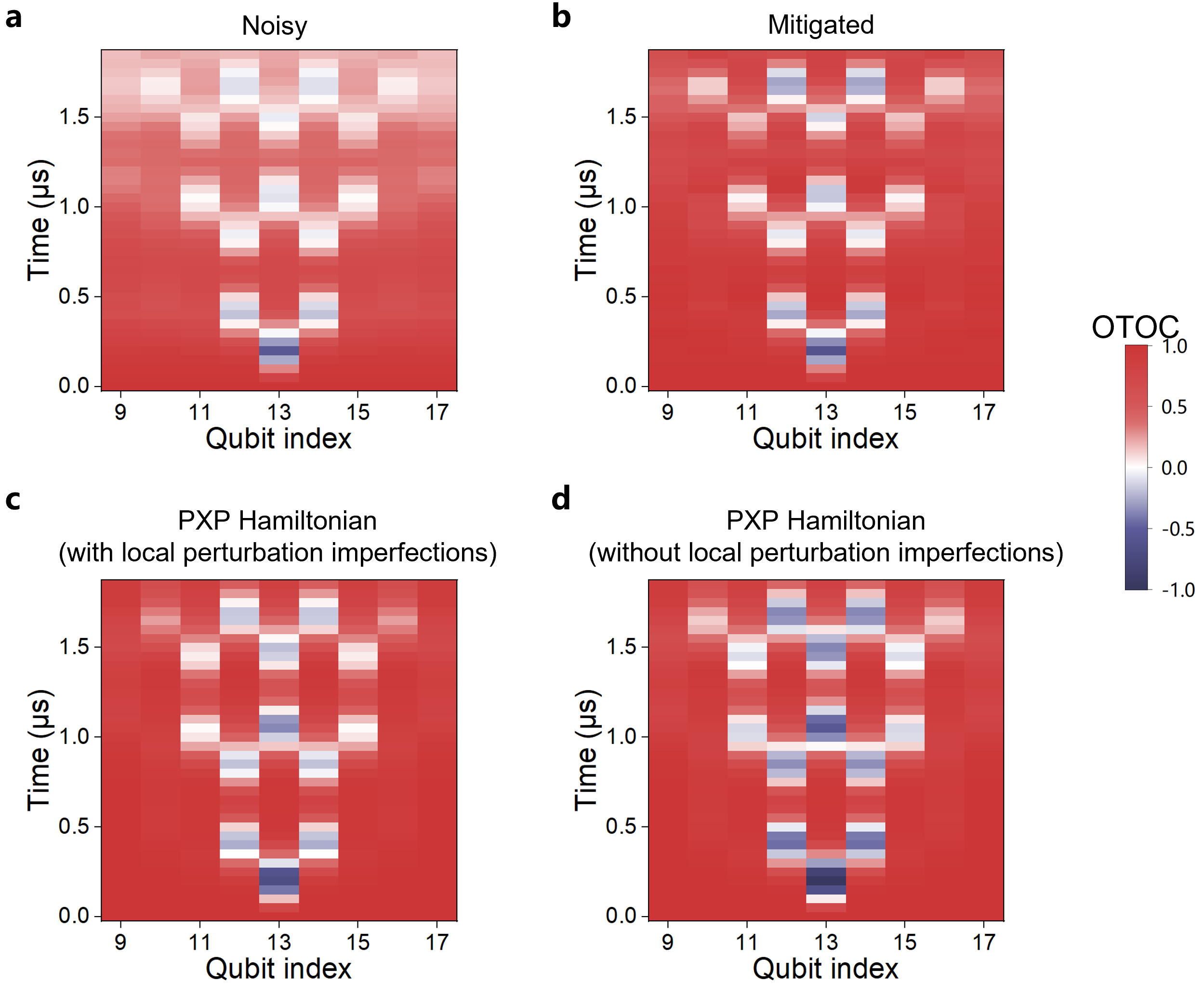}
  \caption{
\textbf{Numerical simulation of error mitigation scheme for ZZ-OTOC.} \textbf{a}, Simulated ZZ-OTOC dynamics for $\ket{\mathbb{Z}_2}$ considering experimental imperfections same as Fig.~\ref{Noise_model}a. \textbf{b}, Mitigated ZZ-OTOCs using simulated IZ-OTOC results, exhibiting enhanced \textit{collapse-and-revival} contrast compared to \textbf{a}.  \textbf{c--d}, Simulated ZZ-OTOC dynamics using the PXP Hamiltonian while accounting for local perturbation imperfections (\textbf{c}) or not (\textbf{d}). The closer resemblance of the mitigated data \textbf{b} to \textbf{c} demonstrates that our error mitigation scheme cannot mitigate local perturbation imperfections. The colour scale represents ZZ-OTOC values from -1.0 (blue) to 1.0 (red). 
	  }
  \label{Mitigate}
\end{figure}

\LL{
We employ an error mitigation scheme inspired by Swingle and Halpern\cite{swingle2018resilience} and Mi \textit{et al.} \cite{mi2021information} to address imperfections in OTOC measurements. Theoretical analysis indicates that under experimental conditions with imperfections, the errors in the measured ZZ-OTOC $F^{m}(W,V)$ can be effectively mitigated using the measured IZ-OTOC $F^{m}(I,V)$  \cite{swingle2018resilience}:
\begin{equation}
\label{equ:mitigation}
F^{c} \approx \frac{F^{m}(W,V)}{F^{m}(I,V)},
\end{equation}
where $F^{c} $ represents the corrected ZZ-OTOC measurement results.}

\LL{
This scheme mitigates the imperfections in forward-and-backward evolution
caused by coherent noise ($\delta{\phi}$, $\delta{\Omega}$, $\delta{\Delta}$) and partially mitigates those from next-nearest-neighbour interactions $V_{i,i+2}$. However, it cannot effectively mitigate errors outside the forward-and-backward evolution, such as imperfections in the local $\sigma^z_i$ perturbation and the detection errors. Consequently, mitigated results are expected to fall between expectations of the noise-free Rydberg Hamiltonian and the ideal PXP model.
}

\LL{The denominator, IZ-OTOC $F^{m}(I,V)$, is crucial in the mitigation protocol. As demonstrated by Mi \textit{et al.}~\cite{mi2021information}, small variations in the IZ-OTOC used as the denominator can dramatically affect corrected results, particularly for near-zero IZ-OTOC values. Recognizing this sensitivity, we conducted a comprehensive analysis of factors potentially affecting OTOC measurements and quantified them (detailed in section~\hyperref[subsection:error model]{4.3}). Notably, numerical simulations show significant edge effects in small-sized chains. Comparing the edge atom's ZZ-OTOC versus IZ-OTOC as the denominator for mitigation in a noisy environment shows that using the edge atom's ZZ-OTOC leads to over-correction (as large as 50\%) and temporal misalignment at critical positions (Fig.~\ref{Fig:Boundary and finite-size effect}d). In contrast, using IZ-OTOC produces results that align with theoretical predictions. To minimize edge effects, we rely on IZ-OTOC measurements instead of the edge atom's ZZ-OTOC for error mitigation.}

\ZP{The numerical simulations are conducted to evaluate the effectiveness of this error mitigation scheme in the context of our experimental imperfections.
Figure.~\ref{Mitigate}a shows the numerically simulated ZZ-OTOC dynamics using the noisy Rydberg Hamiltonian in equation (\ref{eq:Noisy-Hamiltonian}), which incorporates major experimental imperfections (as Fig.~\ref{Noise_model}). The mitigated case (Fig.~\ref{Mitigate}b) closely resembles the PXP Hamiltonian with local perturbation imperfections shown in Fig.~\ref{Mitigate}c, indicating that the mitigation scheme successfully addresses forward-and-backward evolution noise. However, when compared to the ideal PXP Hamiltonian without imperfections (Fig.~\ref{Mitigate}d), the mitigated case shows slightly less pronounced features. This subtle difference can be attributed to the local perturbation imperfections which can not be mitigated.}

\begin{figure}[t]
\centering
\includegraphics[width=1\textwidth]{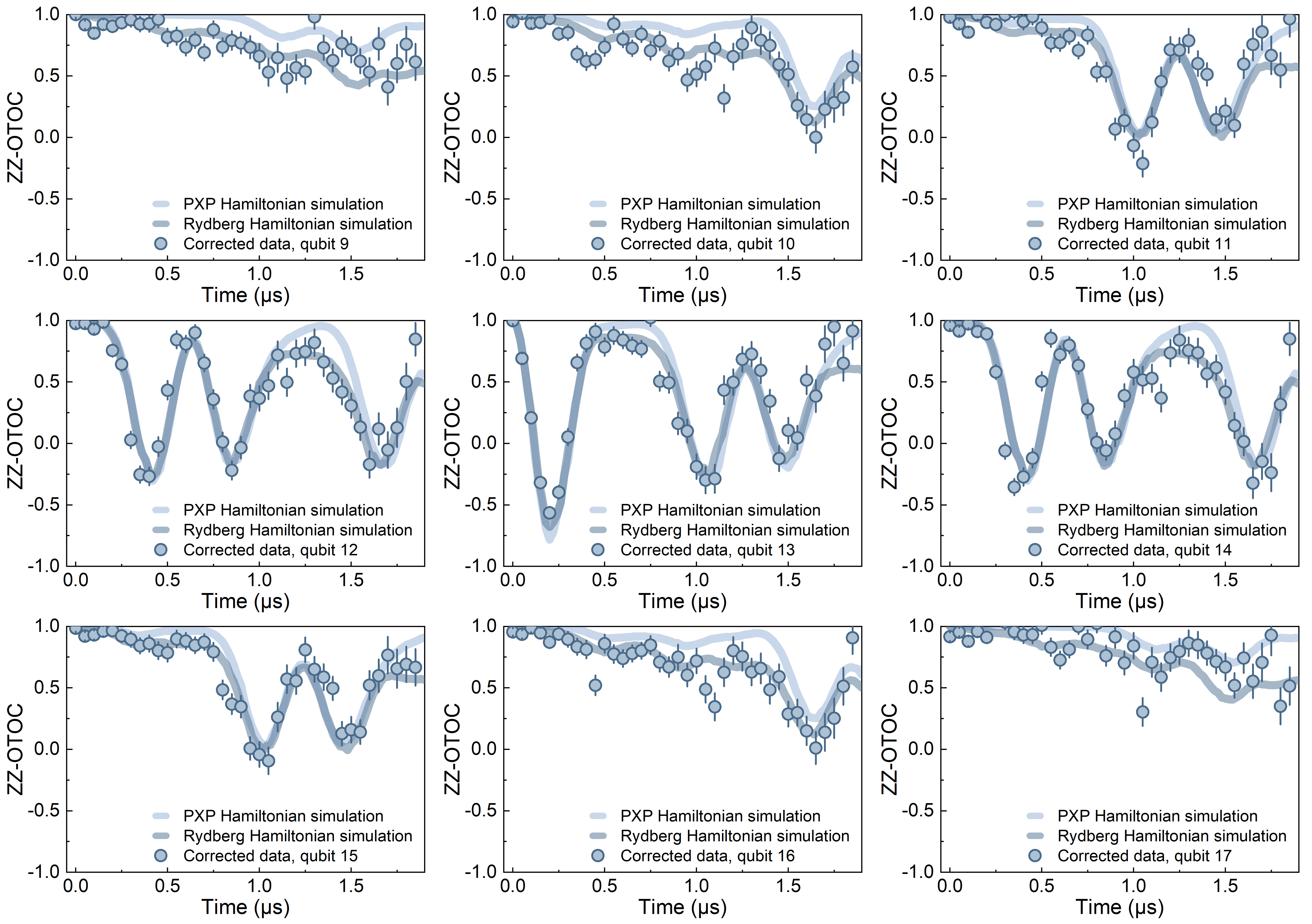}
\caption{\textbf{Mitigating ZZ-OTOC for $\ket{\mathbb{Z}_2}$ state using the simulated IZ-OTOC results.} \LL{
Blue points marked with qubit indices represent the mitigated experimental data using simulated IZ-OTOC results.
Dark blue curves show simulated ZZ-OTOC dynamics for the initial $\ket{\mathbb{Z}_2}$ state with the ideal Rydberg Hamiltonian (equation (\ref{eq:rydberg_hamiltonian}), no gap time during the OTOC evolution).
Light blue curves display simulated ZZ-OTOC dynamics for the initial $\ket{\mathbb{Z}_2}$ state with the ideal PXP Hamiltonian.
The mitigated experimental data show excellent agreement with simulations. }}
\label{Fig:mitigated_data}
\end{figure}

As the first step in error mitigation, we correct the detection errors, i.e., the imperfections in measurement operator $\sigma^z_j$. \LL{Next, we correct the evolution errors.} 
Given the excellent agreement between experimental IZ-OTOC data and numerical simulations of IZ-OTOC (as shown in Fig.~\ref{IZ-OTOC}a), we can effectively use either the experimental IZ-OTOC data or the simulated IZ-OTOC results to mitigate the experimental ZZ-OTOC data shown in Fig.~\ref{Noise_model}c--k. The mitigated results of the initial $\ket{\mathbb{Z}_2}$ state are presented in Extended Data Fig. 2 (using the measured IZ-OTOC data) and Fig.~\ref{Fig:mitigated_data} (using the simulated IZ-OTOC results). 
Figure~\ref{FIG:Ground_Mitigated_data}c shows the mitigated results of the initial $\ket{\mathbf{0}}$ state using the measured IZ-OTOC data (left) and the simulated IZ-OTOC results (right). 
\ZYW{
All the results demonstrate significant improvement in the agreement between mitigated experimental data and theoretical expectations  for the ZZ-OTOC (with imperfections in local perturbation $\sigma^z_i$, light blue curve under ideal PXP Hamiltonian while the dark blue under ideal Rydberg Hamiltonian). This excellent agreement underscores the effectiveness of our error mitigation protocol in addressing the complex noise landscape of the Rydberg quantum simulator.}

\ZYW{This approach effectively overcomes experimental imperfections, particularly those associated with \ZP{forward-and-backward evolution} in OTOC measurements, enabling accurate probe of quantum information dynamics in Rydberg atom quantum simulators.}

\subsection{Error mitigation for Holevo information}

\begin{figure}[t]
\centering
\includegraphics[width=1\textwidth]{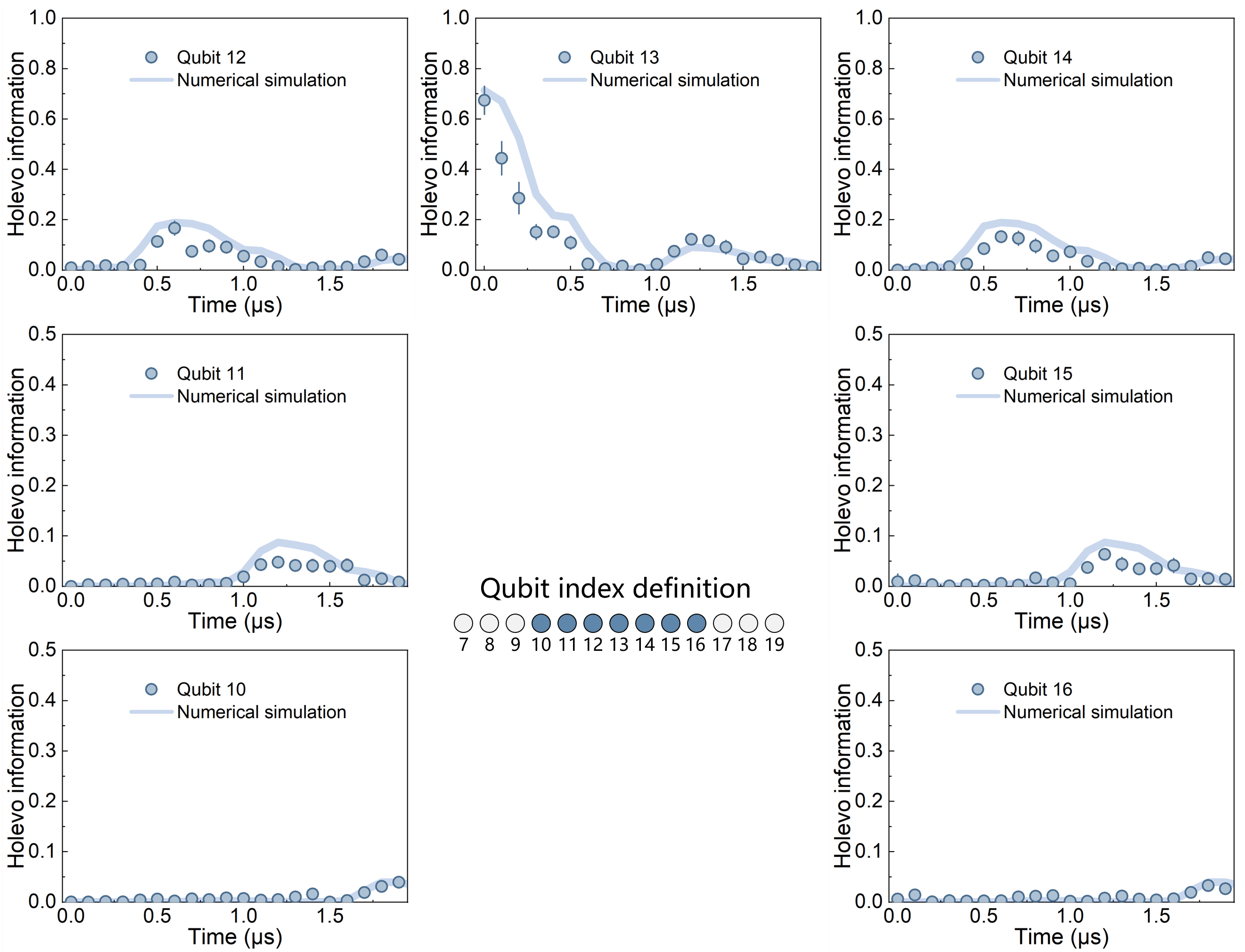}
\caption{\textbf{Experimental data and numerical simulations for Holevo information dynamics.} 
\LL{Holevo information dynamics of the detection-error-corrected experimental data (blue points), compared to the numerical simulation results (solid curves). The inset highlights the index definitions of 7 selected qubits within the 13-qubit chain. Monte Carlo methods are employed in the simulation to account for imperfections during initial state preparation and evolution.}
}
\label{Fig:mitigated_HI}
\end{figure}

\LL{The detection error for Holevo information can be mitigated similarly to how OTOCs are handled. The measured diagonal elements, which are linear transformations of $P(\uparrow)$, are directly corrected for detection errors, and the off-diagonal elements can be extracted from sinusoidal fittings of detection-error-corrected Ramsey oscillations. 
However, evolution errors are more complex and cannot be easily mitigated because they are deeply intertwined with the quantum information dynamics. 
Due to the difficulty in determining whether quantum information initially encoded in a qubit is lost due to evolution errors or transferred to other qubits, it is very challenging to apply traditional error mitigation techniques that focus on compensating for single-qubit decoherence.
Therefore, evolution errors are not mitigated for Holevo information. Instead, they are included in the numerical simulations of Holevo information dynamics, together with state preparation errors, showing good agreement with experimental data (Fig.~\ref{Fig:mitigated_HI}).}

\section{Kinetically constrained dynamics and quantum information collapse-and-revival}

\subsection{Investigation of kinetically constrained dynamics}

\LL{
To characterize the constrained spin dynamics in our Rydberg atom chain, we developed a method to identify and analyse the wavefront of \LL{excitations}. This approach is particularly effective for the $\ket{\mathbb{Z}_2}$ state, where individual spins exhibit periodic but non-sinusoidal oscillations, often with phase differences between neighbouring atoms.
Our wavefront detection method identifies the moments when adjacent atoms have equal Rydberg excitation probabilities, $P_i(\uparrow) = P_{i+1}(\uparrow)$, based on numerical simulations (Fig. 2e,g in the main text). By connecting these time points for each nearest-neighbour atom pair, we construct wavefronts that capture the propagation of excitations throughout the system. This technique allows us to study distinct behaviours for different initial configurations.
}

\LL{
Our simulations indicate that the $\ket{\mathbb{Z}_2}$ state exhibits uniform wavefront propagation throughout the bulk of the spin chain, consistent with the synchronized evolution observed in the experiment. This synchronization arises from the interplay between the PXP constraints and the initial $\ket{\mathbb{Z}_2}$ configuration. In this regime, each spin experiences a similar effective environment due to the alternating pattern of its neighbours, leading to coherent and synchronized rotations across the bulk of the system.
}

\LL{
However, near the edges of the chain, deviations from this synchronized behaviour begin to appear. These boundary effects manifest as distortions in the wavefront shape, reflecting the altered local environment of the outermost atoms. Near the boundaries, the lack of symmetry and the different neighbouring structure cause spins to evolve out of sync with those in the central region. 
This gradual desynchronization, moving from the outer edges toward the center, aligns with the boundary effects described in the main text. 
Figure 2e in the main text provides a spatial map of the wavefront propagation, clearly illustrating the transition from uniform propagation in the bulk to distorted behaviour at the edges.
}

\LL{
For the $\sigma^x_c\ket{\mathbb{Z}_2}$ state with the central spin flipped, we observe rich dynamical behaviour characterized by a clear linear light cone structure, as discussed in the main text. The flipped central spin introduces retardation in adjacent spins' rotation, which propagates outwards as the system evolves.
Inside the light cone, there is an interplay between periodic spin rotations and retardations due to the kinetically constrained dynamics. This results in an arc-shaped, curved wavefront that moves outward from the initial perturbation at the central spin. Figure 2g in the main text illustrates this light cone structure and the corresponding wavefront propagation.
The clear visualization of the light cone and wavefront behaviour offers a valuable tool for understanding the kinetically constrained quantum many-body systems.
}

\subsection{Illustration of quantum information collapse-and-revival in PXP model}

\begin{figure}
  \centering
  \includegraphics[width=\textwidth]{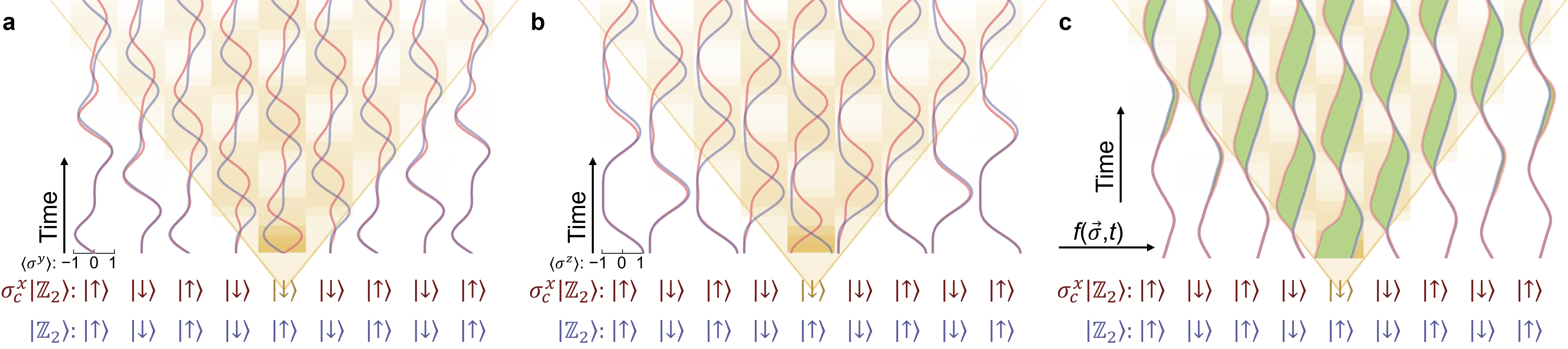}  
  \caption{
\textbf{Illustration of constrained qubit dynamics in PXP model.} 
\LL{
\textbf{a}, Dynamics of $\langle \sigma^y \rangle$ and \textbf{b}, $\langle \sigma^z \rangle$ for each qubit. \textbf{c}, The dynamic indicator $f(\vec{\sigma},t) = \int_{0}^{t}\mathrm{d}\{\mathrm{Arg}[\vec{\sigma}(\tau)]\}\mathrm{d}\tau - \lambda\Omega t$, which represents the total rotation angle of the Bloch vector in the YZ-plane. The blue curves correspond to the initial state $\ket{\mathbb{Z}_2}$, while the red curves represent the initial state $\sigma^x_c \ket{\mathbb{Z}_2}$, where the central spin is flipped. The green-filled intervals in \textbf{c} indicate regions of retardation and distinguishability between the two initial states. Yellow lines connect the divergence points, forming light cones (yellow shaded areas) where the dynamics of $\sigma^x_c \ket{\mathbb{Z}_2}$ are periodically delayed. The checkerboard background in all panels shows the heatmap of the Holevo information dynamics (dark yellow: $\mathbb{X}_j(t)=1$; transparent: $\mathbb{X}_j(t)=0$).
  }}
  \label{Fig:S0}
\end{figure}

\LL{\textit{Collapse-and-revival} is a dynamical phenomenon in quantum systems where observable quantities, like atomic operator expectation values, ``collapse'' into near-zero values before periodically ``reviving''~\cite{fan2023collapse, Michailidis2020}. This effect is most famously observed in systems with discrete quantum states interacting with a quantized field, such as the Jaynes-Cummings model in cavity quantum electrodynamics (QED)~\cite{eberly1980periodic, rempe1987observation,meunier2005rabi}. It serves as clear evidence of quantum coherence and the superposition of quantum states.
It is more readily observed in systems with fewer degrees of freedom, such as single-atom or effectively single-particle systems, where simpler dynamics and longer coherence times make the effect more pronounced. For example, in the Jaynes-Cummings model, a two-level atom interacts with a quantized electromagnetic field mode, producing predictable revival patterns. Similar effects have been seen in superconducting circuits~\cite{kirchmair2013observation} and cold atom systems like Bose-Einstein condensates~\cite{Greiner2002}. In contrast, observing \textit{collapse-and-revival} in strongly-interacting many-body quantum systems presents significant challenges. In these systems, the complex interactions and increased degrees of freedom usually lead to rapid scrambling of quantum information and loss of coherence.}

\LL{As demonstrated in the main text, the \textit{collapse-and-revival} behaviour observed here originates from constrained qubit dynamics. Specifically, the Rydberg blockade effect causes a delayed rotation of spins near the central flipped spin, which creates regions of delayed spin rotation that propagate outwards. Figure~\ref{Fig:S0} highlights the relationship between constrained qubit dynamics and Holevo information, employing three dynamic indicators to characterize two key concepts: retardation and distinguishability.}

\LL{The background heatmaps in Fig.~\ref{Fig:S0} display the numerical simulation results of the Holevo information under the PXP Hamiltonian, which mirrors Fig. 4c in the main text. Figure~\ref{Fig:S0}a,b depict the oscillations in the expectation values of $\langle \sigma^y \rangle$ and $\langle \sigma^z \rangle$, respectively. These oscillations indicate that when the Holevo information approaches zero inside the light cone, the Bloch vectors $\vec{\sigma}$ for each qubit become indistinguishable, sharing the same $\langle \sigma^y \rangle$ and $\langle \sigma^z \rangle$ values.}

\LL{To provide a clearer understanding of the peaks in Holevo information, we introduce a new dynamic indicator:
\begin{equation}
f(\vec{\sigma},t) = \int_{0}^{t}\mathrm{d}\{\mathrm{Arg}[\vec{\sigma}(\tau)]\}\mathrm{d}\tau - \lambda \Omega t,
\end{equation}
where the first term, $\int_{0}^{t}\mathrm{d}\{\mathrm{Arg}[\vec{\sigma}(\tau)]\}\mathrm{d}\tau$, represents the total rotation angle of the Bloch vector $\vec{\sigma}$ in the YZ-plane. This provides a more fundamental perspective on the spin dynamics than the individual expectation values of $\langle \sigma^y \rangle$ and $\langle \sigma^z \rangle$.
To restrict the range of the rotation angle which accumulates almost monotonically over time, we subtract the term $\lambda \Omega t$ in the expression for $f(\vec{\sigma},t)$, where $\Omega$ is the Rabi frequency in the PXP Hamiltonian. The factor $\lambda=1.32$ is extracted from the slope of a linear fit with all the simulation data for the total rotation angle of each qubit with two initial states $\ket{\mathbb{Z}_2}$ and $\sigma_c^x \ket{\mathbb{Z}_2}$. As shown in Fig.~\ref{Fig:S0}c, the retardation can be extracted directly from the difference between the blue and red curves (highlighted by the green-filled intervals), which is the source of distinguishability in the $\langle \sigma^y \rangle$ and $\langle \sigma^z \rangle$ measurements. Within the light cone, the time delay remains nearly constant, but the rotation angle retardation exhibits a periodic \textit{collapse-and-revival} pattern.}

\LL{It is clear that both the retardation and Holevo information follow similar \textit{collapse-and-revival} dynamics. This analysis of constrained qubit dynamics provides insights into the mechanisms behind quantum information propagation and the \textit{collapse-and-revival} behaviour in dynamically constrained systems. The dynamic indicator, $f(\vec{\sigma},t)$, offers a way to visualize and quantify the relationship between spin dynamics and information flow. These results contribute to a better understanding of quantum information behaviour in constrained systems.}

\subsection{Distinguishing scar state oscillations and quantum information collapse-and-revival}

\LL{In our experiment, we observe a clear light-cone structure in the dynamics of both OTOCs and Holevo information, with periodic \textit{collapse-and-revival} behaviour within the light cone. Here, we show that these \textit{collapse-and-revival} dynamics in quantum information are relevant, but not equivalent to the previously discovered oscillations of quantum scar states wavefunction under the PXP Hamiltonian evolution.
The relevance between these two phenomena primarily stems from the fact that the physical mechanism driving both the oscillation of quantum scar states' wavefunctions and the quantum information \textit{collapse-and-revival} observed in our experiment originates from the kinetic constraints in the PXP model.
Their distinction arises because the dynamics of OTOCs and Holevo information are not confined to the scarred subspace but also incorporate contributions from thermal eigenstates. 
For instance, in the measurement of Holevo information dynamics, the $\sigma^x_{c}$ operation, which flips the central spin in the chain, introduces a mixture of scarred subspace and thermal eigenstate bath contributions. Such a mixture precludes attributing the periodic behaviour of Holevo information solely to the eigenstate decomposition of the initial state. Therefore, the observed information backflow within the light cone cannot be simply attributed to the oscillation of quantum scar states wavefunction.}

\begin{figure}
  \centering
  \includegraphics[width=0.7\textwidth]{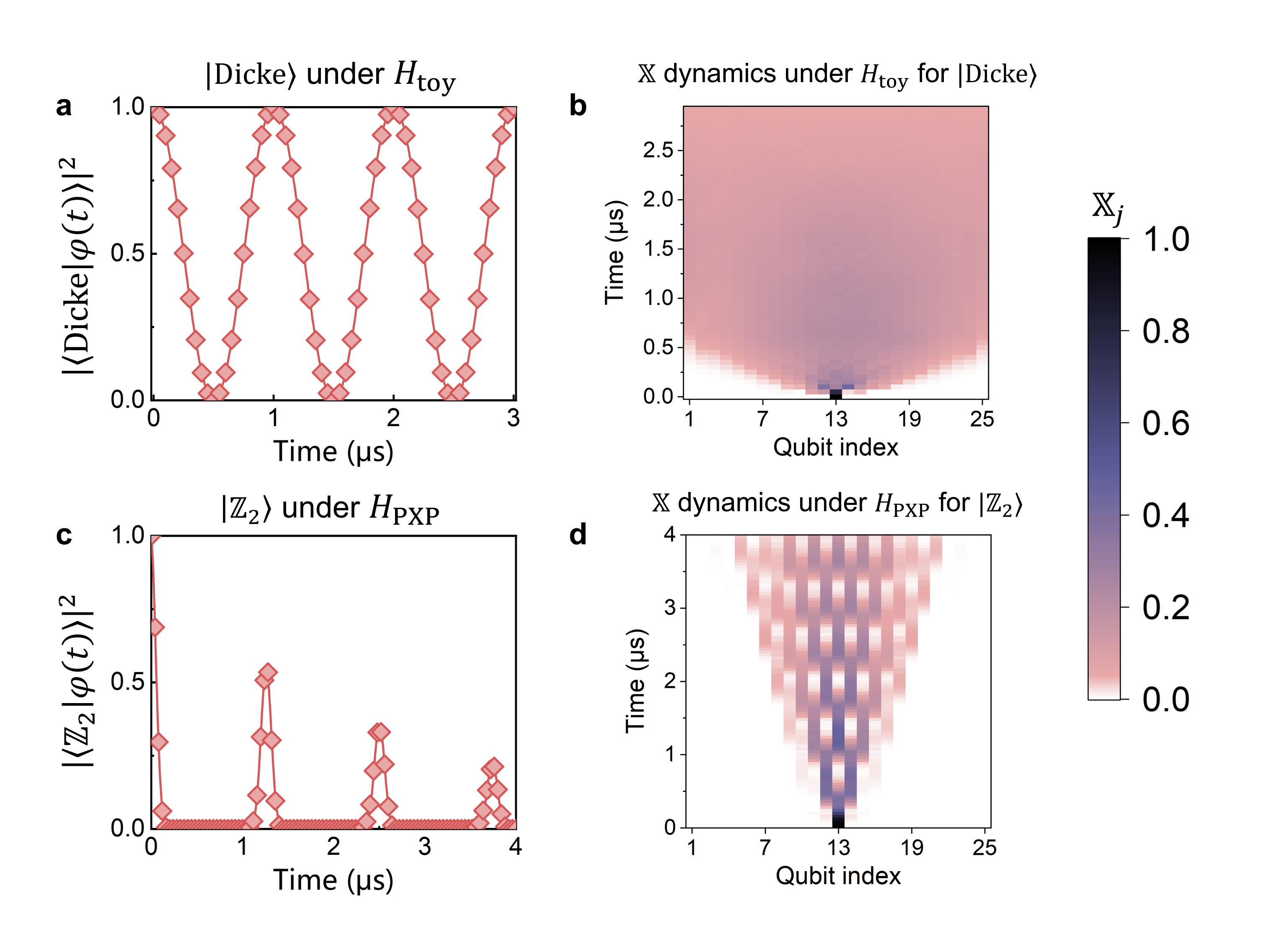}  
  \caption{
    \textbf{Comparison of quantum information dynamics in scar states between the toy model and the PXP model.} \LL{
\textbf{a}, Perfect scarred state wavefunction oscillations under ideal toy model Hamiltonian $H_{\text{toy}}$ evolution.
The simulated wavefunction overlap between the evolved state $\ket{\phi(t)}$ and the initial state $\ket{\text{Dicke}}$ is shown, as a function of evolution time $t$. The state $\ket{\phi(t)}$ is obtained after evolving the initial $\ket{\text{Dicke}}$ state under the ideal toy model Hamiltonian $H_{\text{toy}}$ for time $t$.
\textbf{b}, Numerically simulated spatio-temporal dynamics of the Holevo information for $\ket{\text{Dicke}}$ and $\sigma_c^x\ket{\text{Dicke}}$ initial states, under ideal toy model Hamiltonian $H_{\text{toy}}$ evolution. 
\textbf{c}, Damped scarred state wavefunction oscillations under ideal PXP model Hamiltonian $H_{\text{PXP}}$ evolution.
     The simulated wavefunction overlap between the evolved state $\ket{\phi(t)}$ and the initial state $\ket{\mathbb{Z}_2}$ is shown, as a function of evolution time $t$. The state $\ket{\phi(t)}$ is obtained after evolving the initial $\ket{\mathbb{Z}_2}$ state under the ideal PXP model Hamiltonian $H_{\text{PXP}}$ for time $t$.
    \textbf{d}, Numerically simulated spatio-temporal dynamics of the Holevo information for $\ket{\mathbb{Z}_2}$ and $\sigma_c^x\ket{\mathbb{Z}_2}$ initial states, under ideal PXP model Hamiltonian $H_{\text{PXP}}$ evolution.
  }}
  \label{Fig:SI_toy}
\end{figure}

\LL{
Building on the above discussion, it becomes clear that the quantum information \textit{collapse-and-revival} observed in our experiment is not equivalent to the oscillations of quantum scar states wavefunctions under the Hamiltonian evolution. This raises an intriguing question: more generally, does the presence of quantum scar state oscillations always imply the existence of quantum information \textit{collapse-and-revival}? Furthermore, could there be physical systems where quantum scar state oscillations occur without any accompanying quantum information \textit{collapse-and-revival}? To explore these questions, we turn to a toy model proposed by Choi \textit{et al.}~\cite{choi2019emergent}, described by the Hamiltonian:
}

\begin{equation}
H_{\text{toy}}= \frac{\Omega}{2}\sum_i\sigma^x_i+ \sum_{i}V_{i-1,i+2}P_{i,i+1}  
\end{equation}

\LL{
Here $P_{i,j}= (1-\vec\sigma_{i}\cdot \vec\sigma_{j})/4$ is the projection operator onto the singlet state of spins at sites $i$ and $j$, and $V_{i,j}= \sum_{\mu\nu} J_{ij}^{\mu\nu} \sigma_i^\mu \sigma_j^\nu$ represents an arbitrary long-range interaction between spins at sites $i$ and $j$. Dicke states, expressed as $\ket{s=L/2,S^{x}=m_x}$, are the scarred eigenstates of $H_{\text{toy}}$, as the interaction term does not act on these states ($P_{i,j}\ket{s=L/2,S^{x}=m_x}=0$).
}

\LL{
We perform numerical simulations for this toy model with $L=25$ spins with periodic boundary conditions, using the parameters $\Omega=2\pi\times\SI{1}{\MHz}$ and $V_{i,j}=J(\sigma_i^x\sigma_j^y + \sigma_i^y\sigma_j^x)$ with $J=2\pi\times\SI{2}{\MHz}$. When initialized in the scarred Dicke state $\ket{\text{Dicke}}=\ket{s=L/2,S^{z}=-L/2}= \ket{\downarrow\downarrow\cdots\downarrow}$, the system exhibits perfect scarred state wavefunction oscillations (Fig.~\ref{Fig:SI_toy}a).
Next, we explore quantum information scrambling and transport within the toy model using the scar state. Similar to our study of Holevo information for the $\ket{\mathbb{Z}_2}$ scar state in the PXP model, we applied a central spin flip to the scarred Dicke state in the toy model, denoted as $\sigma^x_c\ket{\text{Dicke}}$. We then simulated the evolution of both $\sigma^x_c\ket{\text{Dicke}}$ and $\ket{\text{Dicke}}$ under the toy model Hamiltonian. From these simulations, we obtained the spatio-temporal evolution of Holevo information, allowing us to investigate how quantum information propagates and scrambles in this system.
Figure~\ref{Fig:SI_toy}b shows the simulated Holevo information dynamics, with clear evidence of a rapid, global scrambling. 
This global scrambling behaviour differs significantly from the kinetically constrained PXP model, where the spatial-temporal \textit{collapse-and-revival} of quantum information is very pronounced (Fig.~\ref{Fig:SI_toy}d).
In the scar state $\ket{\text{Dicke}}$ of the $H_{\text{toy}}$, initially encoded quantum information is quickly lost to the environment without revival.
}

\LL{
These results show that, under the toy model Hamiltonian $H_{\text{toy}}$, while the scarred state $\ket{\text{Dicke}}$ exhibits perfect wavefunction oscillations, no \textit{collapse-and-revival} of quantum information occurs. This suggests that the oscillatory behaviour of quantum scar states does not necessarily coincide with periodic quantum information backflow. The quantum information spatial-temporal \textit{collapse-and-revival} dynamics observed in this work is likely a very unique feature resulting from the kinetic constraints imposed by the Rydberg blockade effect.
}

\nolinenumbers

\bibliographystyle{naturemag}
\bibliography{refs}